\documentclass[%
 reprint,
 superscriptaddress,
 amsmath,amssymb,
 aps,
 prb,
 floatfix,
]{revtex4-2}
\usepackage{graphicx}
\usepackage{dcolumn}
\usepackage{bm}

\usepackage[mathlines]{lineno}

\usepackage{hyperref}
\hypersetup{
  colorlinks   = true,
  citecolor    = blue,
  linkcolor    = blue,
  urlcolor     = blue
}
\makeatletter 
\providecommand{\href@noop}[2]{#2}
\providecommand{\Eprint}[2]{#2}
\makeatother

\usepackage{amssymb,amsmath}
\usepackage{newtxtext,newtxmath}
\usepackage{xspace}
\DeclareFontFamily{OT1}{pzc}{}
\DeclareFontShape{OT1}{pzc}{m}{it}{<-> s * [1.2] pzcmi7t}{}
\DeclareMathAlphabet{\mathpzc}{OT1}{pzc}{m}{it}

\usepackage[T1]{fontenc}
\usepackage[utf8]{inputenc}
\usepackage{amsmath}
\usepackage{url}
\usepackage{hyperref}

\usepackage{siunitx}
\newcommand{\refSIonPeff}{%
    \ref{SI:Peff}
}
\newcommand{\refSIonHeff}{%
    \ref{SI:Heff}
}
\newcommand{\refSIonMCmethod}{%
    \ref{SI:MC}
}
\newcommand{\refSIonMCLF}{%
    \ref{SI:MC-LF}
}
\newcommand{\refSIonMCHF}{%
    \ref{SI:MC-HF}
}
\newcommand{\refSIonTN}{%
    \ref{SI:tensor-network}
}
\newcommand{\refSIonInterlayer}{%
    \ref{SI:Interlayer}
}

\newcommand{\CrystalTNN}{TNN$\cdot$CH$_3$CN\xspace}
\newcommand{\up}{\uparrow}
\newcommand{\dw}{\downarrow}
\newcommand{\ket}[1]{\bigl\lvert{#1}\bigr\rangle}
\newcommand{\bra}[1]{\bigl\langle{#1}\bigr\rvert}

\usepackage[dvipsnames]{xcolor}
\usepackage{xcolor}

\usepackage{titlesec}
\usepackage{ragged2e}

\begin{document}

\title{%
    Designing electronic magnetoelectric matter with organic quantum spin trimers
}%

\author{%
    Yuko Hosokoshi
}%
\email{yhoso@omu.ac.jp}%
\affiliation{%
    Department of Physics, Graduate School of Science, Osaka Metropolitan University, Osaka 558-8585, Japan
}%

\author{%
    Christopher P. Aoyama
}%
\affiliation{%
    Department of Physics, University of Florida, Gainesville, Florida 32611-8440, USA
}%

\author{%
    Zhuowei Zhang
}%
\affiliation{%
    School of Physics and Astronomy, Shanghai Jiao Tong University, Shanghai 200240, China
}%

\author{%
    Toshio Ono
}%
\affiliation{%
    Department of Physics, Graduate School of Science, Osaka Metropolitan University, Osaka 558-8585, Japan
}%

\author{%
    Kosuke Takada
}%
\affiliation{%
    Department of Physical Science, Osaka Prefecture University, Sakai, Osaka 599-8531, Japan
}%

\author{%
    Ayaka Higashiguchi
}%
\affiliation{%
    Department of Physical Science, Osaka Prefecture University, Sakai, Osaka 599-8531, Japan
}%

\author{%
    Seitaro Iisaka
}%
\affiliation{%
    Department of Physical Science, Osaka Prefecture University, Sakai, Osaka 599-8531, Japan
}%

\author{%
    Koudai Yamasaki
}%
\affiliation{%
    Department of Physics, Graduate School of Science, Osaka Metropolitan University, Osaka 558-8585, Japan
}%

\author{%
    Hironori Yamaguchi
}%
\affiliation{%
    Department of Physics, Graduate School of Science, Osaka Metropolitan University, Osaka 558-8585, Japan
}%

\author{%
    Shengzhi Zhang
}%
\affiliation{%
    National High Magnetic Field Laboratory, Florida State University, Tallahassee, Florida 32310, USA
}%

\author{%
    Mohammad Irfan
}%
\affiliation{%
    National High Magnetic Field Laboratory, Florida State University, Tallahassee, Florida 32310, USA
}%
\affiliation{%
    Department of Physics, Florida State University, Tallahassee, Florida 32306, USA
}%

\author{%
    Minseong Lee
}%
\affiliation{%
    National High Magnetic Field Laboratory, Los Alamos National Laboratory, NM 87545, USA
}%
\affiliation{%
    National High Magnetic Field Laboratory, Florida State University, Tallahassee, Florida 32310, USA
}%

\author{%
    Eun Sang Choi
}%
\affiliation{%
    National High Magnetic Field Laboratory, Florida State University, Tallahassee, Florida 32310, USA
}%

\author{%
    Yasuyuki Shimura
}%
\affiliation{%
    Department of Quantum Matter, Hiroshima University, Higashi-Hiroshima 739-8530, Japan
}%
\affiliation{%
    Institute for Solid State Physics, University of Tokyo, Kashiwa 277-8581, Japan
}%

\author{%
    Toshiro Sakakibara
}%
\affiliation{%
    Institute for Solid State Physics, University of Tokyo, Kashiwa 277-8581, Japan 
}%

\author{%
    Zhiyuan Xie
}%
\affiliation{%
    Department of Physics, Renmin University of China, Beijing 100872, China
}%

\author{%
    Hiroki Nakano
}%
\affiliation{%
    Graduate School of Science, University of Hyogo, Kamigori, Hyogo 678-1297, Japan
}%

\author{%
    Yasu Takano
}%
\affiliation{%
    Department of Physics, University of Florida, Gainesville, Florida 32611-8440, USA
}%

\author{%
    Cristian D. Batista
}%
\affiliation{%
    Department of Physics and Astronomy, University of Tennessee, Knoxville, Tennessee 37996, USA
}%
\affiliation{%
    Neutron Scattering Division, Oak Ridge National Laboratory, Oak Ridge, Tennessee 37831, USA
}%

\author{%
    Yoshitomo Kamiya
}%
\email{yoshitomo.kamiya@oist.jp}
\affiliation{%
    Theory of Quantum Matter Unit,
    Okinawa Institute of Science and Technology, Okinawa 904-0495, Japan
}%
\affiliation{%
    School of Physics and Astronomy, Shanghai Jiao Tong University, Shanghai 200240, China
}%

\date{\today}

\begin{abstract}
Magnetoelectric (ME) phenomena are commonly driven by spin–lattice coupling. Here we demonstrate a different route based on frustrated quantum spin trimers that intrinsically intertwine magnetic moments and electric dipoles. Using molecular design principles, we realize a weakly coupled lattice of equilateral $S=1/2$ spin trimers in the organic radical crystal TNN$\cdot$CH$_3$CN. In this material, correlated electronic fluctuations within each trimer generate electric dipoles, while geometrically frustrated intertrimer interactions organize them into collective ME states. Magnetization, thermodynamic, and dielectric measurements reveal multiple magnetic-field-induced phases, including the $1/3$-magnetization plateau marked by pronounced dielectric anomalies. Effective low-energy theories and numerical simulations show that these phenomena are driven by electronically generated trimer dipoles whose collective order is stabilized by frustration relief of the intertrimer interactions, establishing a direct connection between geometric frustration and emergent magnetoelectricity. Our results identify quantum spin trimers as multifunctional building blocks, providing a bottom-up route for designing correlated ME materials from electronically active quantum spin clusters.
\end{abstract}

\maketitle

\section{Introduction}

Designing materials in which electric and magnetic degrees of freedom are intrinsically intertwined is a central challenge in quantum materials research because it enables electric control of magnetism and magnetic control of polarization. In most known magnetoelectric (ME) and multiferroic materials, these phenomena originate from spin--lattice coupling that converts magnetic order into electric polarization~\cite{Khomskii2009,Tokura2014,Cheong2007,Bibes2008,Dong2015,Katsura2005,Sergienko2006,Mostovoy2024}. While highly successful, these mechanisms constrain the microscopic origin and tunability of the resulting responses. Identifying alternative routes toward magnetoelectricity therefore remains an important challenge.

A fundamentally different possibility emerges in frustrated Mott insulators, where correlated electronic motion can directly generate electric polarization. Virtual electronic fluctuations around frustrated plaquettes can redistribute charge and produce purely electronic electric dipoles even in the absence of lattice distortions or spin--orbit coupling~\cite{Bulaevskii2008}. This mechanism suggests that geometric frustration can generate emergent electric degrees of freedom from collective electronic correlations alone.

\begin{figure}
  \centering
  \includegraphics[width=\hsize]{
  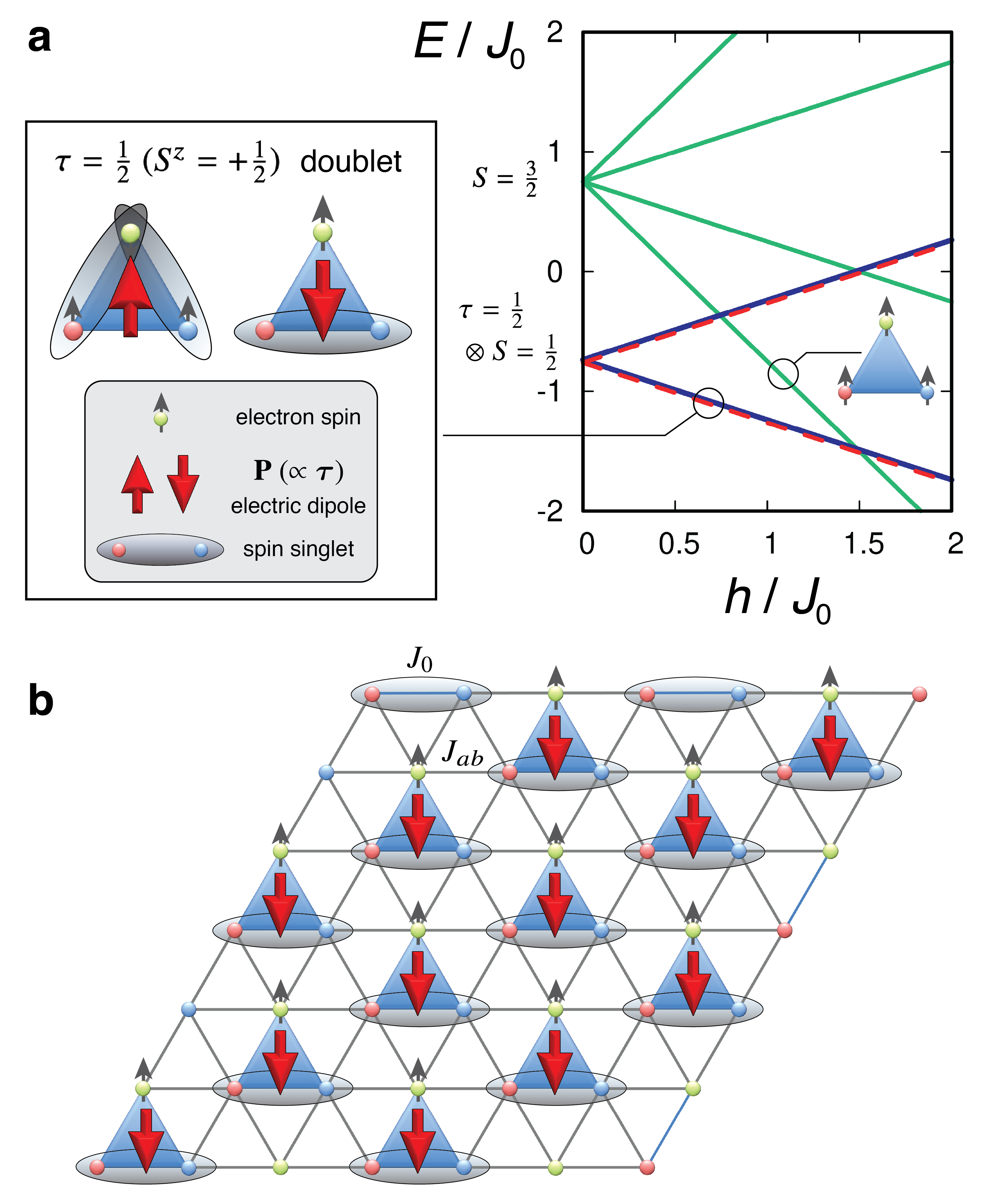
  }
  \caption{%
    \label{FIG1}
    \textbf{Spin, effective orbital degrees of freedom, and the emergent electric dipole in a quantum spin trimer.}
    \textbf{a,}
    A spin-1/2 Heisenberg trimer hosts a fourfold-degenerate $\tau=\frac{1}{2}\otimes S=\frac{1}{2}$ manifold at $h=0$ ($h = g\mu_\mathrm{B}H$, where $g$ is the $g$ factor and $\mu_\mathrm{B}$ is the Bohr magneton.). This low-energy manifold forms a minimal magnetoelectric unit carrying both a magnetic dipole and an electric dipole $\mathbf{P}\propto\bm{\tau}$ (\refSIonPeff). For $0<h<\frac{3}{2}J_0$, a magnetic field selects the $S^z=+\frac{1}{2}$ sector, leaving an effective $\tau=\frac{1}{2}$ doublet (inset). Ovals indicate spin singlets, while thick red arrows represent electric dipoles. The $\tau^x = +\frac{1}{2}$ state (left) can be seen as superposition of two configurations with different spin-singlet patterns.
    \textbf{b,} When the spin sector of the $\tau=\frac{1}{2}\otimes S=\frac{1}{2}$ multiplet is polarized by $h$, the emergent electric dipoles give rise to macroscopic electric polarization in a triangular lattice of weakly coupled trimers through frustration relief via a uniform singlet covering. The dominant intratrimer interaction $J_0 = 4t_0^2/U$ and the weaker intertrimer interaction $J_{ab} = 4t_{ab}^2/U$ are indicated, where $t_0$ and $t_{ab}$ are the corresponding hopping amplitudes in the underlying Hubbard description with the onsite interaction $U$ (\refSIonHeff).
  }%
\end{figure}

Among frustrated units, the equilateral quantum spin trimer represents the minimal building block that intrinsically combines magnetic and electric dipoles. As illustrated in Fig.~\ref{FIG1}{a}, a trimer of three antiferromagnetically coupled $S=1/2$ moments hosts both a magnetic dipole associated with its total spin and an emergent electric dipole arising from bond-dependent spin correlations within the unit~\cite{Bulaevskii2008}. In a simple picture, when two sites host a spin singlet and the third remains unpaired, electronic fluctuations create a subtle imbalance between virtual hopping processes into and out of the singlet-forming pair, inducing a redistribution of electronic charge density and an electric dipole tied to the singlet configuration (\refSIonPeff). The corresponding low-energy manifold consists of two $S=1/2$ doublets distinguished by an orbital pseudospin $\tau=1/2$ encoding the dipole orientation.

When such trimers are weakly coupled on a frustrated lattice, intertrimer interactions collectively organize the singlet configurations and associated dipoles into ME states, which can exhibit ferroelectricity for suitable interlayer couplings~\cite{Kamiya2012a}. In particular, polarizing the magnetic dipoles in the $S=1/2$ sector also drives spontaneous orbital ordering, relieving geometric frustration and stabilizing collective order of electronically generated
electric dipoles in the $1/3$-magnetization plateau (Fig.~\ref{FIG1}{b}). Quantum spin trimers therefore provide multifunctional building blocks in which magnetic and electric degrees of freedom are intrinsically intertwined.

Despite this conceptual promise, no material realization of weakly coupled equilateral quantum spin trimers exhibiting collective ME behavior has been identified so far. The main challenge is to simultaneously realize frustrated equilateral trimers, weak intertrimer couplings, and crystal structures that preserve the predominantly electronic character of the emergent electric dipoles.

\begin{figure*}
  \centering
  \includegraphics[width=0.9\hsize]{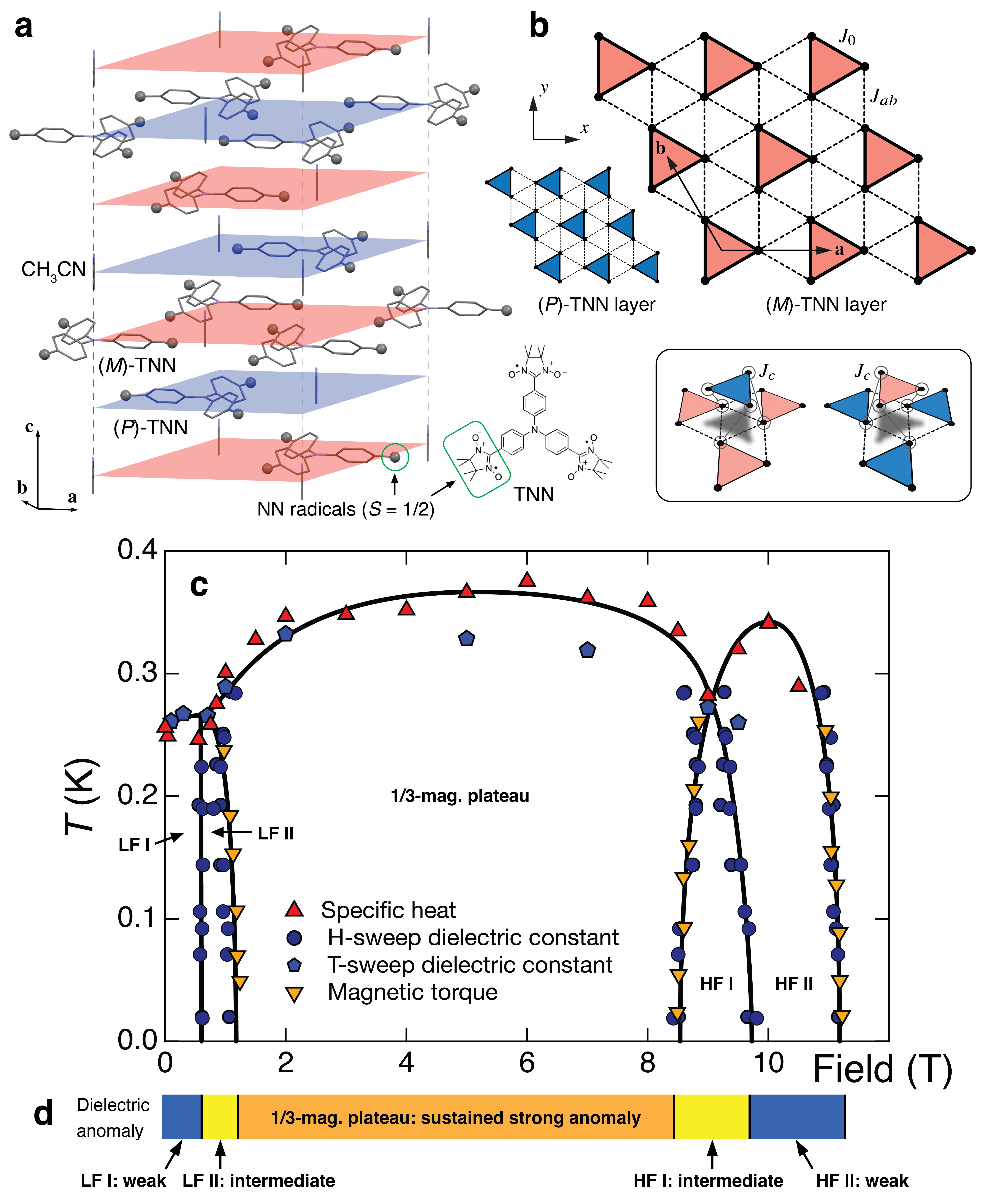}
  \caption{%
    \label{FIG2}
    \textbf{%
        Experimental realization of quantum spin trimers and the measured phase diagram.
    }%
    \textbf{a,}
    Unit cell structure of \CrystalTNN, composed of alternating (\textit{P})- and (\textit{M})-TNN layers of equilateral spin trimers formed by spin-$1/2$ nitronyl nitroxide (NN) radicals. In each NN radical, the unpaired-electron wavefunction is distributed primarily over the O--N--C--N--O moiety~\cite{Zheludev1994}.
    Spheres and short vertical bars in the main panel represent NN radicals and CH$_3$CN molecules, respectively.  
   \textbf{b,}
    Trimerized triangular-lattice model for \CrystalTNN, in which the two trimerization patterns in alternating layers represent the enantiomeric (\textit{P})- and (\textit{M})-TNN layers.
    The inset illustrates the minimal interlayer coupling geometry motivated by molecular orbital overlap, defining the three-dimensional connectivity between neighboring layers. 
    \textbf{c,}
    Magnetoelectric phase diagram of \CrystalTNN, with multiple magnetic-field-induced phases and phase boundaries identified from magnetic, thermodynamic, and dielectric measurements.
    \textbf{d,}
    Qualitative characterization of the field-induced phases based on the observed dielectric anomalies at finite-temperature phase transitions (see Fig.~\ref{FIG3}{e}).
    }%
\end{figure*}

Here we achieve this in the organic radical crystal \CrystalTNN, where equilateral $S=1/2$ spin trimers in nitronyl nitroxide triradicals form a stacked triangular lattice of weakly coupled trimers (space group $R3c$), as shown in Figs.~\ref{FIG2}{a} and \ref{FIG2}{b}. This structure closely realizes the theoretical lattice proposed in Ref.~\cite{Kamiya2012a}. Magnetization, thermodynamic, and dielectric measurements reveal a global ME phase diagram (Fig.~\ref{FIG2}{c}) with multiple magnetic-field-induced phases, including the $1/3$-magnetization plateau marked by pronounced dielectric anomalies. Complementary theoretical analyses identify electronically generated trimer dipoles as the active microscopic degrees of freedom organizing these phenomena: correlated electronic fluctuations within each trimer generate electric dipoles, and frustrated intertrimer interactions drive their collective ordering through frustration relief. Our results establish frustrated quantum spin trimers as a platform for designing correlated ME materials with emergent electronic dipoles.

\section{From Hubbard to emergent electric dipoles}
The key microscopic ingredient of the present work is an emergent electric dipole of purely electronic origin arising in an equilateral half-filled Hubbard trimer in the large-$U$ limit. The triangular geometry provides the minimal frustrated unit in which polar charge fluctuations emerge through odd-order virtual hopping processes. Standard perturbation theory reduces the system to a Heisenberg trimer with $J_0 = 4 t_0^2/U$, whose low-energy spectrum consists of two degenerate $S=1/2$ doublets. We label this additional twofold degeneracy by an orbital pseudospin $\tau=1/2$, giving a fourfold spin--orbital multiplet $\tau=\frac{1}{2}\otimes S=\frac{1}{2}$ (Fig.~\ref{FIG1}{a}). Integrating out the virtual processes yields the effective electric dipole operator
\begin{align}
    \hat{P}^{x(y)}_{\mathrm{eff}}
    = \pm 12\sqrt{3}\, ea 
    \left(\frac{t_0}{U}\right)^3 \hat{\tau}^{x(y)},
    \label{eq:Peff}
\end{align}
at third order in $t_0/U$, where $e$ is the electronic charge and $a$ is the side length of the trimer (\refSIonPeff)~\cite{Bulaevskii2008,Kamiya2012a}. The sign reflects the opposite orientations of the chiral conformers (\textit{P})- and (\textit{M})-TNN in \CrystalTNN (Figs.~\ref{FIG2}{a} and \ref{FIG2}{b}). 

The orbital pseudospins $\hat{\tau}^{x(y)}$ encode the asymmetry of spin correlations within a trimer. Specifically, $\hat{\tau}^{x(y)}$ are constructed from spin scalar products and form the two-dimensional $E$ irreducible representation of the $D_3$ symmetry group:
\begin{align}
    \hat{\tau}^{x} &= \frac{1}{3}\left[
        2 \mathbf{s}_{1}\!\cdot\!\mathbf{s}_{2} 
        - \mathbf{s}_{0}\!\cdot\!(\mathbf{s}_{1} + \mathbf{s}_{2})
    \right], \notag\\
    \hat{\tau}^{y} &= \frac{1}{\sqrt{3}}\,
      \mathbf{s}_{0}\!\cdot\!(\mathbf{s}_{1} - \mathbf{s}_{2}),
\end{align}
where $\mathbf{s}_{0}$--$\mathbf{s}_{2}$ are the spin operators of the trimer.
Thus, the emergent electric polarization $P^{x(y)} \propto \tau^{x(y)}$ originates from the asymmetry of the underlying singlet-bond texture within the trimer.

\begin{figure*}
  \centering
  \includegraphics[width=0.95\hsize]{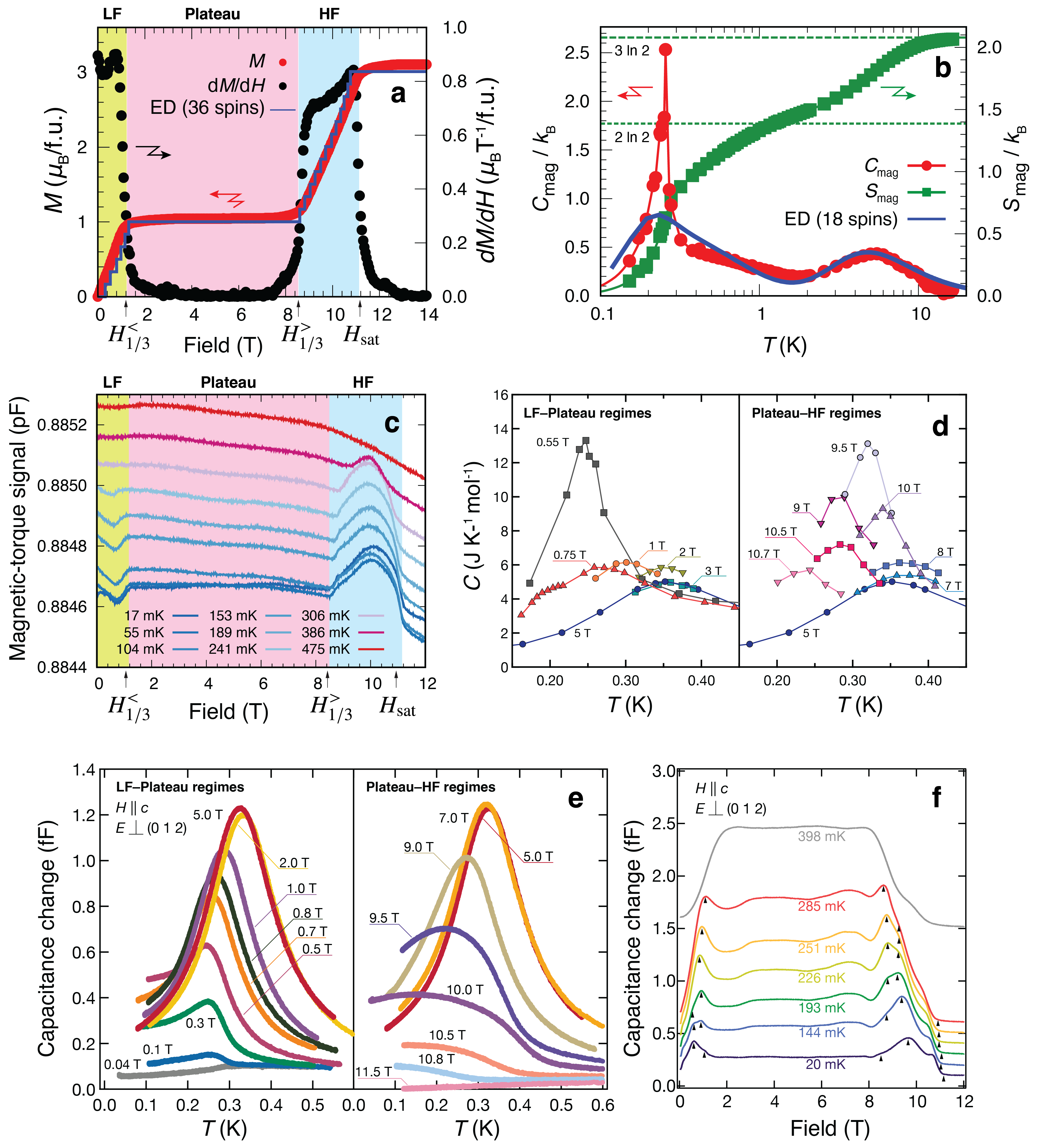}
  \caption{%
    \label{FIG3}
    \textbf{%
    Magnetic, thermodynamic, and dielectric signatures of field-induced magnetoelectric phases of \CrystalTNN.
    }%
    \textbf{a,} 
    Magnetization as a function of $H$ at $T = 0.13$~K, showing the 1/3-magnetization plateau between $H_{1/3}^{<} = 1.2$~T and $H_{1/3}^{>} = 8.5$~T and saturation at $H_\mathrm{sat} = 11.2$~T. Exact diagonalization (ED) data for $J_0 = 7.8$~K and $J_{ab}/J_0 = 0.12$ ($J_c = 0$) are overlaid. The three magnetic-field regimes below the saturation field $H_\mathrm{sat}$---low-field (LF), plateau, and high-field (HF)---are indicated above the panel.
    \textbf{b,}
    Magnetic specific heat $C_\mathrm{mag}$ and entropy $S_\mathrm{mag}$ at $H = 0$, compared with ED data at finite temperatures.
    \textbf{c,} 
    Magnetic-torque response measured with a capacitive cantilever at various temperatures.
    \textbf{d,}
    Temperature dependence of the total specific heat in magnetic fields, shown separately for the LF–plateau and plateau–HF regimes.
    \textbf{e,} 
    Temperature dependence of the dielectric capacitance change for selected magnetic fields $\mathbf{H} \parallel c$, shown separately for the LF–plateau and plateau–HF regimes, measured with an AC electric field $\mathbf{E} \perp (012)$. This geometry provides a finite in-plane electric-field component relative to the $ab$ layers.
    \textbf{f,}
    Magnetic-field dependence of the dielectric capacitance change for $\mathbf{H} \parallel c$ and $\mathbf{E} \perp (012)$.
    Triangles indicate anomalies that were used to construct the phase diagram shown in Fig.~\ref{FIG2}{c}. 
    The capacitance changes are shown relative to the zero-field capacitance at the base temperature.
  }%
\end{figure*}

\section{Experimental results}

\subsection{Realization of equilateral spin trimers}

Single crystals of \CrystalTNN (see~\ref{Methods:synthesis}) realize a weakly coupled lattice of equilateral quantum spin-$1/2$ trimers. The material consists of TNN molecules, a triphenylamine-based nitronyl nitroxide triradical in which three $S=1/2$ radicals form a $C_{3}$-symmetric structure (Fig.~\ref{FIG2}{a})~\cite{Nakano2005}. Structural refinements confirm that \CrystalTNN preserves both the equilateral trimer geometry and the crystalline $C_3$ symmetry (\ref{Methods:synthesis}). The resulting structure closely realizes the stacked triangular lattice of quantum spin trimers proposed in Ref.~\cite{Kamiya2012a}. A minor difference is that enantiomeric (\textit{P})- and (\textit{M})-TNN layers alternate along the $c$ axis, producing a six-layer periodicity rather than simple vertical stacking (Figs.~\ref{FIG2}{a} and \ref{FIG2}{b}).

Magnetization measurements at 0.13~K (\ref{Methods:magnetization}) reveal a pronounced $1/3$-magnetization plateau between $H_{1/3}^{<} = 1.2$~T and $H_{1/3}^{>} = 8.5$~T (Fig.~\ref{FIG3}{a}). This plateau is a hallmark of weakly coupled spin trimers: the magnetic field splits the fourfold-degenerate $S = 1/2$ spin--orbital multiplet, yielding $S^z = 1/2$ per trimer (Fig.~\ref{FIG1}{a}). Outside the plateau, in the low- and high-field (LF and HF) regimes, the nearly linear dependence of the magnetization $M(H)$ reflects intertrimer interactions. The good agreement between the experimental $M(H)$ data and exact diagonalization (ED) results for a single layer (36 spins) supports the quasi-two-dimensional nature of the material.

Magnetic specific heat $C_\mathrm{mag}$ measured at $H = 0$ (\ref{Methods:specific heat}) provides further support for this picture (Fig.~\ref{FIG3}{b}). A Schottky anomaly appears around 5.5\,K, while the magnetic entropy below this anomaly nearly plateaus at $S_\mathrm{mag}/k_\mathrm{B} \sim 2\ln 2$ per trimer, consistent with the fourfold spin--orbital multiplet. $C_\mathrm{mag}$ exhibits a sharp peak at 0.25\,K, signaling a bulk phase transition.

\subsection{Magnetic torque and in-field specific heat}

High-sensitivity torque magnetometry using a capacitive cantilever at 17--475\,mK (\ref{Methods:torque}) reveals clear anomalies at the two critical fields $H_{1/3}^{<}$, $H_{1/3}^{>}$, and the saturation field $H_\mathrm{sat}$ (Fig.~\ref{FIG3}{c}), consistent with the magnetization measurements and providing an unambiguous determination of the field-induced transitions.

Specific heat measurements in magnetic fields $\lesssim 11\,$T provide complementary information (Fig.~\ref{FIG3}{d}). Anomalies as a function of temperature appear not only in the low- and high-field regimes but also within the $1/3$-magnetization plateau. Because the spin sector is fully gapped in the plateau regime, these anomalies indicate ordering of the effective orbital degrees of freedom.

Together, the torque and specific-heat anomalies determine the overall thermodynamic phase boundaries shown in Fig.~\ref{FIG2}{c}.

\subsection{Magnetoelectric responses and the phase diagram}

A distinctive feature of quantum spin trimers is that virtual charge fluctuations coupled to the spin configuration generate emergent electric dipoles of purely electronic origin~\cite{Bulaevskii2008,Kamiya2012a}. Dielectric measurements in magnetic fields are therefore sensitive to the associated ME response (\ref{Methods:dielectric}). Using AC electric fields $\mathbf{E} \perp (012)$, which have a finite in-plane component relative to the $ab$ layers, we probe the in-plane dielectric response expected for this trimer-dipole mechanism. The dielectric response is nearly isotropic with respect to magnetic-field direction, consistent with the weak spin anisotropy expected for organic systems (\ref{Methods:dielectric}).

The temperature dependence of the dielectric constant change $\Delta \varepsilon$, measured through the capacitance change, reveals anomalies in all magnetic-field regimes (Fig.~\ref{FIG3}{e}), closely tracking the specific-heat anomalies. Near zero field, $\Delta \varepsilon(T)$ exhibits only a weak shoulder-like anomaly. With increasing field in the LF regime, this feature evolves into a pronounced peak for $\gtrsim 0.6$\,T and remains prominent throughout the $1/3$-magnetization plateau between $H_{1/3}^< = 1.2$\,T and $H_{1/3}^>  = 8.5$\,T. In the HF regime, the anomaly weakens again into a broad shoulder near saturation for $\gtrsim 9.7$\,T, until almost completely vanishing above $H_\mathrm{sat} = 11.2$\,T.

The isothermal field dependence, $\Delta \varepsilon(H)$, exhibits additional anomalies in both the LF and HF regimes (Fig.~\ref{FIG3}{f}), roughly coinciding with the crossover from weak shoulder-like features to sharper intermediate anomalies in $\Delta \varepsilon(T)$ (see Fig.~\ref{FIG2}{d}), and suggests either phase transitions or crossover phenomena (see Theory). Combining dielectric, specific-heat, torque, and magnetization measurements yields the phase diagram shown in Fig.~\ref{FIG2}{c}. These results establish \CrystalTNN as a weakly coupled trimer system exhibiting multiple field-dependent ME phases, whose microscopic origin is addressed below by effective low-energy theories.

\section{Theory}

The magnetic-field evolution of TNN$\cdot$CH$_3$CN reveals how the low-energy spin trimer degrees of freedom reorganize across distinct correlated regimes in an external magnetic field. As the field progressively quenches the spin sector, the active low-energy degrees of freedom evolve from entangled spin--orbital states to collective electric dipoles and eventually to mobile orbital holes, producing three distinct mechanisms for magnetoelectricity within a single material platform.

\begin{figure*}
  \centering
  \includegraphics[width=\hsize]{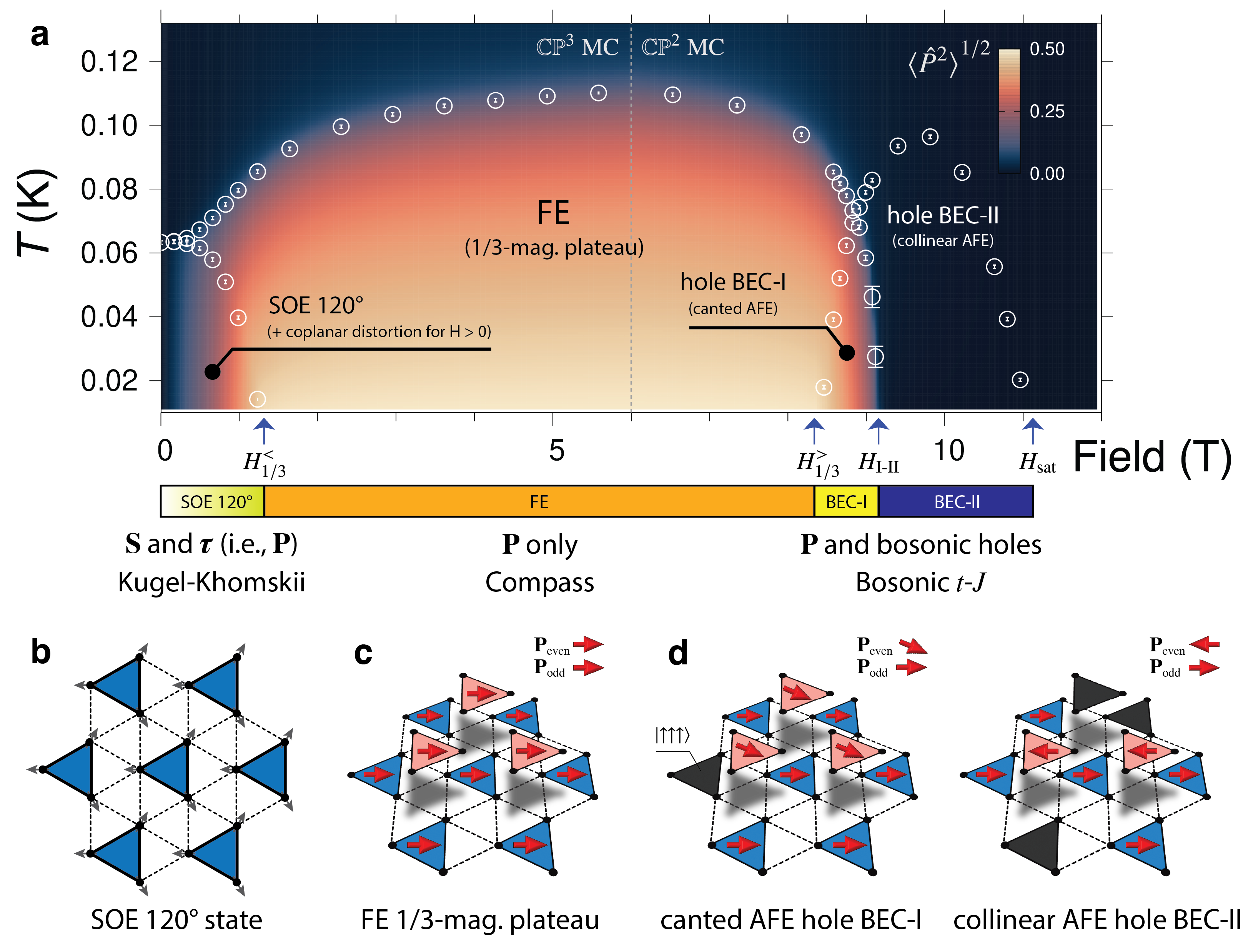}
  \caption{%
    \label{FIG4}
    \textbf{%
        Theoretical characterization of magnetic-field-dependent magnetoelectric (ME) phases in \CrystalTNN.
    }%
    \textbf{a,} 
    Monte Carlo phase diagram showing distinct field-dependent regimes, with the polarization $P \equiv \langle \hat{P}^2\rangle^{1/2}$ displayed in the background. The corresponding low-energy descriptions are summarized below, linking each field regime to its effective degrees of freedom.
    \textbf{b,} Low-field spin--orbital-entangled (SOE) 120$^\circ$ state, where coupled spin--orbital degrees of freedom induce the ME response.
    \textbf{c,} 
    Ferroelectrically-stacked three-dimensional layer-dipole order predicted in the $1/3$-magnetization plateau for the minimal interlayer coupling $J_c$.
    \textbf{d,} High-field Bose--Einstein-condensed (BEC) hole states induced by mobile orbital holes, giving rise to canted and collinear antiferroelectric (AFE) states in hole BEC-I and hole BEC-II phases, respectively, with associated kinetic ME effects.
    The exchange couplings are set to $J_0 = 7.3$~K, $J_{ab} = 1.1$~K, and $J_c = 0.3J_{ab}$.
    }%
\end{figure*}

In the LF regime (Figs.~\ref{FIG4}{a} and \ref{FIG4}{b}), spin and orbital degrees of freedom remain active and become intertwined, generating a spin--orbital-entangled (SOE) state with a linear ME response, where magnetic and electric dipoles fluctuate cooperatively. In the $1/3$-magnetization plateau (Figs.~\ref{FIG4}{a} and \ref{FIG4}{c}), the magnetic field quenches the spin sector, leaving an effective system of interacting electric dipoles whose ordering is driven by geometric frustration. In the HF regime (Figs.~\ref{FIG4}{a} and \ref{FIG4}{d}), fully polarized trimers behave as mobile orbital holes whose kinetic motion competes with orbital exchange interactions, stabilizing field-tunable canted and collinear antiferroelectric (AFE) states.

Projecting the intertrimer Heisenberg interactions $J_{ab}$ and $J_c$ (Fig.~\ref{FIG2}{b}) onto the low-energy manifold of each trimer yields effective models ($\hat{\mathcal H}_{\rm KK}$, $\hat{\mathcal H}_{\rm Comp}$, and $\hat{\mathcal H}_\text{b-\textit{tJ}}$ defined below; see \refSIonHeff), which reproduce the main experimental features and provide a unified microscopic description of the observed phases across the different magnetic-field regimes. Here, $J_c$ denotes the minimal interlayer exchange path motivated by the molecular orbital overlap; possible additional weak interlayer couplings are discussed in \refSIonInterlayer.

\subsection{Spin--orbital--entangled magnetoelectricity}

At low magnetic fields, the spin and orbital sectors remain active and become intertwined through intertrimer interactions, forming a collective SOE state. This regime is described by a Kugel--Khomskii-type Hamiltonian (\refSIonHeff),
\begin{align}
    \hat{\mathcal H}_{\rm KK}
    =
    \sum_{\mathbf r,\bm\delta}
    \hat J_{\bm\delta}(\bm\tau_\mathbf{r},\bm\tau_{\mathbf r+\bm\delta})\,
    \hat{\mathbf S}_{\mathbf r}\cdot
    \hat{\mathbf S}_{\mathbf r+\bm\delta}
    -
    h\sum_{\mathbf r}\hat S^z_{\mathbf r},
    \label{eq:KK}
\end{align}
where the bond operators $\hat J_{\bm\delta}(\bm\tau_\mathbf{r},\bm\tau_{\mathbf r+\bm\delta})$ depend on the orbital pseudospins and therefore couple magnetic and electric degrees of freedom.

To capture the resulting spin--orbital fluctuations on equal footing, we employ generalized classical $\mathbb{CP}^3$ Monte Carlo simulations, which naturally incorporate local spin--orbital entanglement through arbitrary linear combinations of the fourfold-degenerate multiplet (\ref{Methods:CP3/CP2} and \refSIonMCmethod). In contrast, approaches that artificially disentangle spin and orbital sectors produce spurious ground states (\refSIonMCLF).

The resulting phase diagram (Fig.~\ref{FIG4}{a}) reveals a spontaneous SOE order at zero field characterized by the composite order parameter $\hat{\mathbf{S}}_{\mathbf r}(\hat{\tau}^x_{\mathbf r}\pm i\hat{\tau}^y_{\mathbf r})$. In a magnetic field, the intertwined spin and orbital fluctuations generate a linear ME response $P\propto h$  (\refSIonMCLF)~\cite{Dzyaloshinskii1960,Astrov1960}. The ordered state is accompanied by a distorted 120$^\circ$ spin structure consistent with recent $\mu$SR measurements~\cite{PardoSainz2024}. The resulting SOE dome in the theoretical phase diagram (Fig.~\ref{FIG4}{a}) closely follows the experimental low-field phase boundary (Fig.~\ref{FIG2}{c}), indicating that the model captures the essential spin--orbital physics of this regime.

\subsection{Collective trimer-dipole order}
In the $1/3$-magnetization plateau regime, the low-energy degrees of freedom reduce to the orbital pseudospins $\hat{\bm{\tau}}$, which encode electric dipoles [see Eq.~(\ref{eq:Peff})]. This regime therefore realizes a frustrated lattice of interacting electric dipoles, described by an orbital compass model (\refSIonHeff),
\begin{align}
    \hat{\mathcal H}_{\rm Comp}
    =
    \sum_{\mathbf r,\bm\delta}
    J^\tau_{\bm\delta}
    \left(
    \hat{\bm\tau}_{\mathbf r}\cdot \mathbf n_{\mu(\bm\delta)}
    \right)
    \left(
    \hat{\bm\tau}_{\mathbf r+\bm\delta}\cdot \mathbf n_{\mu(\bm\delta)}
    \right),
    \label{eq:Hcomp}
\end{align}
where the bond-dependent easy axes $\mathbf n_{\mu(\bm\delta)}$ are imposed by the trimer geometry.

Microscopically, intertrimer exchange favors orbital configurations that avoid placing unpaired spins on the bonds connecting neighboring trimers~\cite{Kamiya2012a}. This constraint selects particular singlet coverings and thereby aligns the associated electric dipoles, as illustrated schematically in Fig.~\ref{FIG1}{b}. The resulting trimer-dipole order is thus a direct consequence of frustration relief.

\begin{figure}
  \centering
  \includegraphics[width=\hsize]{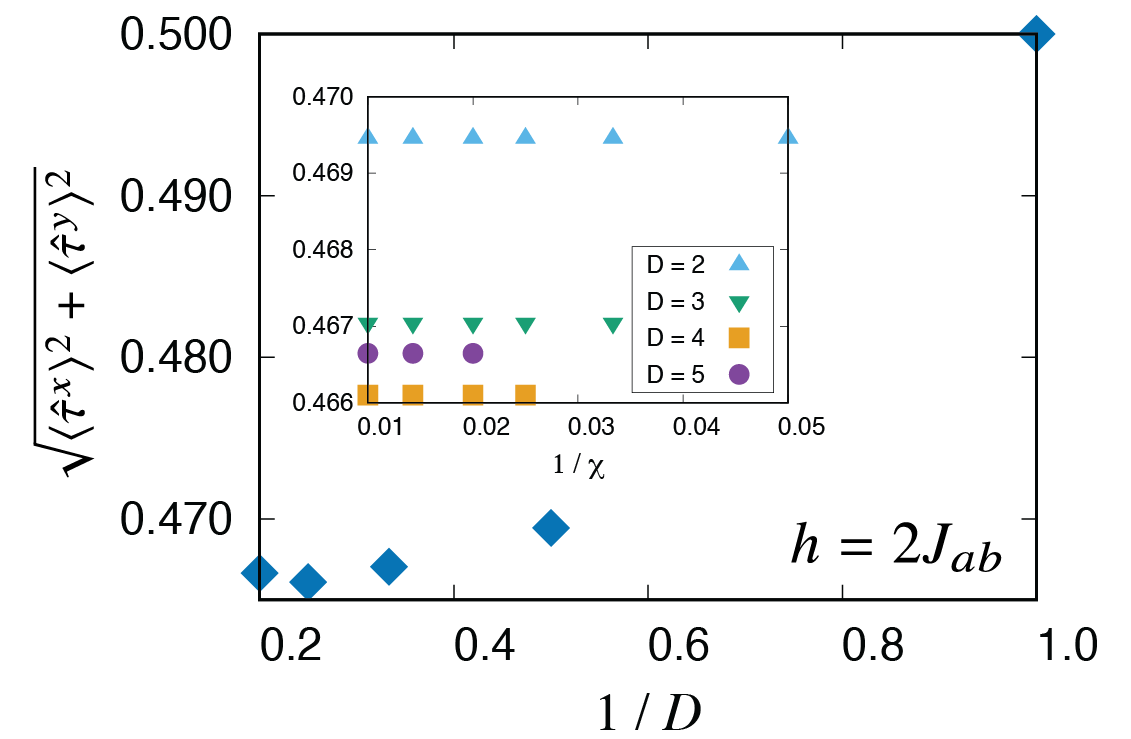}
  \caption{%
    \label{FIG_TN}
    \textbf{%
    Tensor-network confirmation of robust in-plane orbital order in the 1/3-magnetization plateau regime.
    }%
    A finite-$D$ analysis of the orbital order parameter at $h = 2.0J_{ab}$ (inset: finite-$\chi$ analysis), performed using the projected entangled simplex state (PESS) ansatz~\cite{Xie2014} for the 2D trimerized triangular lattice ($J_c = 0$), confirms the robustness of the orbital order. Here, $D$ is the bond dimension of the tensor-network ansatz and $\chi$ is the environment bond dimension. The $D=1$ result corresponds to the polarized product state obtained from the mean-field calculation.
  }%
\end{figure}

To examine this state beyond mean-field theory, we performed tensor-network calculations using the projected entangled simplex-state ansatz~\cite{Xie2014} in the two-dimensional (2D) limit (\ref{Methods:TN}). The results confirm robust in-plane orbital order ($\langle \hat{\tau}^z\rangle\simeq0$) with an order parameter close to the classical limit, $(\langle\hat{\tau}^x\rangle^2 + \langle\hat{\tau}^y\rangle^2)^{\frac{1}{2}} \simeq 0.466$, indicating that quantum fluctuations suppress the trimer-dipole order only weakly (Fig.~\ref{FIG_TN}). 

The predicted trimer-dipole ordering can naturally account for the pronounced dielectric and specific-heat anomalies observed throughout the plateau regime. For the minimal interlayer coupling $J_c$, this 2D dipole order develops into the three-dimensional ferroelectric structure illustrated in Fig.~\ref{FIG4}{c}. At the same time, because $J_c$ is among the smallest energy scales in the material, the actual stacking of layer dipoles may be sensitive to additional microscopic interlayer interactions (\refSIonInterlayer). Irrespective of this stacking pattern, however, the plateau regime demonstrates how geometric frustration can organize electronically generated trimer dipoles into a collective ordered state once the spin sector is quenched by the external magnetic field.

\subsection{Kinetic hole magnetoelectricity}

In the HF regime, the fully polarized trimer state $\ket{\uparrow\uparrow\uparrow}$ enters the low-energy manifold together with the orbital doublet states (Fig.~\ref{FIG1}{a}). 
Because the Pauli principle precludes virtual charge fluctuations in this configuration, it carries no electric dipole and acts as a bosonic orbital hole propagating through the trimer-dipole background. The resulting low-energy description is a bosonic analog of the $t$--$J$ model~\cite{Boninsegni2001,Batista2002,Aoki2009,Zhang.arXiv2409,Harris.arXiv2410},
\begin{align}
    \hat{\mathcal H}_\text{b-\textit{tJ}}
    = \hat{\mathcal H}_{\tau}
    + \hat{\mathcal H}_{\rm hop}
    - \mu_{\rm h} \hat N_{\rm h},
\end{align}
where $\hat{\mathcal H}_{\tau}$ extends the orbital compass model~\eqref{eq:Hcomp}, while $\hat{\mathcal H}_{\rm hop}$ describes the motion of orbital holes through the dipolar background  (\refSIonHeff). The hole chemical potential $\mu_{\rm h}$ varies linearly with the magnetic field and controls the hole density. The competition between exchange and hole kinetics underlies the resulting ME behavior.

The effective orbital exchange interactions $\hat{\mathcal H}_{\tau}$ favor the ferroelectric dipolar order predicted in the plateau phase, whereas the kinetic energy $\hat{\mathcal H}_{\rm hop}$ of mobile orbital holes favors uniform orbital order, analogous to kinetic-energy-driven ordering in Nagaoka ferromagnetism~\cite{Nagaoka1966}. In \CrystalTNN, the uniform orbital alignment corresponds to an AFE state because of the opposite orientations of the chiral conformers [see Eq.~\eqref{eq:Peff}]. Since the mobile quasiparticles are orbital holes directly coupled to electric polarization, this competition between exchange-driven dipolar order and hole kinetics produces a kinetic ME effect.

The resulting phase diagram obtained from $\mathbb{CP}^2$ Monte Carlo simulations (\ref{Methods:CP3/CP2}) is shown in Fig.~\ref{FIG4}{a}. Above the upper plateau boundary, $H > H_{1/3}^>$, orbital-hole condensation stabilizes two successive Bose--Einstein-condensed (BEC) hole phases, hole BEC-I and hole BEC-II (Fig.~\ref{FIG4}{d}). In hole BEC-I, the relative angle $\Delta\phi$ between dipoles in even and odd layers evolves continuously away from the ferroelectric plateau configuration by mutual canting, while hole BEC-II realizes the collinear AFE state favored by hole kinetics (\refSIonMCHF). The resulting peak and subsequent suppression of dielectric fluctuations are consistent with the peak-like structure observed in $\Delta\varepsilon(H)$ (Fig.~\ref{FIG3}{f}). The overall topology of the theoretical high-field phase diagram is also in good agreement with the experimentally observed sequence of overlapping high-field domes (Fig.~\ref{FIG2}{c}).

\section{Discussion}

Our combined experimental and theoretical investigation establishes \CrystalTNN as a weakly coupled equilateral quantum spin-trimer system hosting trimer electric dipoles~\cite{Bulaevskii2008,Kamiya2012a} generated by correlated electronic fluctuations. To the best of our knowledge, this is the first experimental realization in which multiple magnetoelectric (ME) responses can be microscopically understood in terms of collective behavior of such dipoles. The resulting phase diagram reveals how the low-energy trimer degrees of freedom reorganize in a magnetic field, producing a sequence of field-dependent ME regimes within a single microscopic platform (Fig.~\ref{FIG2}{c}). 
The theoretical phase diagram captures the overall topology and magnetic-field scales of the observed phase boundaries (Fig.~\ref{FIG4}{a}), providing a unified microscopic basis for the discussion below. 

A central result of this work is that the magnetic field selectively activates different sectors of the trimer low-energy manifold. The theory identifies three corresponding microscopic mechanisms: a spin--orbital-entangled (SOE) state with a linear ME response in the low-field (LF) regime, collective trimer-dipole order driven by geometric frustration in the $1/3$-magnetization plateau, and a kinetic ME effect generated by mobile orbital holes in the high-field (HF) regime.

In the LF regime, the semidome structure identified experimentally through specific-heat and magnetic-torque anomalies is consistent with the onset of the predicted SOE order. The associated enhancement of the dielectric response follows naturally from the linear ME coupling generated by intertwined spin and orbital fluctuations. The predicted distorted 120$^\circ$ spin structure is also consistent with recent $\mu$SR measurements~\cite{PardoSainz2024}, supporting the microscopic picture emerging from the generalized $\mathbb{CP}^3$ description.

The plateau regime provides the clearest setting for collective trimer-dipole ordering. Experimentally, this regime is marked by pronounced dielectric and thermodynamic anomalies at the phase transition. Theoretically, the orbital compass model stabilizes an ordered state of electronically generated electric dipoles through frustration relief (Fig.~\ref{FIG1}{b})~\cite{Kamiya2012a}, and tensor-network calculations in the two-dimensional (2D) limit confirm that the resulting 2D orbital order remains robust against quantum fluctuations (Fig.~\ref{FIG_TN}). For the minimal interlayer coupling $J_c$ suggested by molecular orbital overlap, this 2D orbital order develops into a three-dimensional (3D) ferroelectric structure, while the actual 3D stacking of layer dipoles may be sensitive to additional interlayer interactions (\refSIonInterlayer). Irrespective of this stacking pattern, however, the plateau demonstrates how geometric frustration can organize electronically generated trimer dipoles into a collective ordered state once the spin sector is quenched.

In the HF regime, the dielectric response $\Delta \varepsilon(H)$ exhibits a pronounced enhancement followed by a rapid suppression around $H = H_{\mathrm{I-II}}$, the critical field separating hole BEC-I and hole BEC-II. Microscopically, these hole BEC phases arise from the competition between exchange-driven dipole order and the kinetic energy of mobile orbital holes associated with locally fully polarized trimers. The resulting evolution of interlayer electric-dipole configurations constitutes a novel kinetic ME effect generated by itinerant orbital holes, producing the overlapping high-field domes in good agreement with the experimental observations.

Taken together, these results show that the experimentally observed anomalies across the magnetic-field phase diagram are naturally organized by different collective manifestations of electronically generated trimer dipoles. In an intuitive picture, these dipoles originate from an asymmetric pattern of singlet-like bonds within each trimer, reflecting the entangled structure of the low-energy trimer multiplet~\cite{Bulaevskii2008,Kamiya2012a}. The magnetic field then reshapes this common trimer-based building block into three distinct ME mechanisms: low-field SOE order, plateau trimer-dipole order, and high-field orbital-hole condensation.

More broadly, our results establish weakly coupled quantum clusters as a versatile platform for designing correlated multifunctional materials with strong and potentially tunable ME coupling arising from electronically generated trimer dipoles. The key ingredients identified in this work naturally generalize beyond trimers to larger cluster motifs and higher-rank multipolar objects~\cite{Haraldsen2016,Khomskii2010,Hayami2018,Zhao2025}. In this perspective, \CrystalTNN provides a prototype for a broader class of materials in which collective ME phenomena are driven by emergent electronic multipoles generated by cluster entanglement.

Recent advances in organic radical chemistry provide substantial control over molecular and supramolecular parameters governing spin frustration in polyradicals, such as the choice of central linker atoms, dihedral angles between radical planes, and crystal packing motifs~\cite{Hosokoshi1999,Blundell2004,Tang2024}. In the present context, these design strategies offer direct routes to tuning both the ratio $t_0/U$, which controls the magnitude of the emergent electric dipole, and the intertrimer interactions that determine the frustrated lattice geometry and the overall scale of the ordering temperature. Our findings therefore point toward a bottom-up strategy for designing quantum multifunctional materials with strong ME coupling driven by emergent electronic degrees of freedom.

\bibliography{ref} 

\renewcommand{\thesection}{\Roman{section}}
\section{Methods}
\renewcommand{\thesection}{Method}
\renewcommand{\thesubsection}{\Alph{subsection}}

\subsection{%
    Sample preparation and X-ray diffraction
    \label{Methods:synthesis}
}%
The organic neutral radical TNN was synthesized by the conventional method~\cite{Ulman1972, Hosokoshi1994} from 2,3-bis(hydroxylamino)-2,3-dimethylbutane and tris(4-formylphenyl)amine. Tris(4-formylphenyl)amine was synthesized in a two-step process using triphenylamine as a starting material via tris(4-bromophenyl)amine. The single crystals of \CrystalTNN were grown from an acetonitrile solution of TNN radicals.

X-ray diffraction data were collected on a Rigaku AFC-8R Mercury CCD RA-Micro7 diffractometer equipped with a Japan Thermal Engineering XR-HR 10K at system temperatures of 293\,K and 23\,K. The crystal structure was solved by direct methods and refined by the full-matrix least-squares technique using SHELX-97. The structural refinement was carried out using anisotropic and isotropic thermal parameters for nonhydrogen and hydrogen atoms, respectively.

\subsection{%
    Magnetization measurements
    \label{Methods:magnetization}
}%

The field dependence of the magnetization in \CrystalTNN was investigated using a force magnetometer, which enables static magnetization measurements~\cite{Sakakibara1994,Shimizu2021}. The superconducting magnet employed in the present study consists of a solenoid coil (main coil) and a gradient coil located inside the solenoid. The main coil generates uniform magnetic fields up to 15 T, whereas the gradient coil, operated with an independent power supply, provides a field gradient of up to 10 T/m at the sample position, superimposed on the uniform field. The magnetic force acting on the sample, $\boldsymbol{F} = (\boldsymbol{M}\cdot\nabla)\boldsymbol{B}$, was detected through variations in the capacitance of a force-sensing capacitor. This technique permits isothermal magnetization measurements under very-low-temperature conditions achieved with a dilution refrigerator. The absolute value of the magnetization $M$ was calibrated by comparison with results obtained using a SQUID magnetometer in fields up to 5 T at 4.2 K.

\subsection{%
    Specific heat measurements
    \label{Methods:specific heat}
}%
The specific heat was measured in magnetic fields up to 11\,T and at temperatures down to 130\,mK, using a superconducting magnet combined with a dilution refrigerator. The measurements were performed using the relaxation method, and details of the calorimeter used in the present experiment are described in \cite{Tsujii2002}. The magnetic field was applied perpendicular to the (012) plane, as in the magnetization measurements. The sample consisted of two co-aligned single crystals, with a total mass of 0.26\,mg. The magnetic specific heat $C_\mathrm{mag}$ of \CrystalTNN was obtained by subtracting the lattice contribution $C_\mathrm{lat}$ from the total specific heat $C_\mathrm{total}$. It was assumed that $C_\mathrm{lat}$ can be approximated by a combined function of the Debye and Einstein models:
\begin{align}
    C_\mathrm{lat} 
    \simeq 9k_\mathrm{B}\left(\frac{T}{\Theta_\mathrm{D}}\right)^3 
    \int_0^{\frac{\Theta_\mathrm{D} }{ T } }\!\!\!\!\mathrm{d}x
    \frac{ x^4 \mathrm{e}^{x}}{ \left( \mathrm{e}^{x} - 1 \right)^2 }
    + n \frac{ 3k_\mathrm{B} \left( \frac{\Theta_\mathrm{E}}{T} \right)^2 \mathrm{e}^{\frac{\Theta_\mathrm{E}}{T}} }{ \left(  \mathrm{e}^{\frac{\Theta_\mathrm{E}}{T}} - 1 \right)^2 },
\end{align}
with $\Theta_\mathrm{D} = 41.3$\,K, $\Theta_\mathrm{E} = 59$\,K, and $n = 2.9$. 
The parameters were determined such that $C_\mathrm{total}$     reproduces the measured specific heat and the entropy of the spin system $S_\mathrm{mag}$ calculated from $C_\mathrm{mag}$ approaches $3  k_\mathrm{B}\ln 2$ at high temperatures.

\subsection{%
    Magnetic torque measurement
    \label{Methods:torque}
}%
The magnetic torque $\boldsymbol{\tau}$ acting on a sample with a magnetic moment $\boldsymbol{M}$ in an external magnetic field $\boldsymbol{B}$ is given by $\boldsymbol{\tau} = \boldsymbol{M}\times \boldsymbol{B}$. When the magnetic moment is not aligned with the applied field, a torque is exerted on the sample. In general, organic radical magnets such as \CrystalTNN exhibit negligible magnetic anisotropy because of the inherently weak spin--orbit interaction. However, in magnetically ordered phases, magnetic dipole-dipole interactions lead to small magnetic anisotropy and hence a magnetic torque. When the sample is attached to a cantilever, the torque exerted by the magnetic field can be detected as the deflection of the cantilever. 
In our case, the cantilever forms a parallel-plate capacitor with a fixed electrode, and the magnetic-torque-induced deflection is detected as a change in capacitance with an Andeen-Hagerling AH 2700A capacitance bridge operated at 5\,kHz with a 15\,V$_\mathrm{rms}$ excitation. The sample used in this study was a single crystal, with a mass of 0.195\,mg, which was glued to the cantilever on its (012) plane with a Wakefield thermal compound.

\subsection{%
    Dielectric measurements 
    \label{Methods:dielectric}
}%

\begin{figure}
  \centering
  \includegraphics[width=0.75\hsize]{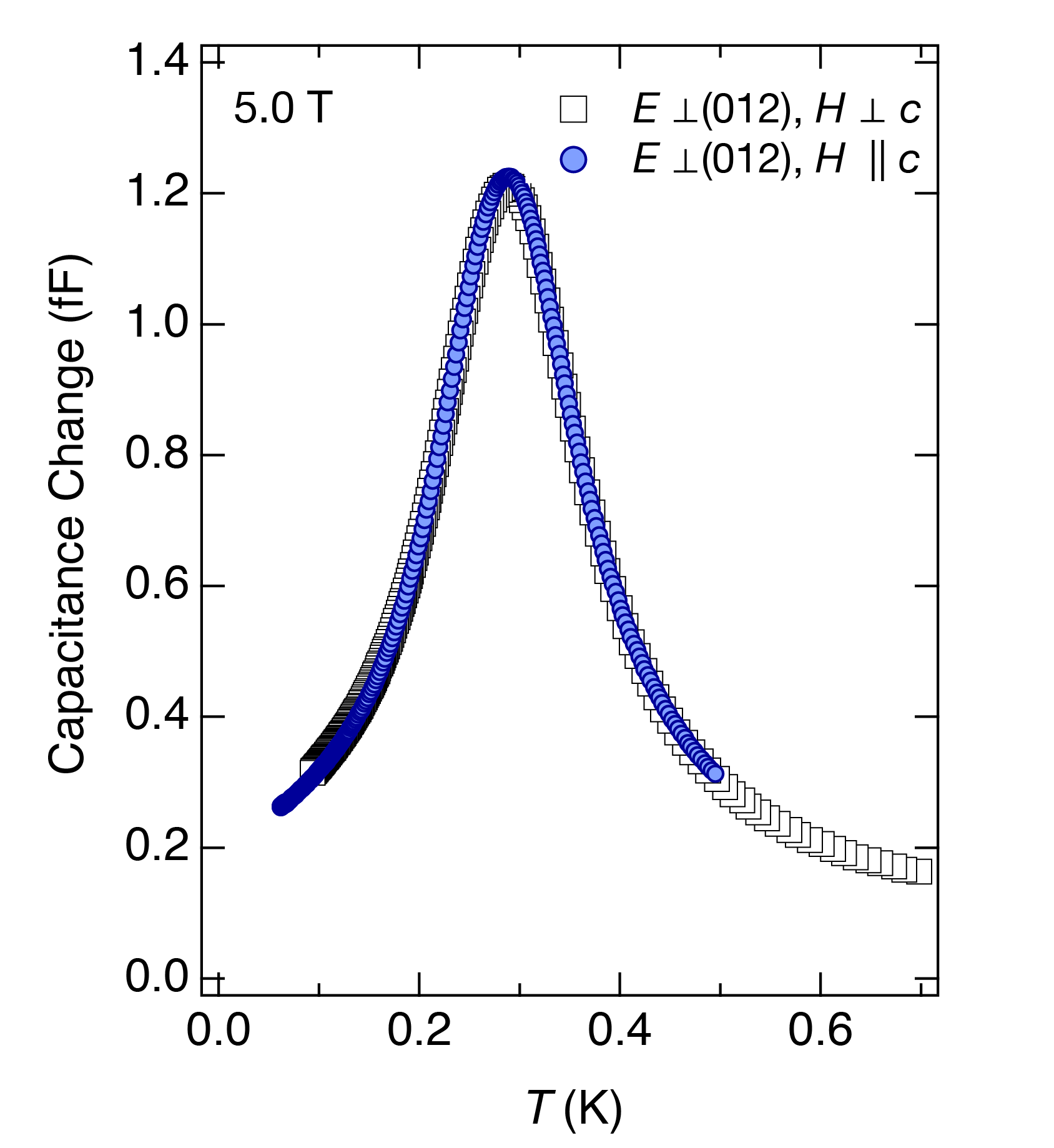}
  \caption{%
    \label{FIG_Isotropy}
    \textbf{%
    Temperature dependence of the dielectric capacitance change for $H \perp c$ and $H \parallel c$.
    }%
    The magnetic field is fixed at $H = 5$~T, and an AC electric field is applied along $\mathbf{E} \perp (012)$.
  }%
\end{figure}

Dielectric measurements were carried out by measuring the capacitance between two electrodes attached to opposite surfaces of the crystal. Two crystals were used for the measurements: Sample 1 was a 3.48\,mg single crystal used as grown without further shaping, while Sample 2 was a 0.44\,mg crystal with two faces cut parallel to the $ab$ plane. For Sample 1, electrodes were attached to the large (012) crystal faces, such that the electric field was applied perpendicular to these faces. This geometry provides a finite electric-field component within the $ab$ layers. For Sample 2, the electrodes were configured to apply the electric field parallel to the $c$ axis. Thin gold films were sputter-deposited as electrodes for Sample 1, whereas silver epoxy was used for Sample 2. The capacitance between the electrodes was measured using an Andeen-Hagerling AH 2700A capacitance bridge by applying an AC electric field at 5\,kHz. The samples with electrodes were glued on a rotating platform of a top-loading probe for a dilution refrigerator. Measurements were conducted in magnetic fields applied both parallel and perpendicular to the $c$ axis for both samples. The perpendicular direction for Sample 1 was $[\bar{1}\bar{2}0]$, whereas the exact perpendicular direction could not be determined for Sample 2. The probe was inserted into the mixing chamber of the dilution refrigerator and cooled down to 20\,mK. Magnetic fields up to 12\,T were applied using a superconducting magnet. The frequency dependence of the capacitance was examined under several magnetic-field and temperature conditions, and the measured capacitance was nearly independent of frequency over the range from 100\,Hz to 25\,kHz.
For comparison between magnetic-field directions, the capacitance change was measured at $H=5$~T for $\mathbf{H}\perp c$ and $\mathbf{H}\parallel c$ with $\mathbf{E}\perp(012)$. The two field orientations show similar temperature dependence, indicating only weak anisotropy with respect to the magnetic-field direction (Fig.~\ref{FIG_Isotropy}).

\subsection{%
    Generalized classical Monte Carlo simulations
    \label{Methods:CP3/CP2}
}%

To investigate the collective behavior of the effective models, we employed generalized classical Monte Carlo simulations formulated on complex projective manifolds. In the low-field regime, the active local Hilbert space corresponds to the four-dimensional spin--orbital manifold of a trimer, motivating a $\mathbb{CP}^3$ description that naturally incorporates spin--orbital entanglement. In contrast, the high-field regime is described by a reduced three-dimensional local manifold associated with orbital doublets and orbital holes, leading to a $\mathbb{CP}^2$ formulation. Similar methods were used for studying systems with multiple degrees of freedom per unit, e.g., spin-$1$ quantum magnets~\cite{Stoudenmire2009,Zhang2021,Seifert2022,Remund2022,Pohle2023} and multi-orbital systems~\cite{Iwazaki2023}. 

The simulations were performed using a combination of Metropolis sampling and over-relaxation updates of normalized local coherent states. Thermodynamic observables, electric polarization, and correlation functions were computed on finite three-dimensional clusters with periodic boundary conditions. The resulting phase diagrams reproduce the experimentally observed sequence of low-field, plateau, and high-field phases, including the evolution from ferroelectric to antiferroelectric interlayer configurations in the high-field regime. Further details of the generalized classical Monte Carlo framework, update procedures, and observables are provided in \refSIonMCmethod.

\subsection{%
    Tensor-network calculations
    \label{Methods:TN}
}

To characterize the 1/3-magnetization plateau regime beyond generalized classical approaches, we performed tensor-network calculations in the two-dimensional limit using the projected entangled simplex state (PESS) ansatz~\cite{Xie2014}. The calculations were implemented on the trimerized triangular lattice while preserving the underlying $C_3$ symmetry of the problem. The energy is evaluated using the corner transfer matrix renormalization group (CTMRG)~\cite{Orus2009,Corboz2014}. The tensor elements are optimized by automatic differentiation~\cite{Zygote.jl-2018, TensorOperations.jl}, following the strategy of Ref.~\cite{Liao2019}, starting from randomly initialized tensors. Further details of the tensor-network calculations are provided in \refSIonTN.

\vspace{10pt}
\renewcommand{\thesection}{\Roman{section}}
\setcounter{subsection}{0}

\begin{acknowledgments}
This work is dedicated to the memory of Lev N.~Bulaevskii and Daniel Khomskii, whose contributions to the field remain an enduring inspiration. 
A portion of this work was performed at the National High Magnetic Field Laboratory (NHMFL), which is supported by National Science Foundation Cooperative Agreement No. DMR-1157490 and the State of Florida.
CPA and YT were supported by the NHMFL UCGP program.
ZZ is grateful to Yangfeng Fu for insightful guidance on tensor network methods.
ZX was supported by the National Natural Science Foundation of China (Grant No. 12274458). 
HN was supported by JSPS KAKENHI (Grant No.~23K11125) and also acknowledges the use of the supercomputing resources of Fugaku provided by RIKEN through the HPCI System Research Projects (Project IDs: hp230114, hp230532, hp230537, hp250164, hp260092, and hp260183), the supercomputer systems at the Institute for Solid State Physics, University of Tokyo, the Supercomputing Division of the Information Technology Center, University of Tokyo, and Pegasus at the Center for Computational Sciences, University of Tsukuba.
YT thanks Aaron Aponick for useful conversations.
YK acknowledges support from JSPS KAKENHI Grant No.~26K07024, and from the Theory of Quantum Matter Unit, Okinawa Institute of Science and Technology Graduate University (OIST).
YK thanks T.~Shimokawa for critically reading the manuscript.
\end{acknowledgments}

\section*{End Notes}
\subsection*{Data Availability.}
All relevant data are available from the corresponding authors upon reasonable request.
\\
\noindent
($\ast$) yhoso@omu.ac.jp
\\
($\dag$) yoshitomo-kamiya@oist.jp

\subsection*{Author Contributions}
YH, 
YT, 
CDB, 
and YK 
conceived the project. 
YH, 
KT, 
SI, 
and HY 
synthesized and characterized the samples.
YH, 
TO, 
KT, 
AH, 
SI, 
HY, 
YS, 
and TS 
performed magnetization measurements.
YH,  
CPA, 
KT, 
HY, 
SZ,  
MI,  
and YT 
performed heat capacity measurements.
YH,  
CPA, 
KY,  
SZ,  
ESC, 
ML,  
and YT 
performed dielectric measurements.
YK 
performed Monte Carlo simulations. 
ZZ, 
ZX, 
and YK 
performed tensor network calculations. 
HN 
performed exact diagonalization calculations. 
YH, 
TO, 
CDB, 
and YK 
wrote the manuscript with contributions from all the authors.

\subsection*{Competing Interests}
The authors declare no competing interests.

\clearpage
\onecolumngrid

\renewcommand{\thesection}{Suppl.~\Alph{section}}
\renewcommand{\thesubsection}{\Alph{section}-\arabic{subsection}}
\renewcommand{\thesubsubsection}{\Alph{section}-\arabic{subsection}-\arabic{subsubsection}}
\renewcommand{\thefigure}{S\arabic{figure}}
\renewcommand{\theequation}{\Alph{section}\arabic{equation}}
\renewcommand{\thetable}{S\arabic{table}}
\renewcommand{\thepage}{\arabic{page}}

\setcounter{page}{1}
\setcounter{section}{0}
\setcounter{figure}{0}
\setcounter{table}{0}
\setcounter{equation}{0}

\renewcommand{\baselinestretch}{1.15}
\fontsize{11pt}{14.5pt}\selectfont
\setlength{\parindent}{1.5em}
\setlength{\parskip}{0pt}

\makeatletter

\renewcommand\section{\@startsection{section}{1}{\z@}%
  {-3.5ex \@plus -1ex \@minus -.2ex}%
  {2.3ex \@plus .2ex}%
  {\normalfont\large\bfseries}}

\@afterindenttrue

\renewcommand\subsection{\@startsection{subsection}{2}{-1.35em}%
  {-3.25ex \@plus -1ex \@minus -.2ex}%
  {1.5ex \@plus .2ex}%
  {\normalfont\large\bfseries}}

\renewcommand\subsubsection{\@startsection{subsubsection}{3}{-1.55em}%
  {-2.5ex \@plus -0.8ex \@minus -.2ex}%
  {1.0ex \@plus .2ex}%
  {\normalfont\normalsize\bfseries}}

\renewcommand{\@makecaption}[2]{%
  \vskip\abovecaptionskip
  \begingroup
  \normalfont\normalsize
  \noindent
  \setlength{\parindent}{0pt}%
  \setlength{\leftskip}{0pt}%
  \setlength{\rightskip}{0pt}%
  \setlength{\parfillskip}{0pt plus 1fil}%
  #1:\hspace{0.5em}#2\par
  \endgroup
  \vskip\belowcaptionskip
}

\makeatother

\vspace*{-0.5em}

\begin{center}
    {\large\bfseries
    Supplementary Information for
    \\
    ``Designing electronic magnetoelectric matter with organic quantum spin trimers''
    }
\end{center}

\section{%
    Emergent electric dipoles in weakly coupled equilateral trimers
    \label{SI:Peff}
}%
\setcounter{equation}{0}

\subsection{Hubbard trimer at half filling}
\begin{figure}[!t]
  \centering
  \includegraphics[width=0.95\hsize]{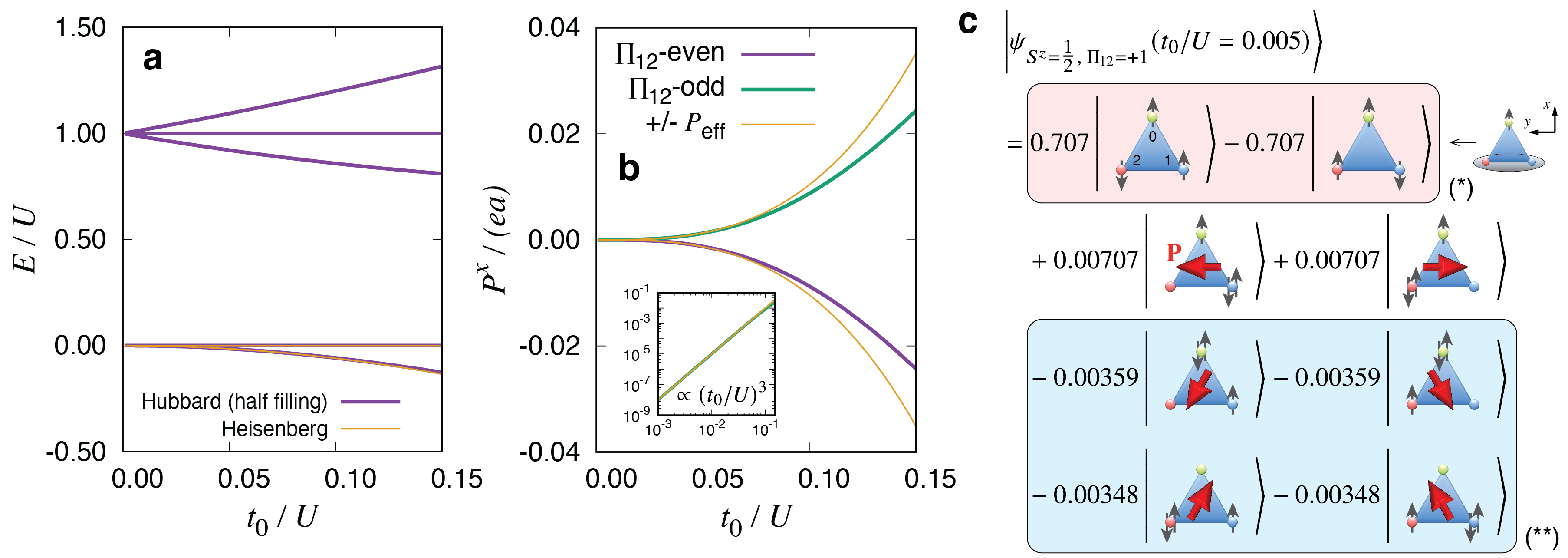}
  \vspace{0pt}
  \caption{%
    \label{FIG_2026_SI1}
    \textbf{%
        Low-energy properties of the Hubbard trimer at half filling.
    }%
    \textbf{(a)}
    Energy diagram for the Hubbard trimer at half filling, compared with the spectrum of the effective Heisenberg model.
    \textbf{(b)}
    Electric dipole moments of the ground states in the strong-coupling regime, compared with the eigenvalues of the effective operator $P_\mathrm{eff}$.
    \textbf{(c)} 
    Graphical representation of a wavefunction in the fourfold-degenerate ground-state manifold with $S^z = \frac{1}{2}$ and $\Pi_{12} = +1$. The first line, marked with (*), corresponds to the unperturbed Heisenberg-trimer ground state. The subleading components, marked with (**), generate local electric dipoles with a net $x$ component. The thick red arrows indicate local electric dipoles associated with virtual charge configurations in each basis component. 
  }%
\end{figure}

Building on the general theory presented in Ref.~\cite{Bulaevskii2008}, we summarize how virtual charge fluctuations generate an effective electric polarization operator for weakly coupled equilateral trimers. We start from an isolated Hubbard trimer at half filling and discuss the electric dipole per trimer from a wavefunction perspective. The Hamiltonian is
\begin{align}
    \hat{\mathcal{H}}^\mathrm{trimer}_\mathrm{Hub}
    = -t_0 \sum_{0 \le \mu \le 2}\sum_{\sigma=\up,\dw}
    \left(\hat{c}_{\mu+1,\sigma}^\dag \hat{c}_{\mu,\sigma}^{} + \mathrm{H.c.}\right) + \frac{U}{2}\sum_{0 \le \mu \le 2}
    \left(\hat{n}_\mu - 1\right)^2,
    \label{eq:SI:trimer:Hub}
\end{align}
where $t_0$ is the intratrimer hopping and $U$ is the onsite Coulomb interaction. In the strong-coupling limit, the model reduces to the Heisenberg trimer,
\begin{align}
   \hat{\mathcal{H}}_{\mathrm{spin}}^{(0)}
    = J_0 \Bigl( 
    \hat{\mathbf{s}}_{0}
     \cdot \hat{\mathbf{s}}_{1}
    + \hat{\mathbf{s}}_{1} \cdot \hat{\mathbf{s}}_{2}
    + \hat{\mathbf{s}}_{2} \cdot \hat{\mathbf{s}}_{0}
    \Bigr) 
    - h \sum_{0 \le \mu \le 2} \hat{s}^z_{\mu},
\end{align}
with $J_{0} = {4t_{0}^2}/{U}$. The low-energy spectrum of the Hubbard trimer is quantitatively reproduced by this effective spin model in the strong-coupling regime, as shown in Fig.~\ref{FIG_2026_SI1}{a}. 

The pseudospin-1/2 operators are defined by
\begin{align}
    \hat{\tau}^{x}_\mathbf{r} &= \frac{1}{3} 
    \left[
    2 \hat{\mathbf{s}}_{\mathbf{r},1} \cdot \hat{\mathbf{s}}_{\mathbf{r},2} - \hat{\mathbf{s}}_{\mathbf{r},0} \cdot \left( \hat{\mathbf{s}}_{\mathbf{r},1} + \hat{\mathbf{s}}_{\mathbf{r},2} \right)
    \right],
    \notag\\
    \hat{\tau}^{y}_\mathbf{r} &= \frac{1}{\sqrt{3}} \hat{\mathbf{s}}_{\mathbf{r},0} \cdot \left(
    \hat{\mathbf{s}}_{\mathbf{r},1} - \hat{\mathbf{s}}_{\mathbf{r},2}
    \right),
    \notag\\
    \hat{\tau}^{z}_\mathbf{r} &= \frac{2}{\sqrt{3}} 
    \left(
    \hat{\mathbf{s}}_{\mathbf{r},0} \cdot 
    \hat{\mathbf{s}}_{\mathbf{r},1} \times \hat{\mathbf{s}}_{\mathbf{r},2} \right).
\end{align}
They commute with both the total spin operator 
$\hat{\mathbf{S}}_{\mathbf{r}} = \sum_{0 \le \mu \le 2} \hat{\mathbf{s}}_{\mathbf{r},\mu}$ and $\hat{\mathcal{H}}_{\mathrm{spin}}^{(0)}$.
The fourfold-degenerate $S=\frac{1}{2}\otimes\tau=\frac{1}{2}$ multiplet of $\hat{\mathcal{H}}_{\mathrm{spin}}^{(0)}$ is given by
\begin{align}
    \ket{S^z = \tfrac{1}{2}, \tau^x = \tfrac{1}{2}} &= \tfrac{1}{\sqrt{6}}\, \Bigl( \ket{\up\up\dw} + \ket{\up\dw\up} - 2 \ket{\dw\up\up} \Bigr),
    \notag\\
    \ket{S^z = \tfrac{1}{2}, \tau^x = -\tfrac{1}{2}} &= 2 \hat{\tau}^y \ket{S^z = \tfrac{1}{2},\tau^x = \tfrac{1}{2}}, 
    \notag\\
    \ket{S^z = -\tfrac{1}{2}, \tau^x = \tfrac{1}{2}} &= \hat{S}^- \ket{S^z = \tfrac{1}{2},\tau^x = \tfrac{1}{2}},
    \notag\\
    \ket{S^z = -\tfrac{1}{2}, \tau^x = -\tfrac{1}{2}} &= \hat{S}^- \ket{S^z = \tfrac{1}{2},\tau^x = -\tfrac{1}{2}}.
    \label{eq:SI:spin--orbital-basis}
\end{align}

To analyze the ground states of $\hat{\mathcal{H}}^\mathrm{trimer}_\mathrm{Hub}$ and compare them with the spin--orbital states, we use the site-permutation symmetry $\hat{\Pi}_{12}$, whose eigenvalues are $\pm1$. Explicitly, $\hat{\Pi}_{12}$ is defined by
$
    \hat{\Pi}_{12} \hat{c}^{(\dag)}_{\mu,\sigma} \hat{\Pi}_{12}^{-1}
    =
    \hat{c}^{(\dag)}_{\Pi_{12}(\mu),\sigma},
$
with $\Pi_{12}(0)=0$, $\Pi_{12}(1)=2$, and $\Pi_{12}(2)=1$. 
The corresponding ground states,
$\ket{\psi_{S^z = \pm \tfrac{1}{2}, \Pi_{12} = \pm 1}(t_0/U)}$,
are adiabatically connected to the fourfold-degenerate $S=\frac{1}{2}\otimes\tau=\frac{1}{2}$ multiplet of $\hat{\mathcal{H}}_{\mathrm{spin}}^{(0)}$:
\begin{align}
 \lim_{t_0/U \to 0}\,
 \ket{\psi_{S^z = \pm \tfrac{1}{2}, \Pi_{12} = \pm 1}(t_0/U)}
 =
 \ket{S^z = \pm \tfrac{1}{2}, \tau^x = \mp \tfrac{1}{2}},
\end{align}
up to phase factors. The expectation values 
$\langle{\hat{P}^x}\rangle$ of the $x$ component of the electric dipole,
\begin{align}
    \hat{\mathbf{P}} = -e \sum_{0 \le \mu \le 2}
    \hat{n}_{\mu} \, \delta\mathbf{r}_\mu,
    \label{eq:SI:P:trimer}
\end{align}
evaluated in the fourfold-degenerate dressed spin--orbital states, scale as $O(t_0^3/U^3)$ for small $t_0/U$ (Fig.~\ref{FIG_2026_SI1}{b}), whereas $\langle{\hat{P}^y}\rangle = 0$ by symmetry.
Here, $\delta\mathbf{r}_\mu$ denotes the position of site $\mu$ relative to the trimer center, and the $x$-$y$ coordinate axes are shown in Fig.~\ref{FIG_2026_SI1}{c}. 

As an illustrative example, the wavefunction for $S^z = \frac{1}{2}$ and $\Pi_{12} = +1$ at $t_0/U = 0.005$ is given by
\begin{align}
    &\ket{\psi_{S^z = \tfrac{1}{2},\, \Pi_{12} = +1}(t_0/U = 0.005)} 
    \notag\\
    &=
    0.707\, c^\dag_{0\uparrow} c^\dag_{1\uparrow} c^\dag_{2\downarrow} \ket{0}
    - 0.707\, c^\dag_{0\uparrow} c^\dag_{1\downarrow} c^\dag_{2\uparrow} \ket{0}
    + 0.00707\, c^\dag_{0\uparrow} c^\dag_{1\uparrow} c^\dag_{1\downarrow} \ket{0}
    + 0.00707\, c^\dag_{0\uparrow} c^\dag_{2\uparrow} c^\dag_{2\downarrow} \ket{0}
    \notag\\
    &\hspace{10pt}
    - 0.00359\, c^\dag_{0\uparrow} c^\dag_{0\downarrow} c^\dag_{1\uparrow} \ket{0}
    - 0.00359\, c^\dag_{0\uparrow} c^\dag_{0\downarrow} c^\dag_{2\uparrow} \ket{0}
    - 0.00348\, c^\dag_{1\uparrow} c^\dag_{2\uparrow} c^\dag_{2\downarrow} \ket{0}
    - 0.00348\, c^\dag_{1\uparrow} c^\dag_{1\downarrow} c^\dag_{2\uparrow} \ket{0}
    \notag\\
    &= \sqrt{2} \left(
    0.707 \ket{\phi_1}
    + 0.00707 \ket{\phi_2}
    - 0.00359 \ket{\phi_3}
    - 0.00348 \ket{\phi_4}
    \right)
    \label{eq:SI:psi}
\end{align}
as schematically shown in Fig.~\ref{FIG_2026_SI1}{c}. The first two terms, collectively denoted by $\ket{\phi_1}$, correspond to the dominant component with one electron at each site. This state is one of the unperturbed Heisenberg-trimer ground states, namely $\ket{S^z = \tfrac{1}{2}, \tau^x = -\tfrac{1}{2}}$, and does not by itself generate an electric dipole. The subleading part of the dressed wavefunction contains virtual charge configurations with one doubly occupied site and one empty site, which generate electronic electric dipoles. In this example, these configurations can be classified into three $\Pi_{12}$-symmetric basis states, $\ket{\phi_2}$, $\ket{\phi_3}$, and $\ket{\phi_4}$, as introduced above. The electric dipoles in $\ket{\phi_2}$ are along the $\pm y$ directions and cancel exactly by symmetry. By contrast, $\ket{\phi_3}$ and $\ket{\phi_4}$ carry net dipoles pointing along $-x$ and $+x$, respectively, whose contributions cancel only imperfectly because their coefficients are not constrained to be equal by any symmetry. The resulting residual imbalance produces a small net electronic electric dipole, as discussed above (Fig.~\ref{FIG_2026_SI1}{c}).

\subsection{Effective operator formalism}
The above wavefunction-based picture can be recast in terms of effective operators~\cite{Bulaevskii2008}, in close analogy with the Schr{\"o}dinger and Heisenberg pictures for time evolution. We introduce a unitary transformation $e^{-\hat{\mathcal{S}}}$, generated by an anti-Hermitian operator $\hat{\mathcal{S}}$, that maps an unperturbed spin state $\ket{\bm{\sigma}}$ to a dressed spin state,
\begin{align}
    \ket{\tilde{\bm{\sigma}}}
    =
    e^{-\hat{\mathcal{S}}}
    \ket{\bm{\sigma}}.
\end{align}
The virtual hopping processes encoded in $e^{-\hat{\mathcal{S}}}$ describe the same kinds of charge fluctuations as those visualized in Fig.~\ref{FIG_2026_SI1}{c}. For any operator $\hat{\mathcal{O}}$, such as the Hubbard Hamiltonian~\eqref{eq:SI:trimer:Hub} or the electric dipole operator~\eqref{eq:SI:P:trimer}, matrix elements within the dressed low-energy space can be written as
\begin{align}
    \bra{\tilde{\bm{\sigma}}} \hat{\mathcal{O}} \ket{\tilde{\bm{\sigma}}'}
    =
    \bra{\bm{\sigma}} e^{\hat{\mathcal{S}}} \hat{\mathcal{O}} e^{-\hat{\mathcal{S}}} \ket{\bm{\sigma}'}
    =
    \bra{\bm{\sigma}} \hat{\mathcal{O}}_\mathrm{eff} \ket{\bm{\sigma}'},
\end{align}
where
\begin{align}
    \hat{\mathcal{O}}_\mathrm{eff} 
    =
    \hat{\mathcal{P}}_\mathrm{spin}
    e^{\hat{\mathcal{S}}} \hat{\mathcal{O}} e^{-\hat{\mathcal{S}}}
    \hat{\mathcal{P}}_\mathrm{spin}
\end{align}
is the corresponding effective operator projected onto the unperturbed spin space by $\hat{\mathcal{P}}_\mathrm{spin}$. By construction, $\hat{\mathcal{O}}_\mathrm{eff}$ can be expressed in terms of spin operators and evaluated perturbatively in $t_0/U$.

We now apply this construction to the charge density operator at site $l$, measured relative to half filling,
\begin{align}
    \delta \hat{n}_l = \hat{n}_{l} - 1.
\end{align}
For an arbitrary lattice, the leading-order contribution to
$\delta \hat{n}_{l,\mathrm{eff}}
\equiv
\hat{\mathcal{P}}_\mathrm{spin}
e^{\hat{\mathcal{S}}}
\delta \hat{n}_l
e^{-\hat{\mathcal{S}}}
\hat{\mathcal{P}}_\mathrm{spin}$
was obtained by Bulaevskii \textit{et al.} as
\begin{align}
    \delta \hat{n}_{l,\mathrm{eff}} 
    =
    \sum_{\{lmn\}}
    \frac{8 t_{lm} t_{mn} t_{nl}}{U^3}
    \left[
    \hat{\mathbf{s}}_l \cdot
    \left(
    \hat{\mathbf{s}}_m + \hat{\mathbf{s}}_n
    \right)
    -
    2 \hat{\mathbf{s}}_m \cdot \hat{\mathbf{s}}_n
    \right]
    +
    \mathcal{O}(t_{ij}^5/U^5),
    \label{eq:SI:neff:general}
\end{align}
where the summation is over three-site loops $\{lmn\}$ that include site $l$ and are formed by nonzero hopping amplitudes. The odd power of $t_{ij}/U$ reflects charge-conjugation symmetry, and such a contribution requires geometric frustration of the underlying lattice~\cite{Bulaevskii2008}.

For an isolated trimer, the only loop of this kind is the triangle itself. This gives
\begin{align}
    \delta \hat{n}_{(\mathbf{r},\mu),\mathrm{eff}} 
    &=
    \frac{8 t_0^3}{U^3}
    \left[
    \hat{\mathbf{s}}_{\mathbf{r},\mu}
    \cdot
    \left(
    \hat{\mathbf{s}}_{\mathbf{r},\mu+1}
    +
    \hat{\mathbf{s}}_{\mathbf{r},\mu+2}
    \right)
    -
    2
    \hat{\mathbf{s}}_{\mathbf{r},\mu+1}
    \cdot
    \hat{\mathbf{s}}_{\mathbf{r},\mu+2}
    \right]
    +
    \mathcal{O}(t_{0}^5/U^5)
    \notag\\
    &=
    -\frac{24 t_0^3}{U^3}
    \hat{\bm{\tau}}_{\mathbf{r}} \cdot \mathbf{n}_\mu
    +
    \mathcal{O}(t_{0}^5/U^5),
\end{align}
where ${\mathbf{n}}_\mu = (\cos\frac{4\mu\pi}{3}, \sin\frac{4\mu\pi}{3}, 0)$, as defined earlier. The corresponding effective electric dipole moment is
\begin{align}
    \hat{\mathbf{P}}_{\mathbf{r},\mathrm{eff}} 
    &=
    -e \sum_{0 \le \mu \le 2}
    \delta \hat{n}_{(\mathbf{r},\mu),\mathrm{eff}}
    \,
    \delta\mathbf{r}_\mu
    \notag\\
    &=
    \frac{24t_0^3 e}{U^3}
    \sum_{0 \le \mu \le 2}
    \left(
    \hat{\bm{\tau}}_{\mathbf{r}} \cdot \mathbf{n}_\mu
    \right)
    \delta\mathbf{r}_\mu
    +
    \mathcal{O}(t_{0}^5/U^5),
\end{align}
where $e$ is the elementary charge and $\delta\mathbf{r}_\mu$ denotes the position of site $\mu$ relative to the trimer center. With the real-space coordinate axes shown in Fig.~\ref{FIG_2026_SI1}{c}, we obtain
\begin{align}
    \hat{P}_{\mathbf{r},\mathrm{eff}}^{x(y)}
    =
    12\sqrt{3} ea 
    \left(\frac{t_0}{U}\right)^3
    \hat{\tau}^{x(y)}_{\mathbf{r}}
    +
    \mathcal{O}(t_{0}^5/U^5).
    \label{eq:SI:Peff:trimer:xy}
\end{align}
Since the trimer lies in a plane, $\hat{P}_{\mathbf{r},\mathrm{eff}}^{z} = 0$. The eigenvalues of $\hat{P}_{\mathbf{r},\mathrm{eff}}^{x}$ in the spin--orbital basis are compared with the direct Hubbard-trimer evaluation of $\langle{\hat{P}^x}\rangle$ in Fig.~\ref{FIG_2026_SI1}{b}. The agreement for small $t_0/U$ confirms the validity of the effective-operator description:
\begin{align}
    \langle{P^x}\rangle_{S^z = \pm \tfrac{1}{2}, P_{12} = \pm 1}
    \simeq
    \bra{S^z = \pm \tfrac{1}{2}, \tau^x = \mp \tfrac{1}{2}}
    P^x_\mathrm{eff}
    \ket{S^z = \pm \tfrac{1}{2}, \tau^x = \mp \tfrac{1}{2}}.
\end{align}
For weakly coupled trimers, we retain the leading intratrimer loop contribution in Eq.~\eqref{eq:SI:neff:general}, whereas loop contributions involving intertrimer hopping are subleading. Therefore, the leading-order expression for $\hat{\mathbf{P}}_{\mathbf{r},\mathrm{eff}}$ retains the same form as in Eq.~\eqref{eq:SI:Peff:trimer:xy}. 

In summary, the low-energy electric polarization of the weakly coupled trimer system described by the half-filled Hubbard model $\hat{\mathcal{H}}_\mathrm{Hub}$ (see below) in the strong-coupling limit can be evaluated entirely within the effective spin description. One first obtains the eigenstates of the effective spin Hamiltonian, $\hat{\mathcal{H}}_\mathrm{spin}$, and then computes the expectation value of $\hat{\mathbf{P}}_{\mathbf{r},\mathrm{eff}}$ within those states.

\section{%
    Effective models
    \label{SI:Heff}
}%
\setcounter{equation}{0}

\subsection{%
    Trimerized Hubbard and Heisenberg models
}%

\begin{figure}[!b]
  \centering
  \includegraphics[width=\hsize]{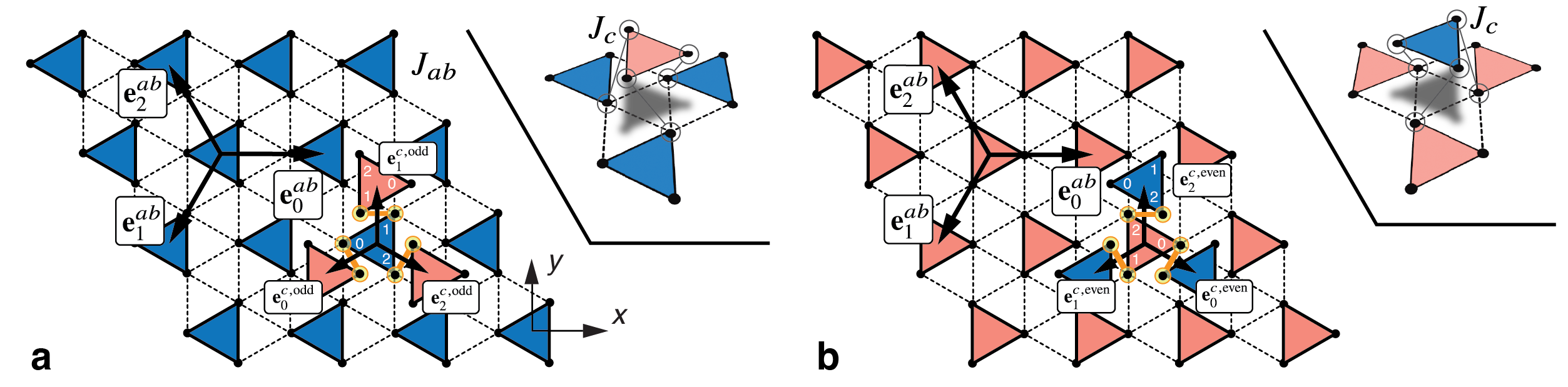}
  \vspace{0pt}
  \caption{%
    \label{FIG_2026_SI2}
    \textbf{%
        Trimerized triangular-lattice planes in the (a) odd and (b) even layers.
    }%
    The insets illustrate the interlayer connectivity. 
  }%
\end{figure}

We consider the half-filled Hubbard model on the trimerized stacked triangular lattice shown in Fig.~\ref{FIG_2026_SI2}, with the alternating stacking of ({\it P})- and ({\it M})-TNN layers. The Hamiltonian is
\begin{align}
  &\hat{\mathcal{H}}_\mathrm{Hub}
  = -t_0 \sum_{\mathbf{r}}\sum_{0 \le \mu \le 2}\sum_{\sigma=\uparrow,\downarrow}
  \left( 
  \hat{c}_{\mathbf{r},\mu,\sigma}^{\dag} \hat{c}_{\mathbf{r},\mu+1,\sigma}^{\;} + \mathrm{H.c.}
  \right)
  + \frac{U}{2} \sum_{\mathbf{r}}\sum_{0 \le \mu \le 2} \left(
    \hat{n}_{\mathbf{r},\mu}  - 1\right)^2
  \notag\\
  &~~- t_{ab}
  \biggl[
  \sum_{\mathbf{r} \in \text{even}}
  \sum_{\mu \neq \nu}
  \sum_{\sigma=\uparrow,\downarrow}
  \left(
    \hat{c}_{\mathbf{r},\mu,\sigma}^{\dag}\hat{c}_{\mathbf{r} + \mathbf{e}_{\mu}^{ab}\!\!\!,\,\,\nu,\sigma}^{\;}
    + \mathrm{H.c.}
  \right)
  + 
  \sum_{\mathbf{r} \in \text{odd}}
  \sum_{\mu \neq \nu}
  \sum_{\sigma=\uparrow,\downarrow}
  \left(
    \hat{c}_{\mathbf{r},\mu,\sigma}^{\dag}\hat{c}_{\mathbf{r} - \mathbf{e}_{\mu}^{ab}\!\!\!,\,\nu,\sigma}^{\;}
    + \mathrm{H.c.}
  \right)
  \biggr]
  \notag\\
  &~~- t_{c}\,
  \biggl[
  \sum_{\mathbf{r} \in \text{even}}
  \sum_{0 \le \mu \le 2}
  \sum_{\sigma=\uparrow,\downarrow}
  \left(
    \hat{c}_{\mathbf{r},\mu,\sigma}^{\dag}\hat{c}_{\mathbf{r} + \mathbf{e}_{\mu}^{c,\mathrm{even}}\!\!\!\!\!\!\!\!\!,\,\,\,\,\,\,\mu,\sigma}^{\;}
    + \mathrm{H.c.}
  \right)
  +
  \sum_{\mathbf{r} \in \text{odd}}
  \sum_{0 \le \mu \le 2}
  \sum_{\sigma=\uparrow,\downarrow}
  \left(
    \hat{c}_{\mathbf{r},\mu,\sigma}^{\dag} \hat{c}_{\mathbf{r} + \mathbf{e}_{\mu}^{c,\mathrm{odd}}\!\!\!\!\!\!\!\!,\,\,\,\,\,\mu,\sigma}^{\;}
    + \mathrm{H.c.}
  \right)
  \biggr],
\end{align}
where $t_0$, $t_{ab}$, and $t_c$ are the intra\-trimer, intralayer intertrimer, and interlayer hopping amplitudes, respectively. We assume $U \gg t_0 \gg t_{ab} > t_c$ throughout. The displacement vectors $\mathbf{e}_{\mu}^{ab}$, $\mathbf{e}_{\mu}^{c,\,\mathrm{even}}$, and $\mathbf{e}_{\mu}^{c,\,\mathrm{odd}} = \mathbf{e}_{(\mu+1)\bmod 3}^{c,\,\mathrm{even}}$ are illustrated in Fig.~\ref{FIG_2026_SI2}.

In the strong-coupling limit, $\hat{\mathcal{H}}_\mathrm{Hub}$ reduces to the spin-1/2 Heisenberg Hamiltonian $\hat{\mathcal{H}}_\mathrm{spin}$ by using second-order perturbation theory:
\begin{align}
  &\hat{\mathcal{H}}_\mathrm{spin}
  = J_0 \sum_{\mathbf{r}}\sum_{0 \le \mu \le 2}
  \hat{\mathbf{s}}_{\mathbf{r},\mu} \cdot \hat{\mathbf{s}}_{\mathbf{r},\mu+1}
  + J_{ab}\,
  \Biggl(\,
  \sum_{\mathbf{r} \in \text{even}}
  \sum_{\mu \neq \nu}
  \hat{\mathbf{s}}_{\mathbf{r},\mu} \cdot \hat{\mathbf{s}}_{\mathbf{r}+\mathbf{e}_{\mu}^{ab}\!\!\!,\,\,\nu} 
  + \sum_{\mathbf{r} \in \text{odd}}
  \sum_{\mu \neq \nu}
  \hat{\mathbf{s}}_{\mathbf{r},\mu} \cdot \hat{\mathbf{s}}_{\mathbf{r} - \mathbf{e}_{\mu}^{ab}\!\!\!,\,\nu}
  \Biggr)
  \notag\\
  &
  + J_{c}\,
  \Biggl(\,
  \sum_{\mathbf{r} \in \text{even}}
  \sum_{0 \le \mu \le 2}
  \hat{\mathbf{s}}_{\mathbf{r},\mu} \cdot \hat{\mathbf{s}}_{\mathbf{r} + \mathbf{e}_{\mu}^{c,\mathrm{even}}\!\!\!\!\!\!\!\!\!,\,\,\,\,\,\,\mu}
  +
  \sum_{\mathbf{r} \in \text{odd}}
  \sum_{0 \le \mu \le 2}
  \hat{\mathbf{s}}_{\mathbf{r},\mu} \cdot \hat{\mathbf{s}}_{\mathbf{r} + \mathbf{e}_{\mu}^{c,\mathrm{odd}}\!\!\!\!\!\!\!\!,\,\,\,\,\,\mu}
  \Biggr)
  - h \sum_{\mathbf{r}}\sum_{0 \le \mu \le 2}
  \hat{s}^z_{\mathbf{r},\mu},
  \label{SI:eq:Hspin}
\end{align}
where $J_0 = 4t_0^2/U$, $J_{ab} = 4t_{ab}^2/U$, and $J_c = 4t_c^2/U$. The reduced magnetic field is $h = g\mu_B H$. Constant energy offsets are omitted hereafter, unless otherwise mentioned.

\subsection{%
    Low-energy effective models in different field regimes
}%
\subsubsection{%
    Heisenberg spin trimer
}%
We first consider an isolated spin trimer,
\begin{align}
   \hat{\mathcal{H}}_{\mathrm{spin}}^{(0)}(h; \mathbf{r})
    = J_0 \Bigl( 
    \hat{\mathbf{s}}_{\mathbf{r},0}
     \cdot \hat{\mathbf{s}}_{\mathbf{r},1}
    + \hat{\mathbf{s}}_{\mathbf{r},1} \cdot \hat{\mathbf{s}}_{\mathbf{r},2}
    + \hat{\mathbf{s}}_{\mathbf{r},2} \cdot \hat{\mathbf{s}}_{\mathbf{r},0}
    \Bigr) 
    - h \sum_{0 \le \mu \le 2} \hat{s}^z_{\mathbf{r},\mu}.
\end{align}
For $0<h<h_{\mathrm{sat},0}=\frac{3}{2}J_0$, the ground state is the orbital doublet $\ket{S^z = \tfrac{1}{2}, \tau^x = \pm \tfrac{1}{2}}$, as mentioned above. At $h=h_{\mathrm{sat},0}$, a level crossing occurs with the fully polarized state $\ket{\up\up\up}$, yielding magnetization plateaus $M(h=0)=0$, $M(0<h<h_{\mathrm{sat},0})=\frac{1}{2}$, and $M(h>h_{\mathrm{sat},0})=M_{\mathrm{sat}}=\frac{3}{2}$.

\subsubsection{%
    Kugel--Khomskii (KK) Hamiltonian in the low-field (LF) regime
}%
The LF effective Hamiltonian is obtained by projecting $\hat{\mathcal{H}}_\mathrm{spin}$~\eqref{SI:eq:Hspin} onto the $S=\frac{1}{2} \otimes \tau=\frac{1}{2}$ subspace. Within this low-energy space,
\begin{align}
  \hat{\mathcal{P}}_\mathrm{LF}\, \hat{\mathbf{s}}_{\mathbf{r},\mu} \hat{\mathcal{P}}_\mathrm{LF}
  &= 
  \frac{1}{3}
  \hat{\mathbf{S}}_{\mathbf{r}}\, \left(1 - 4 \hat{\bm{\tau}}_{\mathbf{r}} \cdot {\mathbf{n}}_\mu\right),
  \label{eq:SI:spin}
\end{align}
where ${\mathbf{n}}_\mu = (\cos\frac{4\mu\pi}{3}, \sin\frac{4\mu\pi}{3}, 0)$ and $\hat{\mathcal{P}}_\mathrm{LF}$ denotes the projection operator. 

The resulting KK Hamiltonian is
\begin{align}
  \hat{\mathcal{H}}_{\mathrm{KK}}
  &=
  \hat{\mathcal{P}}_\mathrm{LF}
  \left(
  \hat{\mathcal{H}}_\mathrm{spin}
  -
  \hat{\mathcal{H}}_\mathrm{LF}^{(0)}
  \right)
  \hat{\mathcal{P}}_\mathrm{LF}
  \notag\\
  &=
  \sum_{\mathbf{r}\in \text{even}}
  \sum_{0 \le \mu \le 2}
  \left(
     \hat{\mathcal{H}}_{\textrm{KK}}^{ab,\mu}
     (\mathbf{r},\mathbf{e}_{\mu}^{ab})
     +
     \hat{\mathcal{H}}_{\textrm{KK}}^{c,\mu}
     (\mathbf{r},\mathbf{e}_{\mu}^{c,\mathrm{even}})
  \right)
  \notag\\
  &~~~
  +
  \sum_{\mathbf{r}\in \text{odd}}
  \sum_{0 \le \mu \le 2}
  \left(
  \hat{\mathcal{H}}_{\textrm{KK}}^{ab,\mu}
  (\mathbf{r}, -\mathbf{e}_{\mu}^{ab})
  +
  \hat{\mathcal{H}}_{\textrm{KK}}^{c,\mu}
  (\mathbf{r}, \mathbf{e}_{\mu}^{c,\mathrm{odd}})
  \right)
  -
  h \sum_\mathbf{r} \hat{S}^z_{\mathbf{r}},
  \label{eq:SI:Heff-LF}
\end{align}
where $\hat{\mathcal{H}}_\mathrm{LF}^{(0)}
=
\sum_\mathbf{r}
\hat{\mathcal{H}}_{\mathrm{spin}}^{(0)}(h;\mathbf r)$.
The intralayer interaction is
\begin{align}
  \hat{\mathcal{H}}_{\textrm{KK}}^{ab,\mu}
  (\mathbf{r}, \bm{\delta}^{ab}_{\mathbf{r},\mu})
  &=
  \frac{2 J_{ab}}{9}
  \left(
  1
  -
  4 \hat{\bm{\tau}}_{\mathbf{r}}\cdot{\mathbf{n}}_\mu
  \right)
  \left(
  1
  +
  2 \hat{\bm{\tau}}_{\mathbf{r} + \bm{\delta}^{ab}_{\mathbf{r},\mu}}
  \cdot
  {\mathbf{n}}_\mu
  \right)
  \hat{\mathbf{S}}_{\mathbf{r}}
  \cdot
  \hat{\mathbf{S}}_{\mathbf{r} + \bm{\delta}^{ab}_{\mathbf{r},\mu}}
  \notag\\
  &\equiv
  \hat{J}_{\bm{\delta}^{ab}_{\mathbf{r},\mu}}(\hat{\bm{\tau}}_{\mathbf{r}},\hat{\bm{\tau}}_{\mathbf{r} + \bm{\delta}^{ab}_{\mathbf{r},\mu}})
  \hat{\mathbf{S}}_{\mathbf{r}}
  \cdot
  \hat{\mathbf{S}}_{\mathbf{r} + \bm{\delta}_{\mathbf{r},\mu}},
  \label{eq:SI:Heff-LF1}
\end{align}
where $\bm{\delta}^{ab}_{\mathbf r,\mu} = \mathbf e_\mu^{ab}$ and $\bm{\delta}^{ab}_{\mathbf r,\mu} = -\mathbf e_\mu^{ab}$ for even and odd layers, respectively. The interlayer interaction is
\begin{align}
  \hat{\mathcal{H}}_{\textrm{KK}}^{c,\mu}
  (\mathbf{r}, \bm{\delta}^{c}_{\mathbf{r},\mu})
  &=
  \frac{J_{c}}{9}
  \left(
  1
  -
  4 \hat{\bm{\tau}}_{\mathbf{r}}\cdot{\mathbf{n}}_\mu
  \right)
  \left(
  1
  -
  4 \hat{\bm{\tau}}_{\mathbf{r} + \bm{\delta}^{c}_{\mathbf{r},\mu}}
  \cdot
  {\mathbf{n}}_\mu
  \right)
  \hat{\mathbf{S}}_{\mathbf{r}}
  \cdot
  \hat{\mathbf{S}}_{\mathbf{r} + \bm{\delta}^{c}_{\mathbf{r},\mu}}
  \notag\\
  &\equiv
  \hat{J}_{\bm{\delta}^{c}_{\mathbf{r},\mu}}(\hat{\bm{\tau}}_{\mathbf{r}},\hat{\bm{\tau}}_{\mathbf{r} + \bm{\delta}^{c}_{\mathbf{r},\mu}})
  \hat{\mathbf{S}}_{\mathbf{r}}
  \cdot
  \hat{\mathbf{S}}_{\mathbf{r} + \bm{\delta}^{c}_{\mathbf{r},\mu}},
  \label{eq:SI:Heff-LF2}
\end{align}
where $\bm{\delta}^{c}_{\mathbf r,\mu} = \mathbf e_\mu^{c,\,\mathrm{even}}$ and $\mathbf e_\mu^{c,\,\mathrm{odd}}$ for even and odd layers, respectively.

\subsubsection{%
    Orbital compass Hamiltonian in the plateau regime
}%
The effective Hamiltonian $\mathcal{H}_\mathrm{comp}$ in the plateau regime is obtained by further projecting $\hat{\mathcal{H}}_\mathrm{KK}$ onto the two-dimensional $\tau=\frac{1}{2}$ subspace with $S = S^z = \frac{1}{2}$. This corresponds to replacing $\mathbf{S} \mapsto (0,0,1/2)$ for all trimers in Eq.~\eqref{eq:SI:Heff-LF}.
All terms linear in the pseudospin operators cancel by symmetry, yielding the orbital compass Hamiltonian,
\begin{align}
   \hat{\mathcal{H}}_{\mathrm{Comp}}
   &= 
   \sum_{\mathbf{r} \in \mathrm{even}}
   \sum_{0 \le \mu < 2}
   \left(
    J_{ab}^\tau \hat{\bm{\tau}}_{\mathbf{r}}\cdot{\mathbf{n}}_\mu \, \hat{\bm{\tau}}_{\mathbf{r} + \mathbf{e}_{\mu}^{ab}}\cdot{\mathbf{n}}_\mu
    + J_{c}^\tau \hat{\bm{\tau}}_{\mathbf{r}}\cdot{\mathbf{n}}_\mu \, \hat{\bm{\tau}}_{\mathbf{r} + \mathbf{e}_{\mu}^{c,\mathrm{even}}}\cdot{\mathbf{n}}_\mu
   \right)
   \notag\\
   &~~~
   + \sum_{\mathbf{r} \in \mathrm{odd}}
   \sum_{0 \le \mu < 2}
   \left(
    J_{ab}^\tau \hat{\bm{\tau}}_{\mathbf{r}}\cdot{\mathbf{n}}_\mu \, \hat{\bm{\tau}}_{\mathbf{r} - \mathbf{e}_{\mu}^{ab}}\cdot{\mathbf{n}}_\mu
    + J_{c}^\tau \hat{\bm{\tau}}_{\mathbf{r}}\cdot{\mathbf{n}}_\mu \, \hat{\bm{\tau}}_{\mathbf{r} + \mathbf{e}_{\mu}^{c,\mathrm{odd}}}\cdot{\mathbf{n}}_\mu
   \right),
   \label{Eq:Heff:Plateau}
\end{align}
where $J^{\tau}_{ab} = -\frac{4}{9} J_{ab}$ and $J^{\tau}_{c} = \frac{4}{9} J_c$.

\subsubsection{%
    Bosonic {\it t}-{\it J} Hamiltonian in the high-field (HF) regime
}%
The HF effective Hamiltonian is obtained by projecting $\hat{\mathcal{H}}_\mathrm{spin}$~\eqref{SI:eq:Hspin} onto the three-dimensional subspace spanned by the orbital doublet $\ket{S^z = \tfrac{1}{2}, \tau^x = \pm \tfrac{1}{2}}$ and $\ket{\up\up\up}$. We regard $\ket{\up\up\up}$ as the orbital vacancy state and define
\begin{align}
    \hat{b}^\dag_{\pm} = \ket{S^z = \tfrac{1}{2}, \tau^x = \pm\frac{1}{2}}\bra{\up\up\up},~~
    \hat{b}_{\pm} = (\hat{b}^\dag_{\pm})^\dag,
\end{align}
as the creation and annihilation operators of the $S = S^z = \frac{1}{2}$ orbital states. We also define $\hat{n}_b = \sum_{\tau=\pm} \hat{b}^{\dag}_{\tau} \hat{b}_{\tau}$ and $\hat{n}_\mathrm{h} = \hat{\mathcal{P}}_\mathrm{HF}(1 - \hat{n}_b)\hat{\mathcal{P}}_\mathrm{HF}$ as the effective particle and hole density operators, respectively, where $\hat{\mathcal{P}}_\mathrm{HF}$ is the projection operator. Similarly, we introduce
\begin{align}
    \hat{b}^\dag_{\mathbf{n}_0} &= \hat{b}^\dag_{+},
    \notag\\
    \hat{b}^\dag_{\mathbf{n}_1} &= -\frac{1}{2}\hat{b}^\dag_{+} + \frac{\sqrt{3}}{2}\hat{b}^\dag_{-},
    \notag\\
    \hat{b}^\dag_{\mathbf{n}_2} &= -\frac{1}{2}\hat{b}^\dag_{+} - \frac{\sqrt{3}}{2}\hat{b}^\dag_{-},
    \label{SI:eq:bdag}
\end{align}
which create bosons with pseudospins polarized along ${\mathbf{n}}_\mu = (\cos\frac{4\mu\pi}{3}, \sin\frac{4\mu\pi}{3}, 0)$. 

With these definitions, the projected spin operators are given by
\begin{align}
  \hat{\mathcal{P}}_\mathrm{HF}\, \hat{s}^z_{\mathbf{r},\mu} \hat{\mathcal{P}}_\mathrm{HF}
  &= \frac{1 - 4 \hat{\bm{\tau}}_\mathbf{r} \cdot \mathbf{n}_\mu}{6}\hat{n}_{b,\mathbf{r}}
  + \frac{1}{2}\hat{n}_{\mathrm{h},\mathbf{r}},
  \notag\\
  \hat{\mathcal{P}}_\mathrm{HF}\, \hat{s}^-_{\mathbf{r},\mu} \hat{\mathcal{P}}_\mathrm{HF}
  &= -\frac{2}{\sqrt{6}} \hat{b}^\dag_{\mathbf{r},\mathbf{n}_\mu}.
  \label{eq:SI:spin:HF}
\end{align}
The resulting HF effective Hamiltonian is
\begin{align}
    \hat{\mathcal{H}}_{\text{b-\textit{tJ}}} 
    &= \hat{\mathcal{P}}_\mathrm{HF}\, (\hat{\mathcal{H}}_\mathrm{spin} - \hat{\mathcal{H}}_\text{HF}^{(0)}) \hat{\mathcal{P}}_\mathrm{HF}
    \notag\\
    &= 
    \sum_{\mathbf{r}\in \text{even}}
    \sum_{0 \le \mu \le 2} 
    \left(
        \hat{\mathcal{H}}_\text{b-\textit{tJ}}^{ab,\mu}(\mathbf{r},\mathbf{e}_{\mu}^{ab}) 
        + \hat{\mathcal{H}}_\text{b-\textit{tJ}}^{c,\mu}(\mathbf{r},\mathbf{e}_{\mu}^{c,\mathrm{even}})
    \right)
    \notag\\
    &\hspace{20pt}
    + \sum_{\mathbf{r}\in \text{odd}}
    \sum_{0 \le \mu \le 2} 
    \left( 
        \hat{\mathcal{H}}_\text{b-\textit{tJ}}^{ab,\mu}(\mathbf{r}, -\mathbf{e}_{\mu}^{ab}) 
        + \hat{\mathcal{H}}_\text{b-\textit{tJ}}^{c,\mu}(\mathbf{r}, \mathbf{e}_{\mu}^{c,\mathrm{odd}})
    \right)
    - \left(h - \frac{3}{2}J_0\right) \sum_\mathbf{r} \hat{n}_{\mathrm{h},\mathbf{r}},
  \label{SI:eq:Heff:HF}
\end{align}
where $\hat{\mathcal{H}}_\mathrm{HF}^{(0)} = \sum_\mathbf{r}\hat{\mathcal{H}}^\mathrm{trimer}_\mathrm{spin}(h=\frac{3}{2}J_0, \mathbf{r})$ is the Hamiltonian for decoupled trimers at the level crossing magnetic field.
The intralayer interaction is
\begin{align}
  \hat{\mathcal{H}}_\text{b-\textit{tJ}}^{ab,\mu}(\mathbf{r}, \bm{\delta}^{ab}_{\mathbf{r},\mu}) 
  &= J_{ab} 
  \left(
  -\frac{2}{3}\hat{\bm{\tau}}_\mathbf{r} \cdot \mathbf{n}_\mu
  - \frac{1}{3}\hat{n}_{b,\mathbf{r}}
  + \frac{1}{2}
  \right)
  \left(
  \frac{2}{3}\hat{\bm{\tau}}_{\mathbf{r} + \bm{\delta}^{ab}_{\mathbf{r},\mu}} \cdot \mathbf{n}_\mu
  - \frac{2}{3}\hat{n}_{b,\mathbf{r} + \bm{\delta}^{ab}_{\mathbf{r},\mu}}
  + 1
  \right)
  \notag\\
  &\hspace{180pt}
  -\frac{J_{ab}}{3} \left(
  \hat{b}^\dag_{\mathbf{n}_\mu,\mathbf{r}} \hat{b}^{\;}_{\mathbf{n}_\mu,\mathbf{r} + \bm{\delta}^{ab}_{\mathbf{r},\mu}} + \text{H.c.}
  \right)
  \notag\\[3pt]
  &\equiv
  \hat{\mathcal{H}}_{\tau}^{ab,\mu}(\mathbf{r}, \bm{\delta}^{ab}_{\mathbf{r},\mu}) 
  + \hat{\mathcal{H}}_\text{hop}^{ab,\mu}(\mathbf{r}, \bm{\delta}^{ab}_{\mathbf{r},\mu}),
\end{align}
where the first term is an extension of the compass interaction in Eq.~\eqref{Eq:Heff:Plateau} and the second term is a kinetic term. 
Summing over the bond index $\mu$ yields the compact form for the extended compass term:
\begin{align}
    \sum_{0 \le \mu \le 2} \hat{\mathcal{H}}_{\tau}^{ab,\mu}(\mathbf{r},\bm{\delta}^{ab}_{\mathbf{r},\mu}) 
    &= \frac{J_{ab}}{9} \sum_{0 \le \mu \le 2}
    \tilde{\bm{\tau}}_\mathbf{r}
    \begin{pmatrix}
       -4 (n_\mu^x)^2 & -4 n_\mu^x n_\mu^y & 0 & 4 n_\mu^x \\[2.5pt]
       -4 n_\mu^x n_\mu^y & -4 (n_\mu^y)^2 & 0 & 4 n_\mu^y \\[2.5pt]
       0 & 0 & 0 & 0 \\[2.5pt]
       -2n_\mu^x & -2 n_\mu^y & 0 & 2
    \end{pmatrix}
    \tilde{\bm{\tau}}_{\mathbf{r} + \bm{\delta}^{ab}_{\mathbf{r},\mu}} 
    - \frac{J_{ab}}{3} \sum_{0 \le \mu \le 2}
    \left( \hat{n}_{b,\mathbf{r}} + \hat{n}_{b,{\mathbf{r} + \bm{\delta}^{ab}_{\mathbf{r},\mu}}} \right)
    + \mathrm{const.}
    \notag\\
    &=
    \sum_{0 \le \mu \le 2}
    \tilde{\bm{\tau}}^{\mathrm{T}}_{\mathbf{r}}\,
    {\mathcal{J}}_{ab}^{(\mu)}\,
    \tilde{\bm{\tau}}_{\mathbf{r} + \bm{\delta}^{ab}_{\mathbf{r},\mu}}
    - \frac{J_{ab}}{3} \sum_{0 \le \mu \le 2}
    \left( \hat{n}_{b,\mathbf{r}} + \hat{n}_{b,{\mathbf{r} + \bm{\delta}^{ab}_{\mathbf{r},\mu}}} \right)
    + \mathrm{const.},
\end{align}
where
\begin{align}
    \tilde{\bm{\tau}}_{\mathbf{r}} = \begin{pmatrix}
        \hat{\tau}_{\mathbf{r}}^x \\
        \hat{\tau}_{\mathbf{r}}^y \\
        \hat{\tau}_{\mathbf{r}}^z \\
        \hat{n}_{b,\mathbf{r}}
    \end{pmatrix}   
\end{align}
bundles the pseudospin and density operators and
\begin{align}
    \mathcal{J}_{ab}^{(0)}
    &=
    \frac{J_{ab}}{9} 
    \begin{pmatrix}
       -4 & 0 & 0 & 4 \\[2.5pt]
       0 & 0 & 0 & 0 \\[2.5pt]
       0 & 0 & 0 & 0 \\[2.5pt]
       -2 & 0 & 0 & 2
    \end{pmatrix},
    ~~
    \mathcal{J}_{ab}^{(1)}
    =
    \frac{J_{ab}}{9} 
    \begin{pmatrix}
       -1 & -\sqrt{3} & 0 & -2 \\[2.5pt]
       -\sqrt{3} & -3 & 0 & -2\sqrt{3} \\[2.5pt]
       0 & 0 & 0 & 0 \\[2.5pt]
       1 & \sqrt{3} & 0 & 2
    \end{pmatrix},
    ~~
    \mathcal{J}_{ab}^{(2)}
    =
    \frac{J_{ab}}{9} 
    \begin{pmatrix}
       -1 & \sqrt{3} & 0 & -2 \\[2.5pt]
       \sqrt{3} & -3 & 0 & 2\sqrt{3} \\[2.5pt]
       0 & 0 & 0 & 0 \\[2.5pt]
       1 & -\sqrt{3} & 0 & 2
    \end{pmatrix}.
\end{align}
The interlayer interaction is
\begin{align}
  \hat{\mathcal{H}}_\text{b-\textit{tJ}}^{c,\mu}(\mathbf{r}, \bm{\delta}^c_{\mathbf{r},\mu}) 
  &= J_{c} 
  \left(
  -\frac{2}{3}\hat{\bm{\tau}}_{\mathbf{r}} \cdot \mathbf{n}_\mu
  - \frac{1}{3}\hat{n}_{b,\mathbf{r}}
  + \frac{1}{2}
  \right)
  \left(
  -\frac{2}{3}\hat{\bm{\tau}}_{\mathbf{r} + \bm{\delta}^c_{\mathbf{r},\mu}} \cdot \mathbf{n}_\mu
  - \frac{1}{3}\hat{n}_{b,\mathbf{r} + \bm{\delta}^c_{\mathbf{r},\mu}}
  + \frac{1}{2}
  \right)
  \notag\\
  &\hspace{180pt}+ \frac{J_{c}}{3} \left(
  \hat{b}^\dag_{\mathbf{n}_\mu,\mathbf{r}}
  \hat{b}^{\;}_{\mathbf{n}_\mu, \mathbf{r} + \bm{\delta}^c_{\mathbf{r},\mu}} + \text{H.c.}
  \right)
  \notag\\[3pt]
  &\equiv
  \hat{\mathcal{H}}_{\tau}^{c,\mu}(\mathbf{r}, \bm{\delta}^c_{\mathbf{r},\mu}) 
  + \hat{\mathcal{H}}_\text{hop}^{c,\mu}(\mathbf{r}, \bm{\delta}^c_{\mathbf{r},\mu}),
\end{align}
and summing the first term over $\mu$ yields
\begin{align}
    \sum_{0 \le \mu \le 2} \hat{\mathcal{H}}_{\tau}^{c,\mu}(\mathbf{r},\bm{\delta}^c_{\mathbf{r},\mu}) 
    &= \frac{J_{c}}{9} \sum_{0 \le \mu \le 2}
    \tilde{\bm{\tau}}_\mathbf{r}
    \begin{pmatrix}
       4 (n_\mu^x)^2 & 4 n_\mu^x n_\mu^y & 0 & 2 n_\mu^x \\[2.5pt]
       4 n_\mu^x n_\mu^y & 4 (n_\mu^y)^2 & 0 & 2 n_\mu^y \\[2.5pt]
       0 & 0 & 0 & 0 \\[2.5pt]
       2n_\mu^x & 2 n_\mu^y & 0 & 1
    \end{pmatrix}
    \tilde{\bm{\tau}}_{\mathbf{r} + \bm{\delta}^c_{\mathbf{r},\mu}} 
    - \frac{J_{c}}{6} \sum_{0 \le \mu \le 2}
    \left( \hat{n}_{b,\mathbf{r}} + \hat{n}_{b,\mathbf{r} + \bm{\delta}^c_{\mathbf{r},\mu}} \right)
    + \mathrm{const.},
\end{align}
where
\begin{align}
    \mathcal{J}_{c}^{(0)}
    &=
    \frac{J_{c}}{9} 
    \begin{pmatrix}
       4 & 0 & 0 & 2 \\[2.5pt]
       0 & 0 & 0 & 0 \\[2.5pt]
       0 & 0 & 0 & 0 \\[2.5pt]
       2 & 0 & 0 & 1
    \end{pmatrix},
    ~~
    \mathcal{J}_{c}^{(1)}
    =
    \frac{J_{c}}{9} 
    \begin{pmatrix}
       1 & \sqrt{3} & 0 & -1 \\[2.5pt]
       \sqrt{3} & 3 & 0 & -\sqrt{3} \\[2.5pt]
       0 & 0 & 0 & 0 \\[2.5pt]
       -1 & -\sqrt{3} & 0 & 1
    \end{pmatrix},
    ~~
    \mathcal{J}_{c}^{(2)}
    =
    \frac{J_{c}}{9} 
    \begin{pmatrix}
       1 & -\sqrt{3} & 0 & -1 \\[2.5pt]
       -\sqrt{3} & 3 & 0 & \sqrt{3} \\[2.5pt]
       0 & 0 & 0 & 0 \\[2.5pt]
       -1 & \sqrt{3} & 0 & 1
    \end{pmatrix}.
\end{align}

By combining these contributions, we find
\begin{align}
    \hat{\mathcal H}_\text{b-\textit{tJ}}
    = \hat{\mathcal H}_{\tau}
    + \hat{\mathcal H}_{\rm hop}
    - \mu_{\rm h} \hat N_{\rm h},~~
    \hat N_{\rm h} = \sum_\mathbf{r} \hat{n}_{\mathrm{h},\mathbf{r}},
\end{align}
with $\mu_\mathrm{h} = h - \frac{3}{2}J_0 - 2J_{ab} - J_{c}$,
\begin{align}
    \hat{\mathcal{H}}_\tau
    &= 
    \sum_{\mathbf{r}\in \text{even}}
    \sum_{0 \le \mu \le 2} \left(
        \tilde{\bm{\tau}}^{\mathrm{T}}_{\mathbf{r}}\,
        {\mathcal{J}}_{ab}^{(\mu)}\,
        \tilde{\bm{\tau}}_{\mathbf{r} + \bm{e}^{ab}_{\mu}}
        +
        \tilde{\bm{\tau}}^{\mathrm{T}}_{\mathbf{r}}\,
        {\mathcal{J}}_{c}^{(\mu)}\,
        \tilde{\bm{\tau}}_{\mathbf{r} + \bm{e}^{c,\mathrm{even}}_{\mu}}
    \right)
    \notag\\
    &~~~
    +
    \sum_{\mathbf{r}\in \text{odd}}
    \sum_{0 \le \mu \le 2} \left(
        \tilde{\bm{\tau}}^{\mathrm{T}}_{\mathbf{r}}\,
        {\mathcal{J}}_{ab}^{(\mu)}\,
        \tilde{\bm{\tau}}_{\mathbf{r} - \bm{e}^{ab}_{\mu}}
        +
        \tilde{\bm{\tau}}^{\mathrm{T}}_{\mathbf{r}}\,
        {\mathcal{J}}_{c}^{(\mu)}\,
        \tilde{\bm{\tau}}_{\mathbf{r} + \bm{e}^{c,\mathrm{odd}}_{\mu}}
    \right),
\end{align}
and
\begin{align}
    \hat{\mathcal H}_{\rm hop}
    &=
    \sum_{\mathbf{r}\in \text{even}}
    \sum_{0 \le \mu \le 2} \left(
        \hat{\mathbf{b}}^{\dag}_{\mathbf{r}}\,
        {\mathcal{K}}_{ab}^{(\mu)}\,
        \hat{\mathbf{b}}^{}_{\mathbf{r} + \bm{e}^{ab}_{\mu}}
        +
        \hat{\mathbf{b}}^{\dag}_{\mathbf{r}}\,
        {\mathcal{K}}_{c}^{(\mu)}\,
        \hat{\mathbf{b}}^{}_{\mathbf{r} + \bm{e}^{c,\mathrm{even}}_{\mu}}
    \right)
    \notag\\
    &~~~
    +
    \sum_{\mathbf{r}\in \text{odd}}
    \sum_{0 \le \mu \le 2} \left(
        \hat{\mathbf{b}}^{\dag}_{\mathbf{r}}\,
        {\mathcal{K}}_{ab}^{(\mu)}\,
        \hat{\mathbf{b}}^{}_{\mathbf{r} - \bm{e}^{ab}_{\mu}}
        +
        \hat{\mathbf{b}}^{\dag}_{\mathbf{r}}\,
        {\mathcal{K}}_{c}^{(\mu)}\,
        \hat{\mathbf{b}}^{}_{\mathbf{r} + \bm{e}^{c,\mathrm{odd}}_{\mu}}
    \right) + \mathrm{H.c.},
\end{align}
where $\hat{\mathbf{b}}^{\dag}_{\mathbf{r}} = (\hat{b}^\dag_{\mathbf{r},+}, \hat{b}^\dag_{\mathbf{r},-})$ and
\begin{align}
    \mathcal{K}_{ab}^{(0)}
    = -\frac{J_{ab}}{3}
    \begin{pmatrix}
    1 & 0 \\
    0 & 0
    \end{pmatrix},
    ~~
    \mathcal{K}_{ab}^{(1)}
    = 
    -\frac{J_{ab}}{3}
    \begin{pmatrix}
    \frac{1}{4} & -\frac{\sqrt{3}}{4} \\[2pt]
    -\frac{\sqrt{3}}{4} & \frac{3}{4}
    \end{pmatrix},
    ~~
    \mathcal{K}_{ab}^{(2)}
    =
    -\frac{J_{ab}}{3}
    \begin{pmatrix}
    \frac{1}{4} & \frac{\sqrt{3}}{4} \\[2pt]
    \frac{\sqrt{3}}{4} & \frac{3}{4}
    \end{pmatrix},
\end{align}
for the intralayer hopping and $\mathcal{K}_{c}^{(\mu)} = -\tfrac{J_c}{J_{ab}} \mathcal{K}_{ab}^{(\mu)}$ for the interlayer hopping.

\section{%
    Generalized classical \texorpdfstring{$\mathbb{CP}^{n-1}$}{CP(n-1)} Monte Carlo (MC) simulations: Algorithm
    \label{SI:MC}
}%
\setcounter{equation}{0}
\subsection{Variational formalism}
In $\mathbb{CP}^{n-1}$ MC simulations for an arbitrary integer $n \ge 2$, we introduce a cluster-based direct-product wavefunction ansatz,
\begin{align}
    \ket{\Psi[\Omega]} = \prod_{\mathbf{r}} \ket{\mathbf{z}(\Omega_\mathbf{r})}_\mathbf{r},
    \label{eq:Methods:coherent-state}
\end{align}
with the local state $\ket{\mathbf{z}(\Omega_\mathbf{r})}_\mathbf{r}$ as an arbitrary linear combination of chosen orthonormal basis states ${\ket{\psi^\alpha}_{\mathbf{r}}}$ with $0 \le \alpha \le n - 1$:
\begin{align}
    \ket{\mathbf{z}(\Omega_\mathbf{r})}_{\mathbf{r}} = \sum_{\alpha = 0}^{n - 1} z_{\mathbf{r}}^\alpha(\Omega_\mathbf{r}) \ket{\psi^\alpha}_{\mathbf{r}}.
    \label{eq:Methods:coherent-state:local}
\end{align}
Similar methods were used for studying systems with multiple degrees of freedom per unit, e.g., spin-1 quantum magnets~\cite{Stoudenmire2009,Zhang2021,Seifert2022,Remund2022,Pohle2023} and multi-orbital systems~\cite{Iwazaki2023}. In the present work, the orthonormal basis set in the LF regime can be chosen as $\ket{S^z = \pm \frac{1}{2}, \tau^x = \pm \frac{1}{2}}$ ($n = 4$), whereas that in the HF regime can be chosen as $\ket{S^z = \frac{1}{2}, \tau^x = \pm \frac{1}{2}}$ and $\ket{\uparrow\uparrow\uparrow}$ ($n = 3$). $\Omega_\mathbf{r}$ denotes a set of variational parameters that specify $\mathbf{z}_\mathbf{r} = (z_{\mathbf{r}}^0, z_{\mathbf{r}}^1, \dots, z_{\mathbf{r}}^{n-1})$ with $\left\lvert{\mathbf{z}_\mathbf{r}}\right\rvert^2 = 1$:
\begin{align}
    z^\alpha_\mathbf{r} &= \left(\prod_{\alpha + 1 \le \beta \le n - 1} \sin\theta_{\mathbf{r}}^{\beta}\right) \cos\theta^\alpha_{\mathbf{r}} e^{i\chi_{\mathbf{r}}^{\alpha}},
\end{align}
where $\theta^0 \equiv 0$ by convention, $0 \le \theta^\alpha \le \pi/2$ ($1 \le \alpha \le n - 1$), and $0 \le \chi^\alpha < 2\pi$ ($0 \le \alpha \le n - 1$). In addition, $\chi^0 \equiv 0$ can be fixed in MC simulations as the overall phase factor. 
$\ket{\mathbf{z}(\Omega_\mathbf{r})}_{\mathbf{r}}$
is also known as the SU($n$) coherent state, but the method applies even without SU($n$) symmetry; we therefore refer to it as a $\mathbb{CP}^{n-1}$ coherent state, emphasizing the manifold structure. 

When we consider a low-energy effective Hamiltonian $\hat{\mathcal{H}}_\mathrm{eff}$, the effective free energy $\mathcal{F}_\mathrm{eff}$ is defined as $e^{-\mathcal{F}_\mathrm{eff}/T} = \mathrm{Tr}\, \hat{\mathcal{P}} e^{-\hat{\mathcal{H}}_\mathrm{eff}/T}$. 
Here, $\hat{\mathcal{P}} = \otimes_\mathbf{r} \hat{\mathcal{P}}_\mathbf{r}$ is a projection operator that factorizes over clusters (trimers), where $\hat{\mathcal{P}}_\mathbf{r} = \sum_{0 \le \alpha \le n - 1}\ket{\psi^\alpha}_\mathbf{r}\bra{\psi^\alpha}_\mathbf{r}$ is the local projection operator. By using the coherent states, the resolution of the local projection operator can be obtained as
\begin{align}
    \hat{\mathcal{P}}_\mathbf{r} &= \frac{n!}{2\pi^n} \int \left(\prod_{0 \le \alpha \le n-1}d^2 z_\alpha\right) \delta(\lvert{\mathbf{z}}\rvert - 1)
    \ket{\mathbf{z}}_\mathbf{r} \bra{\mathbf{z}}_\mathbf{r}
    \notag\\
    &\equiv  \int d\Omega\, \ket{\mathbf{z}(\Omega)}_\mathbf{r} \bra{\mathbf{z}(\Omega)}_\mathbf{r},
    \label{eq:SI:resolution}
\end{align}
where 
$\Omega \equiv \{\theta^{}_{1 \le \alpha \le n - 1}, \chi^{}_{0 \le \beta \le n - 1}\}$ 
and
\begin{align}
    d \Omega 
    &= 
    \prod_{1 \le \alpha \le n - 1} \sin^{2\alpha - 1}\theta_{\alpha}\cos\theta_{\alpha} d\theta_{\alpha}
    \times 
    \prod_{0 \le \beta \le n - 1} d \chi_\beta.
    \label{eq:SI:measure}
\end{align}
By using Eq.~\eqref{eq:SI:resolution}, we find
\begin{align}
    e^{-\mathcal{F}_\mathrm{eff}/T}
    &= \int \left( \prod_{\mathbf{r}} d\Omega_\mathbf{r} \right)
    \bra{\Psi[\Omega]} e^{-\hat{\mathcal{H}}_\mathrm{eff}/T} \ket{\Psi[\Omega]}
    \notag\\
    &\ge \int \left( \prod_{\mathbf{r}} d\Omega_\mathbf{r} \right)
    \exp\left(-\frac{\bra{\Psi[\Omega]} \hat{\mathcal{H}}_\mathrm{eff} \ket{\Psi[\Omega]}}{T}
    \right),
\end{align}
which corresponds to neglecting temporal fluctuations of the coherent states on each cluster. This corresponds to the variational treatment of the free energy~\cite{Chaikin-Lubensky}, and
\begin{align}
    H_\mathrm{eff} = \bra{\Psi[\Omega]} \hat{\mathcal{H}}_\mathrm{eff} \ket{\Psi[\Omega]}
    \label{eq:SI:Heff}
\end{align}
defines a generalized classical Hamiltonian for the chosen ansatz.
Below, we discuss the two types of updates used in our study.

\subsection{Metropolis update}
By using
\begin{align}
    \xi_\alpha = \sin^{2\alpha} \theta_\alpha \in [0,1],~~1 \le \alpha \le n - 1,
\end{align}
we can rewrite the measure~\eqref{eq:SI:measure} of the coherent-state integral per site as
\begin{align}
    d \Omega 
    &= \frac{1}{2^{n-1}(n-1)!}
    \prod_{1 \le \alpha \le n - 1} d \xi_\alpha
    \times 
    \prod_{0 \le \beta \le n - 1} d \chi_\beta.
\end{align}
Hence, the $(\xi,\chi)$ scheme is convenient for Metropolis MC, because a trial state can be generated by drawing the parameters uniformly and randomly from their respective intervals. The energy difference is then computed, and the trial state is accepted or rejected according to the standard Metropolis criterion.

\subsection{Over-relaxation update}
In the over-relaxation method, we perform a local update that does not change the classical energy $H_\mathrm{eff}$~\eqref{eq:SI:Heff}. By performing a random walk within the constant-energy manifold in the configuration space, the system is expected to escape from local minima more quickly than the process with only Metropolis updates. Although this method is discussed in the literature~\cite{Iwazaki2023}, we include the following discussion for completeness.

When we fix all local states but one (say, located at $\mathbf{r}$, which is selected either sequentially or randomly), $H_\mathrm{eff}$ depends only on $\mathbf{z}_\mathbf{r} = \mathbf{z}(\Omega_\mathbf{r})$. 
We collect all the interaction terms in the (quantum) Hamiltonian $\hat{\mathcal{H}}_\mathrm{eff}$ that act on the local Hilbert space and express them by using on-site generators of SU($n$),
$\hat{\mathcal{S}}_l(\mathbf{r}) = \sum_{\alpha,\beta} S_l^{\alpha\beta} \ket{\psi^\alpha}_\mathbf{r} \bra{\psi^\beta}_\mathbf{r}$, 
where $l$ is the generator index ($1 \le l \le n^2 - 1$). In this work, such terms include two-site interactions,
\begin{align}
    \hat{\mathcal{H}}_\mathrm{eff}^{(2)}(\mathbf{r}, \mathbf{r}') = 
    \sum_{l,m} h^{(2)}_{lm}(\mathbf{r}, \mathbf{r}') 
    \hat{\mathcal{S}}_l(\mathbf{r}) \hat{\mathcal{S}}_m(\mathbf{r}'),~~
    \mathbf{r}' \ne \mathbf{r},
\end{align}
and on-site interactions,
\begin{align}
    \hat{\mathcal{H}}_\mathrm{eff}^{(1)}(\mathbf{r})
    = -\sum_{l} h^{(1)}_{l}(\mathbf{r}) 
    \hat{\mathcal{S}}_l(\mathbf{r}),
\end{align}
where $h^{(1)}_{l}$ and $h^{(2)}_{lm}$ are the expansion coefficients. By taking the coherent state expectation values,
\begin{align}
    \left\langle{ \hat{\mathcal{H}}_\mathrm{eff}^{(2)}(\mathbf{r}, \mathbf{r}') }\right\rangle
    =
    \sum_{l,m} 
    \sum_{0 \le \alpha,\beta,\gamma,\delta \le n - 1}
    h^{(2)}_{lm}(\mathbf{r}, \mathbf{r}') 
    S_l^{\alpha\beta} S_m^{\gamma\delta}
    \left(z_\mathbf{r}^{\alpha}\right)^\ast
    z_\mathbf{r}^{\beta}
    \left(z_{\mathbf{r}'}^{\gamma}\right)^\ast
    z_{\mathbf{r}'}^{\delta},
\end{align}
and
\begin{align}
    \left\langle{ \hat{\mathcal{H}}_\mathrm{eff}^{(1)}(\mathbf{r}) }\right\rangle
    =
    -\sum_{l} 
    \sum_{0 \le \alpha,\beta \le n - 1}
    h^{(1)}_{l}(\mathbf{r}) 
    S_l^{\alpha\beta} 
    \left(z_\mathbf{r}^{\alpha}\right)^\ast
    z_\mathbf{r}^{\beta},
\end{align}
we find the contribution to $H_\mathrm{eff}$ that depends on $\mathbf{z}_\mathbf{r} = \mathbf{z}(\Omega_\mathbf{r})$ as
\begin{align}
    H_{\mathrm{eff},\mathrm{local}}(\mathbf{r})
    &= 
    -\sum_{0 \le \alpha,\beta \le n - 1}
    \left(z_\mathbf{r}^{\alpha}\right)^\ast
    h_\mathrm{eff}^{\alpha\beta}(\mathbf{r}) 
    z_\mathbf{r}^{\beta},
    \label{eq:Weiss}
\end{align}
where
\begin{align}
    h_\mathrm{eff}^{\alpha\beta}(\mathbf{r}) 
    =
    \sum_{l} 
    \left(
       h^{(1)}_{l}(\mathbf{r}) 
       -
       \sum_{\mathbf{r}' \ne \mathbf{r}}
       \sum_{m} 
       \sum_{0 \le \gamma,\delta \le n - 1}
       h^{(2)}_{lm}(\mathbf{r}, \mathbf{r}') 
       S_m^{\gamma\delta}
       \left(z_{\mathbf{r}'}^{\gamma}\right)^\ast
    z_{\mathbf{r}'}^{\delta}
    \right)
    S_l^{\alpha\beta},
\end{align}
is a Hermitian matrix representing the generalized Weiss field. 

By diagonalizing $h_\mathrm{eff}^{\alpha\beta}(\mathbf{r})$, we obtain real eigenvalues $\lambda_{\alpha}$ and eigenvectors $\mathbf{u}_{\alpha}$ ($0 \le \alpha \le n - 1$), which constitute a unitary matrix $U = (\mathbf{u}_{0} \dots \mathbf{u}_{n-1})$. With this transformation, we obtain
\begin{align}
    H_{\mathrm{eff},\mathrm{local}}(\mathbf{r})
    = \sum_{0 \le \alpha \le n - 1}
    \lambda_{\alpha}
    \lvert{\zeta^\alpha_\mathbf{r}}\rvert^2,~~
    \zeta_\mathbf{r}^\alpha = (U^\dag \mathbf{z}_\mathbf{r})^\alpha
    = \sum_{0 \le \beta \le n - 1} \left(u_{\alpha}^{\beta}\right)^\ast z_\mathbf{r}^\beta.
\end{align}
Thus, updating $ \left(\zeta_\mathbf{r}^\alpha\right)' = e^{i\varphi_\alpha} \zeta_\mathbf{r}^\alpha$ for $0 \le \alpha \le n - 1$ with a set of arbitrary phase factors $\{\varphi^{}_{0 \le \alpha \le n - 1}\}$ does not change $H_{\mathrm{eff},\mathrm{local}}(\mathbf{r})$, and hence $H_\mathrm{eff}$, providing an over-relaxation update. The corresponding update for $\mathbf{z}_\mathbf{r}$ is
\begin{align}
    \mathbf{z}_\mathbf{r}'
    = U \bm{\zeta}_\mathbf{r}'
    = U
    \begin{pmatrix}
        e^{i\varphi_0} & & & & \\
        & e^{i\varphi_1} & & & \\
        & & e^{i\varphi_2} & & \\
        & & & \ddots & \\
        & & & & e^{i\varphi_{n-1}}
    \end{pmatrix}
    U^\dag \mathbf{z}_\mathbf{r}.
\end{align}

\section{%
    Generalized classical Monte Carlo (MC) results in the low-field (LF) regime: \texorpdfstring{$\mathbb{CP}^1\times\mathbb{CP}^1$}{CP1xCP1} and \texorpdfstring{$\mathbb{CP}^3$}{CP3} theories
    \label{SI:MC-LF}
}%
\setcounter{equation}{0}

\subsection{%
    Variational ansatz with and without spin--orbital entanglement
}%
To clarify the role of spin--orbital entanglement (SOE) in the LF regime, we compare two trimer-based generalized classical descriptions of $\hat{\mathcal{H}}_{\mathrm{KK}}$. The first is a factorized $\mathbb{CP}^1\times\mathbb{CP}^1 \simeq S^2 \times S^2$ coherent-state ansatz, in which the spin and orbital degrees of freedom are treated as independent classical vectors of length $S=\tau=\frac{1}{2}$:
\begin{align}
     \ket{\Psi^{\mathbb{CP}^1\times\mathbb{CP}^1}\{\Omega_S, \Omega_\tau\}} 
     = \bigotimes_\mathbf{r} \left[
     \mathcal{R}_{S,\mathbf{r}}^{\mathrm{SU(2)}}(\Omega_{S,\mathbf{r}})
     \otimes
     \mathcal{R}_{\tau,\mathbf{r}}^{\mathrm{SU(2)}}(\Omega_{\tau,\mathbf{r}})
     \right]
     \ket{S^z = \tfrac{1}{2}, \tau^x = \tfrac{1}{2}}_{\mathbf{r}}.   
     \label{SI:eq:CP1xCP1}
\end{align}
Here, $\mathcal{R}_{S,\mathbf{r}}^{\mathrm{SU(2)}}$ and $\mathcal{R}_{\tau,\mathbf{r}}^{\mathrm{SU(2)}}$ denote SU(2) rotations in the spin and orbital spaces of a trimer at $\mathbf{r}$, parametrized by Euler angles $\Omega_{S,\mathbf{r}}$ and $\Omega_{\tau,\mathbf{r}}$, respectively. The corresponding classical Hamiltonian is
\begin{align}
    H^{\mathbb{CP}^1\times\mathbb{CP}^1}_{\mathrm{eff}}
    =
    \bra{\Psi^{\mathbb{CP}^1\times\mathbb{CP}^1}}
    \hat{\mathcal{H}}_\mathrm{KK}
    \ket{\Psi^{\mathbb{CP}^1\times\mathbb{CP}^1}}.
    \label{SI:eq:Heff:CP1xCP1}
\end{align}
This ansatz provides a minimal classical description of the trimer gap between the $S=\frac{1}{2}\otimes\tau=\frac{1}{2}$ and $S=\frac{3}{2}$ sectors and has been used previously for trimerized spin systems~\cite{Ferrero2003,Kamiya2012a,Tanaka2021}. However, because spin and orbital variables are factorized, SOE is absent by construction.

To include SOE, we instead use the $\mathbb{CP}^3$ coherent-state ansatz on the four-dimensional spin--orbital manifold,
\begin{align}
     \ket{\Psi_{\mathbb{CP}^3}\{\Omega_{\mathbb{CP}^3}\}}
     =
     \bigotimes_{\mathbf{r}}
     \mathcal{R}_{\mathbf{r}}^{\mathrm{SU(4)}}(\Omega_{\mathbb{CP}^3,\mathbf{r}})
     \ket{S^z = \tfrac{1}{2},\tau^x = \tfrac{1}{2}}_{\mathbf{r}}.
\end{align}
Here, $\mathcal{R}_{\mathbf{r}}^{\mathrm{SU(4)}}$ acts on the local $S=\frac{1}{2}\otimes\tau=\frac{1}{2}$ Hilbert space, and $\Omega_{\mathbb{CP}^3,\mathbf{r}}$ parametrizes a general element of $\mathbb{CP}^3$. The corresponding generalized classical Hamiltonian is
\begin{align}
    H_{\mathrm{eff},\mathbb{CP}^3}
    =
    \bra{\Psi_{\mathbb{CP}^3}}
    \hat{\mathcal{H}}_\mathrm{KK}
    \ket{\Psi_{\mathbb{CP}^3}}.
\end{align}

\subsection{Bulk electric polarization}
The effective in-plane electric dipole of a trimer is proportional to the orbital pseudospin, $\hat{P}^{x(y)}_\mathrm{eff} \propto \pm \hat{\tau}^{x(y)}$, where the sign reflects the opposite orientations of the (\textit{P})- and (\textit{M})-TNN molecules in \CrystalTNN. As discussed in the main text and in \ref{SI:Peff}, the proportionality constant scales as $(t_0/U)^3$. Setting this constant to unity when presenting numerical results, the bulk electric polarization is evaluated as
\begin{align}
    \hat{P}^{x(y)}_\mathrm{eff}
    =
    \frac{1}{N_\mathrm{trimer}}
    \left(
    \sum_{\mathbf{r} \in \textrm{even}} \hat{\tau}^{x(y)}_{\mathbf{r}}
    -
    \sum_{\mathbf{r} \in \textrm{odd}} \hat{\tau}^{x(y)}_{\mathbf{r}}
    \right),
\end{align}
where $N_\mathrm{trimer}$ is the total number of trimers.

\subsection{%
    Diagnostic $\mathbb{CP}^1\times\mathbb{CP}^1$ MC results
}%
\begin{figure}
  \centering
  \includegraphics[width=\hsize]{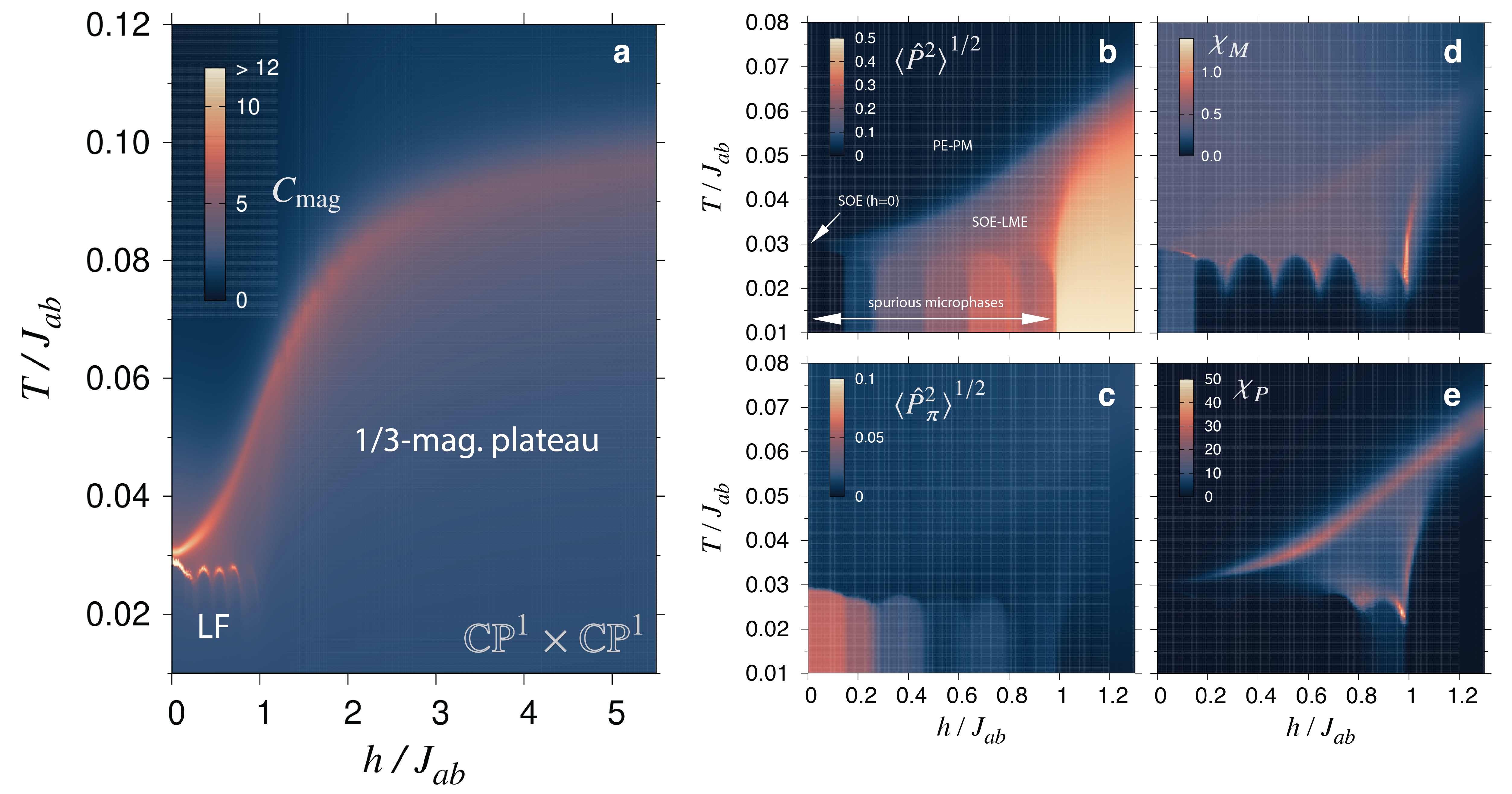}
  \vspace{0pt}
  \caption{%
    \label{FIG_2026_SI3}
    \textbf{%
        Generalized classical $\mathbb{CP}^1\times\mathbb{CP}^1$ Monte Carlo (MC) results.
    }%
    \textbf{(a)} 
    Magnetic specific heat $C_\mathrm{mag}$.
    The system size is $12 \times 12 \times 6$ with periodic boundary conditions. 
    \textbf{(b)}
    Electric polarization $P$.
    \textbf{(c)}
    Antiferroelectric polarization $P_\pi$.
    \textbf{(d)}
    Magnetic susceptibility $\chi_{M}^{}$.
    \textbf{(e)}
    Electric susceptibility $\chi_{P}^{}$.
    We take $J_{ab}$ as the unit of energy and set $J_c = 0.3J_{ab}$. 
  }%
\end{figure}

We first summarize the $\mathbb{CP}^1\times\mathbb{CP}^1$ MC results as a diagnostic comparison. The purpose is not to establish this factorized ansatz as the final LF description, but to show how neglecting SOE qualitatively changes the phase diagram. The simulations use standard Metropolis and over-relaxation updates on $N_{ab} \times N_{ab} \times N_c$ clusters with periodic boundary conditions, where $N_\mathrm{trimer}=N_{ab}^2N_c$. We set $J_c=0.3J_{ab}$ to reflect the quasi-two-dimensionality of \CrystalTNN.
       
Figure~\ref{FIG_2026_SI3} summarizes the resulting phase diagram. The overall trend that ferroelectric and antiferroelectric correlations are favored at higher and lower magnetic fields, respectively, is broadly consistent with experiment. However, the factorized $\mathbb{CP}^1\times\mathbb{CP}^1$ ansatz also produces several features that are not observed experimentally: the LF transition temperatures are strongly suppressed, and the low-temperature phase diagram contains a cascade of field-induced microphases. These phases have large magnetic unit cells, many of which may correspond to incommensurate modulations. In any case, their energies in the zero-temperature limit are higher than those obtained in the $\mathbb{CP}^3$ simulations discussed below. We therefore regard these microphases as artifacts of imposing a classical product structure between spin and orbital variables.

\begin{figure}[!t]
  \centering
  \includegraphics[width=0.85\hsize]{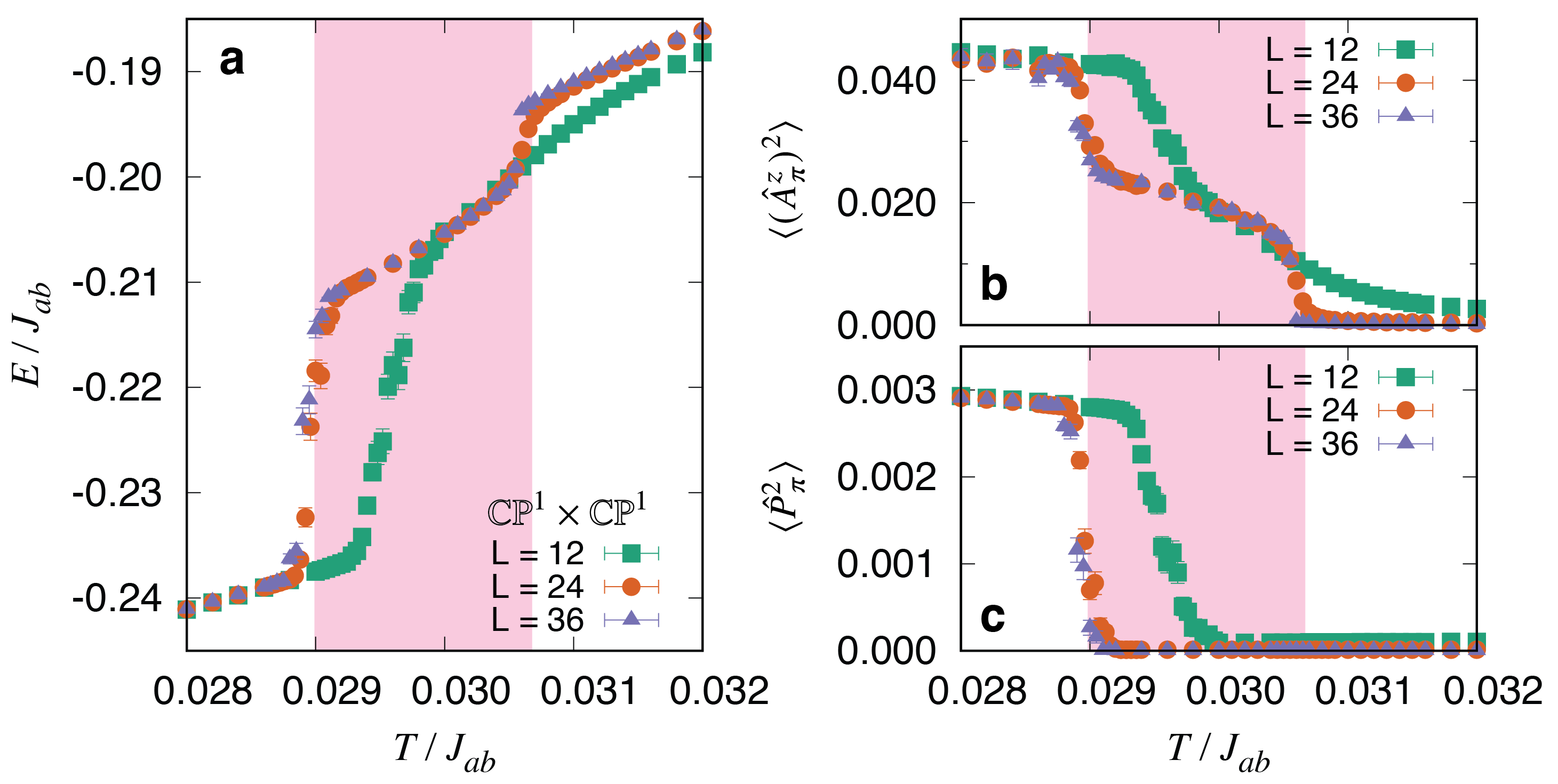}
  \vspace{0pt}
  \caption{%
    \label{FIG_2026_SI4}
    \textbf{
    Generalized classical $\mathbb{CP}^1\times\mathbb{CP}^1$ MC simulation results at $h = 0$.
    }
    \textbf{(a)}
    Temperature dependence of the internal energy density $E$.
    \textbf{(b)}
    Antiferromagnetic order parameter $\langle{(\hat{A}^z_\pi)^2}\rangle$, where
    $\hat{A}^z_\pi = N_\mathrm{trimer}^{-1} \bigl(\sum_{\mathbf{r} \in \textrm{even}} \hat{A}^z_{\mathbf{r}} - \sum_{\mathbf{r} \in \textrm{odd}} \hat{A}^z_{\mathbf{r}}\bigr)$.
    \textbf{(c)}
    Antiferroelectric order parameter
    $\langle{\hat{P}^2_\pi}\rangle$, with $\hat{P}^2_\pi = (\hat{P}^{x}_\pi)^2 + (\hat{P}^{y}_\pi)^2$
    and $\hat{P}^{x(y)}_\pi = N_\mathrm{trimer}^{-1} \sum_{\mathbf{r}} \hat{\tau}^{x(y)}_{\mathbf{r}}$.
    These results indicate a narrow intermediate-$T$ SOE phase at $0.0290(5) \le T / J_{ab} < 0.0307(2)$.
    We take $J_{ab}$ as the unit of energy and set $J_c = 0.3J_{ab}$. 
  }%
\end{figure}

One useful remnant of the $\mathbb{CP}^1\times\mathbb{CP}^1$ analysis is that it identifies the composite character of the SOE order. At $h=0$, there is a narrow intermediate-temperature phase (Fig.~\ref{FIG_2026_SI4}). The corresponding order parameter is neither purely spin nor purely orbital, but the composite combination $\hat{\mathbf{S}}(\hat{\tau}^x \pm i\hat{\tau}^y)$. This order physically corresponds to the 120$^\circ$ state, as the composite order parameter is obtained by projecting the trimer antiferromagnetic operator $\hat{\mathbf{A}}_{\mathbf{r}}$ onto the LF manifold,
\begin{align}
  \hat{\mathcal{P}}_\mathrm{LF}\, \hat{\mathbf{A}}_\mathbf{r} \hat{\mathcal{P}}_\mathrm{LF}
  =
  -2 \hat{\mathbf{S}}_\mathbf{r}
  \left(\hat{\tau}^x_\mathbf{r}+i\hat{\tau}^y_\mathbf{r}\right),
\end{align}
where
\begin{align}
    \hat{\mathbf{A}}_{\mathbf{r}}
    =
    \hat{\mathbf{s}}_{\mathbf{r},0}
    + e^{-2\pi i/3}\hat{\mathbf{s}}_{\mathbf{r},1}
    + e^{2\pi i/3}\hat{\mathbf{s}}_{\mathbf{r},2}.
\end{align}
The corresponding staggered order parameter,
\begin{align}
    \hat{\mathbf{A}}_\pi
    =
    \frac{1}{N_\mathrm{trimer}}
    \left(
    \sum_{\mathbf{r} \in \textrm{even}} \hat{\mathbf{A}}_{\mathbf{r}}
    -
    \sum_{\mathbf{r} \in \textrm{odd}} \hat{\mathbf{A}}_{\mathbf{r}}
    \right),
    \label{eq:SI:A}
\end{align}
is analyzed in Fig.~\ref{FIG_2026_SI4}{b}. At finite magnetic field, this SOE order induces electric polarization through the linear magnetoelectric (LME) effect, giving rise to a ferroelectric intermediate-temperature phase. Correspondingly, the electric susceptibility $\chi_P = d\langle P\rangle / dE$ shows a divergent peak at the transition, whereas the magnetic susceptibility $\chi_M=d\langle M^z\rangle/dh$ exhibits only a nondivergent anomaly (Figs.~\ref{FIG_2026_SI3}{d} and \ref{FIG_2026_SI3}{e}).

\subsection{%
    Generalized classical $\mathbb{CP}^3$ MC results
}%

\begin{figure}[!t]
  \centering
  \includegraphics[width=0.9\hsize]{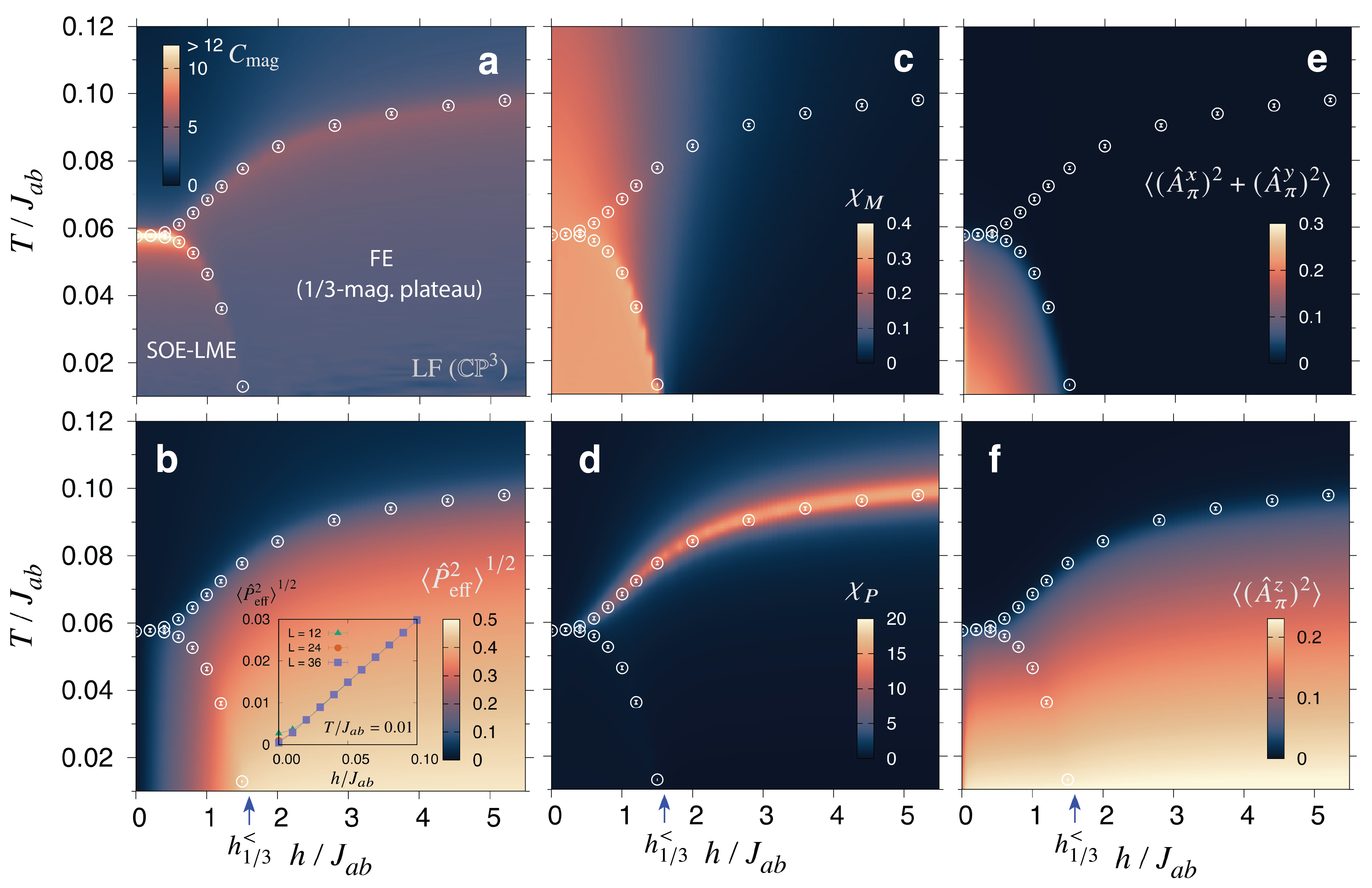}
  \caption{%
    \label{FIG_2026_SI5}
    \textbf{
    Generalized classical $\mathbb{CP}^3$ Monte Carlo (MC) results in the low-field and plateau regimes.
    }
    \textbf{(a)} Magnetic specific heat $C_\mathrm{mag}$.
    \textbf{(b)} Electric polarization $P = \langle{\hat{P}^2}\rangle^{1/2}$, with the inset demonstrating linear magnetoelectric (LME) behavior induced by spin--orbital-entangled (SOE) order.
    \textbf{(c)}
    Magnetic susceptibility $\chi_{M}^{}$.
    \textbf{(d)}
    Electric susceptibility $\chi_{P}^{}$.    
    \textbf{(e)}
    Average of the squared $xy$ components of $\hat{\mathbf{A}}_\pi$, $\langle{(\hat{A}^x_\pi)^2 + (\hat{A}^y_\pi)^2}\rangle$.
    \textbf{(f)}
    Average of the squared $z$ component of $\hat{\mathbf{A}}_\pi$, $\langle{(\hat{A}^z_\pi)^2}\rangle$.
    The system size for the intensity plot is $(N_{ab},N_c)=(12,6)$.
    We take $J_{ab}$ as the unit of energy and set $J_c = 0.3J_{ab}$. 
  }%
\end{figure}

The $\mathbb{CP}^3$ simulations provide the LF phase diagram used in the main text. At $h=0$, analysis of $\langle \hat{\mathbf{A}}_\pi^2\rangle$ and the Binder ratio confirms the SOE phase, which persists down to $T=0$. At finite field, the microphases found in the factorized $\mathbb{CP}^1\times\mathbb{CP}^1$ description disappear. Instead, the SOE phase extends to $h>0$ through the LME effect, producing the SOE-LME phase shown in Fig.~\ref{FIG_2026_SI5}. The transition into the SOE-LME phase is determined from the Binder ratio associated with the transverse components of $\hat{\mathbf{A}}_\pi$,
\begin{align}
    U_{\mathbf{A}_\pi}^{xy} = 
    \frac{
    \left\langle
    \left[
    (\hat{A}^x_\pi)^2+(\hat{A}^y_\pi)^2
    \right]^2
    \right\rangle
    }{
    \left\langle
    (\hat{A}^x_\pi)^2+(\hat{A}^y_\pi)^2
    \right\rangle^2
    }.
\end{align}
The ferroelectric $1/3$-magnetization plateau phase is instead characterized by a nonzero longitudinal component of $\hat{\mathbf{A}}_\pi$ and vanishing transverse components. We therefore use the Binder ratio
\begin{align}
    U_{\mathbf{A}_\pi}^{z} = 
    \frac{
    \langle{(\hat{A}^z_\pi)^4}\rangle
    }{
    \langle{(\hat{A}^z_\pi)^2}\rangle^2
    }
\end{align}
to determine the transition into the plateau phase.

Both the SOE-LME phase at $h > 0$ and the $1/3$-magnetization plateau phase exhibit ferroelectricity, and the electric susceptibility $\chi_P$ shows a pronounced peak at the transition upon cooling from high temperatures (Fig.~\ref{FIG_2026_SI5}{d}). The two phases can also be distinguished by the magnetic susceptibility $\chi_M$ in the zero-temperature limit: $\chi_M(T\to0)$ remains finite in the SOE-LME phase but vanishes in the plateau phase (Fig.~\ref{FIG_2026_SI5}{c}).

\section{%
    Generalized classical \texorpdfstring{$\mathbb{CP}^2$}{CP2} Monte Carlo (MC) results in the high-field (HF) regime
    \label{SI:MC-HF}    
}%
\setcounter{equation}{0}

\subsection{%
    Summary of generalized classical $\mathbb{CP}^2$ MC simulation results
}%

\begin{figure}[!t]
  \centering
  \includegraphics[width=0.9\hsize]{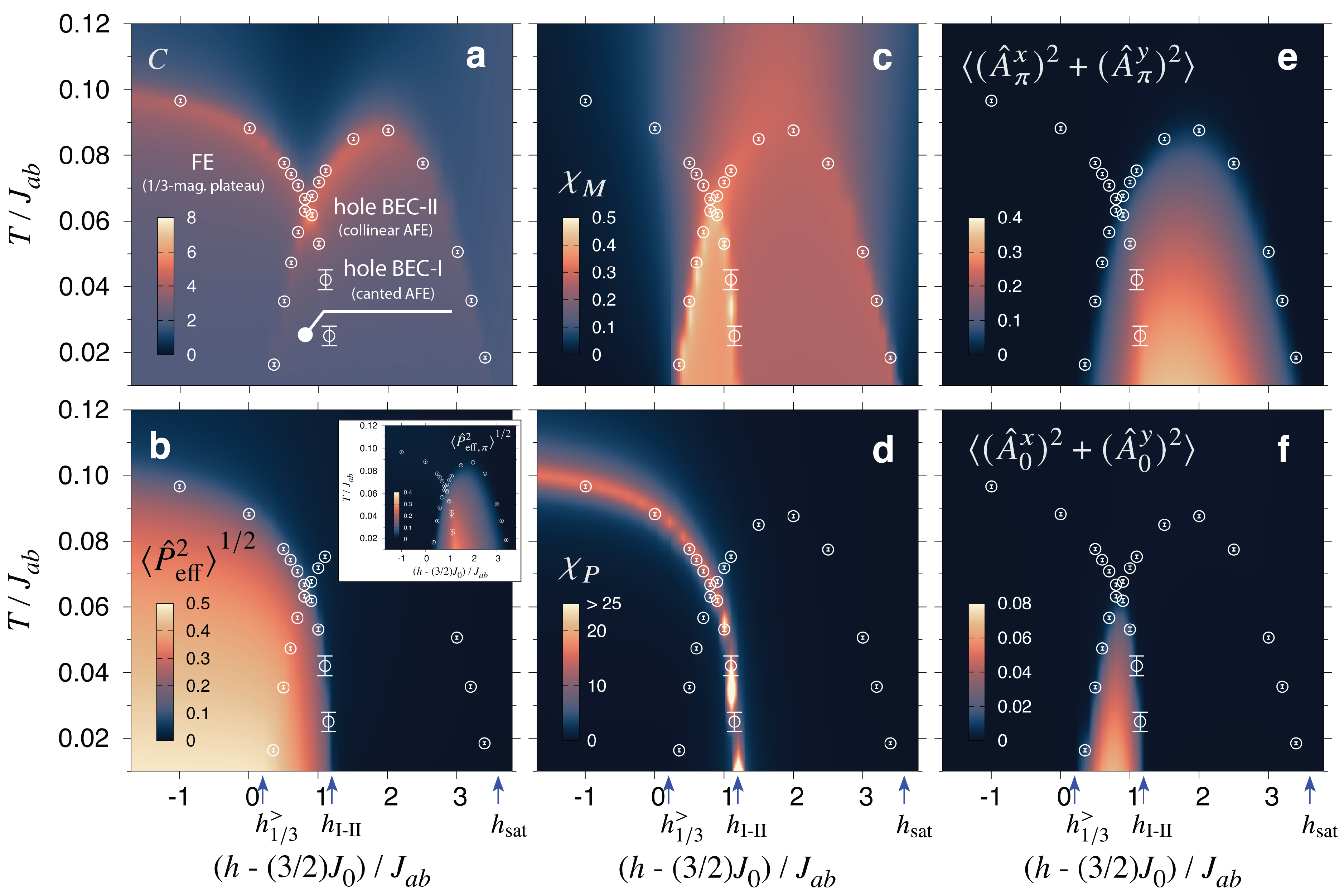}
  \caption{%
    \label{FIG_2026_SI6}
    \textbf{
        Generalized classical $\mathbb{CP}^2$ Monte Carlo (MC) results in the high-field regime.
    }
    \textbf{(a)} Magnetic specific heat $C_\mathrm{mag}$.
    \textbf{(b)} Electric polarization $P = \langle{\hat{P}^2}\rangle^{1/2}$, with the inset showing the antiferroelectric (AFE) order parameter $P_\pi = \langle{\hat{P}_\pi^2}\rangle^{1/2}$ characterized by a $\pi$ phase shift between even and odd layers.  
    \textbf{(c)}
    Magnetic susceptibility $\chi_{M}^{}$.
    \textbf{(d)}
    Electric susceptibility $\chi_{P}^{}$.
    \textbf{(e)}
    Average squared transverse component of $\hat{\mathbf{A}}_\pi$, $\langle{(\hat{A}^x_\pi)^2 + (\hat{A}^y_\pi)^2
    }\rangle$.
    \textbf{(f)}
    Average squared transverse component of $\hat{\mathbf{A}}_0$, $\langle{(\hat{A}^x_0)^2 + (\hat{A}^y_0)^2
    }\rangle$.
    The system size for producing the intensity plot is $(N_{ab}, N_{c}) = (12, 6)$.
    We take $J_{ab}$ as the unit of energy and set $J_c = 0.3J_{ab}$.
  }%
\end{figure}

In the HF regime, the local low-energy space is three-dimensional, and we use the corresponding $\mathbb{CP}^2$ coherent-state ansatz,
\begin{align}
     \ket{\Psi_{\mathbb{CP}^2}\{\Omega_{\mathbb{CP}^2}\}} = \bigotimes_{\mathbf{r}} \mathcal{R}_{\mathbf{r}}^{\mathrm{SU(3)}}(\Omega_{\mathbb{CP}^2,\mathbf{r}})  \ket{S^z = \tfrac{1}{2},\tau^x = \tfrac{1}{2}}_{\mathbf{r}}.
\end{align}
Here, $\mathcal{R}_{\mathbf{r}}^{\mathrm{SU(3)}}$ is an SU(3) transformation acting on the orbital-hole Hilbert space spanned by $\ket{S^z = \tfrac{1}{2}, \tau^x = \pm\tfrac{1}{2}}$ and $\ket{\up\up\up}$ on the trimer at $\mathbf{r}$. The variational parameter $\Omega_{\mathbb{CP}^2,\mathbf{r}}$ specifies an element of the two-dimensional complex projective space $\mathbb{CP}^2$. The corresponding generalized classical Hamiltonian is defined as
\begin{align}
    H_{\mathrm{eff},\mathbb{CP}^2} \equiv \bra{\Psi_{\mathbb{CP}^2}} \hat{\mathcal{H}}_\text{b-\textit{tJ}}
    \ket{\Psi_{\mathbb{CP}^2}}.
\end{align}
As in the $\mathbb{CP}^3$ simulations, both Metropolis and over-relaxation updates are employed in the MC simulations based on $H_{\mathrm{eff},\mathbb{CP}^2}$.

The condensate order parameter $\langle{\hat{b}^\dag_{\pm}}\rangle$ 
of the hole BEC phases is associated with the antiferromagnetic order parameter $\hat{\mathbf{A}}$ as follows:
\begin{align}
  \hat{\mathcal{P}}_\mathrm{HF}\, \left(\hat{A}_{\mathbf{r}}^x - i\hat{A}_{\mathbf{r}}^y\right) \hat{\mathcal{P}}_\mathrm{HF}
  &= -\frac{3}{\sqrt{6}} \left( \hat{b}^\dag_{\mathbf{r},+} - i\hat{b}^\dag_{\mathbf{r},-}\right).
  \label{eq:A:HF}
\end{align}
Accordingly, the hole BEC-I and hole BEC-II phases can be interpreted as antiferromagnetically ordered phases with nonzero transverse ordered components. More specifically, the canted antiferroelectric (AFE) hole BEC-I phase can be characterized by the coexistence of nonzero order parameters $\langle{\hat{A}_\pi^{x(y)}}\rangle$~\eqref{eq:SI:A} and $\langle{\hat{A}_0^{x(y)}}\rangle$, where
\begin{align}
    \hat{A}^{x(y)}_0 = \frac{1}{N_\mathrm{trimer}}\sum_\mathbf{r} \hat{A}^{x(y)}_{\mathbf{r}}.
\end{align}
In the collinear AFE hole BEC-II phase, on the other hand, only $\langle{\hat{A}^{x(y)}_\pi}\rangle$ is nonzero with vanishing $\langle{\hat{A}^{x(y)}_0}\rangle$ (Fig.~\ref{FIG_2026_SI6}).

In terms of susceptibilities, the magnetic susceptibility $\chi_M$ is proportional to the hole compressibility, remaining finite as $T \to 0$ within the hole BEC phases (Fig.~\ref{FIG_2026_SI6}{c}). The electric susceptibility $\chi_P$ exhibits a divergence upon cooling from high temperatures into either the plateau phase or the canted AFE hole BEC-I phase, both corresponding to ordered phases with nonzero net polarization (Fig.~\ref{FIG_2026_SI6}{d}).

\subsection{%
    Analysis of the kinetic magnetoelectric effect in the zero-temperature limit
}%
The competition between the orbital-exchange interaction and hole kinetic energy can be captured by a mean-field theory formulated in the $T \to 0$ limit of the $\mathbb{CP}^2$ framework. Motivated by the $\mathbb{CP}^2$ MC results at low temperatures, we adopt the following simplified variational wavefunction $\ket{\Psi_\mathrm{HF}^{T=0}} = \bigotimes_{\ell} \ket{\Psi_{\mathrm{HF}, \ell}^{T=0}}$ with
\begin{align}
    \ket{\Psi_{\mathrm{HF},\ell}^{T=0}} 
    &= \!\!\!\!\bigotimes_{ \mathbf{r}\in\text{$\ell$th layer}}
    \left[
    \cos\frac{\theta}{2} 
    \left(
        \cos\frac{\phi^{}_\ell}{2}     
        \ket{S^z = \tfrac{1}{2},\tau^x = \tfrac{1}{2}}_\mathbf{r}
        + \sin\frac{\phi^{}_\ell}{2} 
        \ket{S^z = \tfrac{1}{2},\tau^x = -\tfrac{1}{2}}_\mathbf{r}
    \right)
    + \sin\frac{\theta}{2} e^{i\chi^{}_\ell}
    \ket{\up\up\up}_\mathbf{r}
    \right],
\end{align}
where $0 \le \ell < N_c$ is the layer index.
This ansatz gives $\langle{\tau^z_\ell}\rangle = 0$ for all $\ell$ and the magnetization per trimer is uniform, 
\begin{align}
    M^z = \langle{M^z_\ell}\rangle = 1 - \frac{\cos\theta}{2},~\forall\ell.  
\end{align}
$\phi_\ell$ defines the azimuthal angle of the in-plane pseudospin,
\begin{align}
    \langle{\tau^x_\ell}\rangle
    &=
    \frac{1}{2}\cos^2\frac{\theta}{2} \cos\phi_\ell,
    \notag\\
    \langle{\tau^y_\ell}\rangle
    &=
    \frac{1}{2}\cos^2\frac{\theta}{2} \sin\phi_\ell,
\end{align}
and $\chi_\ell$ parametrizes the phase of the hole condensate,
\begin{align}
    \langle{b^\dag_{+,\ell}}\rangle
    &=
    \frac{1}{2}\sin\theta\cos\frac{\phi_\ell}{2} e^{i\chi_\ell},
    \notag\\
    \langle{b^\dag_{-,\ell}}\rangle
    &=
    \frac{1}{2}\sin\theta\sin\frac{\phi_\ell}{2} e^{i\chi_\ell}.
\end{align}
The variational ground-state energy per trimer is
\begin{align}
    E_{\mathrm{MF}} 
    &= \frac{1}{N_\mathrm{trimer}} \bra{\Psi_\mathrm{HF}^{T=0}} \hat{\mathcal{H}}_\text{b-\textit{tJ}} \ket{\Psi_\mathrm{HF}^{T=0}}
    \notag\\
    &= \frac{1}{N_c} \sum_{\ell}
    \left(g_1(M^z) + g_2(M^z) \cos\left(\phi_{\ell} - \phi_{\ell+1}\right) + g_3(M^z) \cos\left(\chi_\ell - \chi_{\ell + 1}\right) \cos\frac{\phi_\ell - \phi_{\ell + 1}}{2}\right),
\end{align}
where
\begin{align}
    g_1(M^z) &= \frac{3}{2}\left(M^z - \frac{1}{2}\right)^2 J_{ab} + \frac{(M^{z})^2}{3} J_{c} - M^z h,
    \notag\\
    g_2(M^z) &= \frac{1}{6} 
        \left(M^z - \frac{3}{2}\right)^2 J_c,
    \notag\\
    g_3(M^z) &= -\left(M^z - \frac{1}{2}\right)
        \left(M^z - \frac{3}{2}\right) J_c.
\end{align}

The pseudospin (or electric dipole) configuration for a given value of $M^z$ is determined by the competition between $g_2(M^z)$ and $g_3(M^z)$. For $J_c > 0$, both terms are non-negative across the entire HF range ($\frac{1}{2} \le M^z \le \frac{3}{2}$). The $g_2$ term arises from the interlayer effective orbital interaction, favoring $\phi_\ell - \phi_{\ell + 1} = \pi$ for $J_c > 0$. This antiferro-orbital state is stabilized within the 1/3-magnetization plateau, since $g_3 = 0$ in this situation ($M^z = \frac{1}{2}$). This state corresponds to ferroelectric order because of the opposite orientations of the (\textit{P})- and (\textit{M})-TNN molecules in \CrystalTNN, as mentioned in the main text. Once the system leaves the plateau, the $g_3$ term becomes nonzero, representing the kinetic energy of mobile holes arising from interlayer hopping. The kinetic $g_3$ term is not minimized at $\phi_\ell-\phi_{\ell+1}=\pi$, regardless of the sign of $\cos(\chi_\ell-\chi_{\ell+1})$, and therefore competes with the orbital-exchange $g_2$ term.

We propose a two-sublattice ansatz, with $\phi_\ell=\phi_{\rm even}$ and $\chi_\ell=\chi_{\rm even}$ for even $\ell$, and $\phi_\ell=\phi_{\rm odd}$ and $\chi_\ell=\chi_{\rm odd}$ for odd $\ell$. For $\tfrac{1}{2} < M^z < \tfrac{9}{10}$, the solution satisfies $\phi_\text{even} - \phi_\text{odd} = \pm \Delta\phi(M^z)$ with
\begin{align}
    \Delta\phi(M^z) &= 2 \cos^{-1} \left( -\frac{g_3}{4g_2} \right)
    = 2 \cos^{-1} \left( \frac{3}{2}\frac{M^z - \frac{1}{2}}{M^z - \frac{3}{2}} \right).
\end{align}
This solution describes a continuous evolution of the relative angle between the electric dipole moments of even and odd layers in the canted AFE hole BEC-I phase, representing the kinetic magnetoelectric effect. For $\tfrac{9}{10} \le M^z < \tfrac{3}{2}$, the solution is $\phi_\ell = \text{const.}$, which corresponds to the collinear AFE hole BEC-II phase.

\section{%
    Tensor-network calculation at the
    \texorpdfstring{\mbox{1/3-}magnetiza\-tion}{1/3-magnetization}
    plateau
    \label{SI:tensor-network}
}%
\setcounter{equation}{0}

\begin{figure}
  \centering
  \includegraphics[width=0.5\hsize]{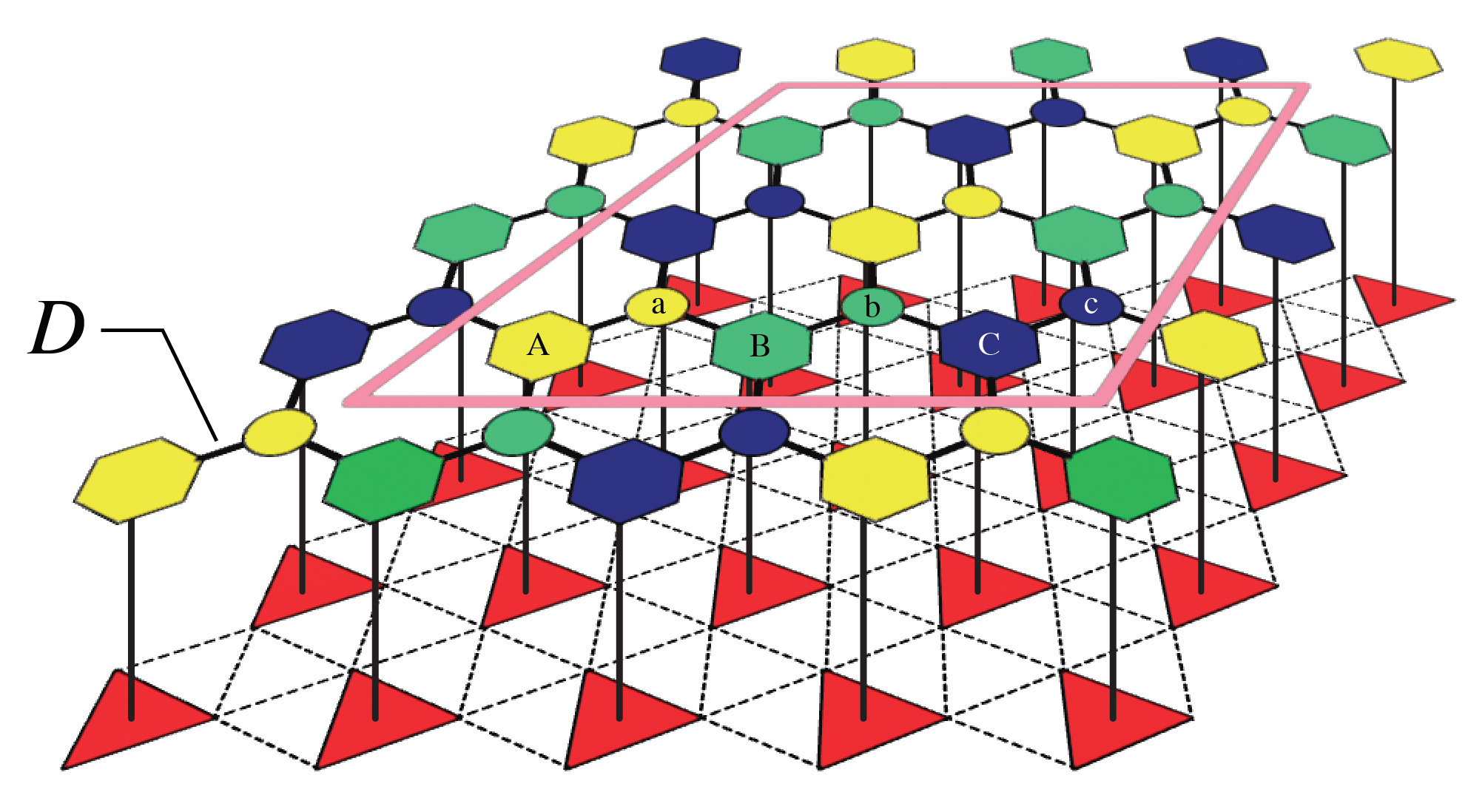}
  \vspace{0pt}
  \caption{%
    \label{FIG_2026_SI_TN}
    \textbf{%
        Projected entangled simplex state (PESS) ansatz used for the 2D trimerized triangular lattice. 
    }%
    The hexagons $A$--$C$ and circles $a$--$c$ denote tensors with and without physical indices (projecting tensors simplex tensors), respectively, all with bond dimension $D$. The $3\times3$ cell defines a subsystem preserving $C_3$ symmetry used in the corner transfer matrix calculations~\cite{Orus2009,Corboz2014}, where the environment bond dimension is $\chi$. 
  }%
\end{figure}

We use tensor-network calculations to examine the stability of the orbital order in the $1/3$-magnetization plateau regime beyond the generalized classical treatment. The variational state is represented by a projected entangled simplex state (PESS)~\cite{Xie2014}, and we adopt the 9-PESS ansatz with a $3\times3$ trimer unit cell appropriate for the trimerized triangular lattice, as shown in Fig.~\ref{FIG_2026_SI_TN}. Here, each trimer is represented by a physical index associated with the local low-energy spin--orbital manifold. The energy is evaluated using the corner transfer matrix renormalization group (CTMRG)~\cite{Orus2009,Corboz2014}. The tensor elements are optimized by automatic differentiation~\cite{Zygote.jl-2018, TensorOperations.jl}, following the strategy of Ref.~\cite{Liao2019}, starting from randomly initialized tensors. The environment bond dimension $\chi$ is first chosen large enough to ensure convergence of the energy during optimization. Specifically, we use $\chi=20,30,50,70$ for $D=2,3,4,5$, respectively. After the optimized tensors are obtained, we further check the convergence of observables with respect to $\chi$.

\begin{figure}
  \centering
  \includegraphics[width=\hsize]{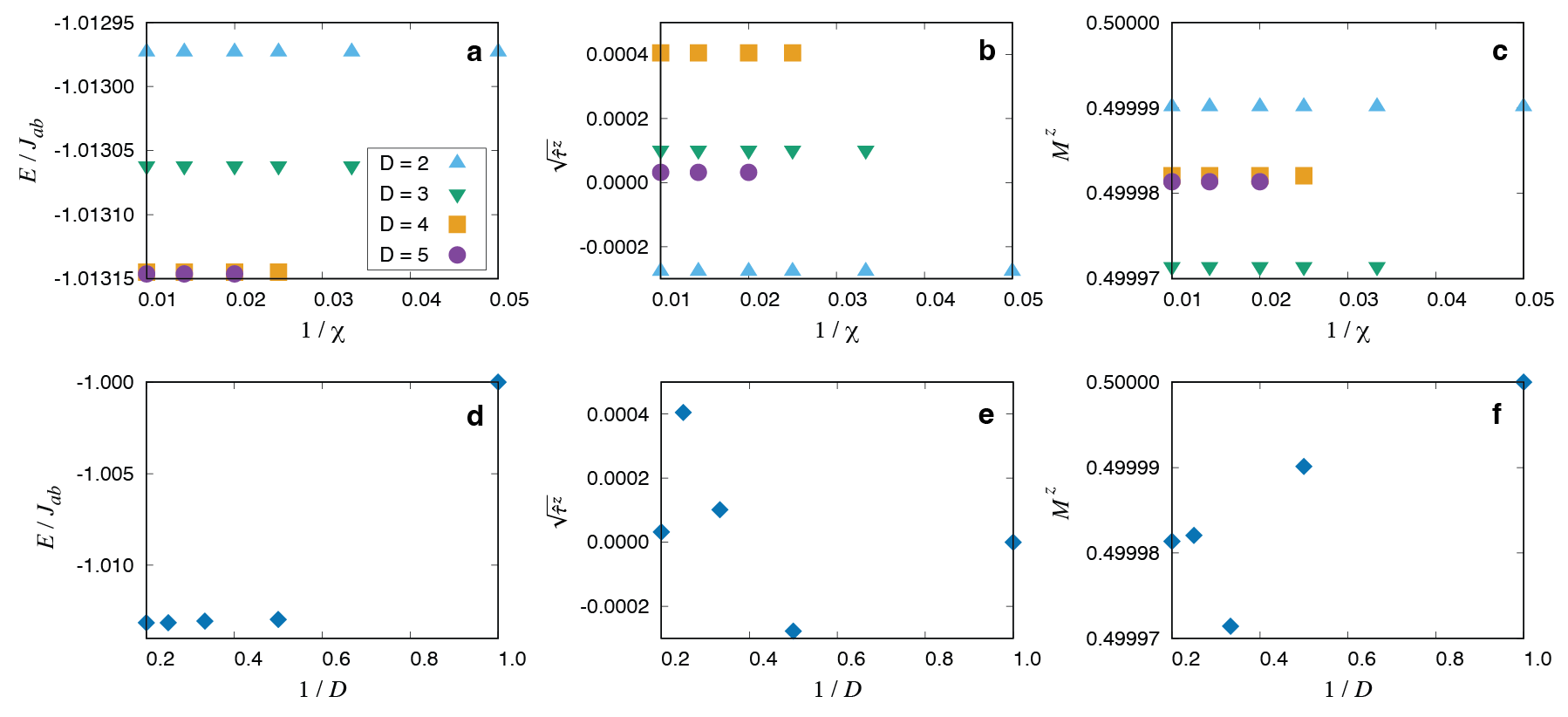}
  \vspace{0pt}
  \caption{%
    \label{FIG_2026_SI7}
    \textbf{%
        Convergence of observables in the tensor-network simulation.
    }%
    Convergence with respect to $1/\chi$ for different choices of $D$:
    \textbf{(a)} Energy per trimer,
    \textbf{(b)} $\langle{\hat{\tau}^z}\rangle$, and
    \textbf{(c)} magnetization.
    Convergence with respect to $1/D$:
    \textbf{(d)} Energy per trimer,
    \textbf{(e)} $\langle{\hat{\tau}^z}\rangle$, and
    \textbf{(f)} magnetization. The $D=1$ result corresponds to the polarized product state obtained from the mean-field calculation.
    All calculations are performed at $h=2J_{ab}$.
  }%
\end{figure}

The optimized $D=5$ state yields
\begin{align}
    E/J_{ab} = -1.0131463,
    \qquad
    \sqrt{
    \langle{\hat{\tau}^x}\rangle^2 
    + \langle{\hat{\tau}^y}\rangle^2
    } = 0.4666,
    \qquad
    \langle{\hat{\tau}^z}\rangle \simeq 0,
\end{align}
per trimer. The small changes between the $D=4$ and $D=5$ results indicate that the chosen bond dimensions are sufficient for the present calculation (Fig.~\ref{FIG_2026_SI7}).

\section{%
    Additional interlayer interactions and their consequences in the 1/3-magnetization plateau
    \label{SI:Interlayer}
}%
\begin{figure}
  \centering
  \includegraphics[width=0.95\hsize]{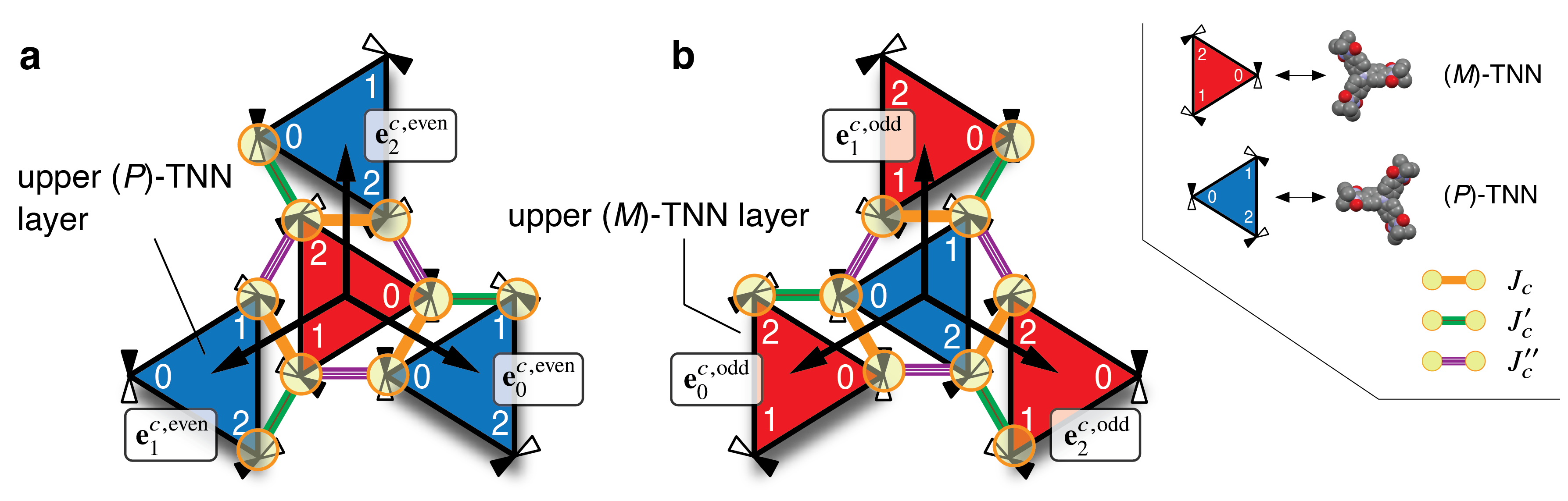}
  \vspace{0pt}
  \caption{%
    \label{FIG_2026_SI8}
    \textbf{
    Modified interlayer interactions.
    }
    \textbf{(a)}
    Interlayer interactions for a trimer in an even, (\textit{M})-TNN, layer coupled to three sites in the adjacent upper layer.
    \textbf{(b)}
    Interlayer interactions for a trimer in an odd, (\textit{P})-TNN, layer coupled to three sites in the adjacent upper layer.
    The bow-tie symbols made of open and filled triangles represent nitronyl nitroxide (NN) radicals with nonzero dihedral angles relative to the $ab$ plane, where the open (filled) triangles are directed towards the adjacent upper (lower) layers. 
  }%
\end{figure}

\setcounter{equation}{0}
We consider the effect of additional interlayer interactions $J_c'$ and $J_c''$ (Fig.~\ref{FIG_2026_SI8}). In the main text, we only consider $J_c$ by neglecting $J_c'$ and $J_c''$, because the tilted molecular planes of the nitronyl nitroxide (NN) radicals suggest the largest wavefunction overlap along the exchange path of $J_c$ (Fig.~\ref{FIG_2026_SI8}). An important question is whether the ferroelectricity in the 1/3-magnetization plateau phase for $J'_c = J''_c = 0$ remains robust upon including these interactions. For a trimer at $\mathbf{r}$ in an even layer, the modified interlayer interaction with trimers in the adjacent upper layer is
\begin{align}
  \hat{\mathcal{H}}^\mathrm{c,\,even-odd}_{\mathbf{r}}
 &= \sum_{0 \le \mu \le 2} \hat{\mathbf{s}}_{\mathbf{r},\mu} \cdot 
 \left(
 J_c \hat{\mathbf{s}}_{\mathbf{r}+\mathbf{e}_{\mu}^{c,\mathrm{even}}\!\!\!\!\!\!\!\!\!,\,\,\,\,\,\,\,\mu} 
 + J_c' \hat{\mathbf{s}}_{\mathbf{r}+\mathbf{e}_{\mu}^{c,\mathrm{even}}\!\!\!\!\!\!\!\!\!,\,\,\,\,\,\,\,\mu+1}
 + J_c'' \hat{\mathbf{s}}_{\mathbf{r}+\mathbf{e}_{\mu+2}^{c,\mathrm{even}}\!\!\!\!,\,\,\mu+2}
 \right),
\end{align}
where the intratrimer site indices $\mu$ are defined modulo three (Fig.~\ref{FIG_2026_SI8}{a}). Similarly, for a trimer in an odd layer, the corresponding interaction is
\begin{align}
  \hat{\mathcal{H}}^\mathrm{c,\,odd-even} _{\mathbf{r}}
  &= \sum_{0 \le \mu \le 2} \hat{\mathbf{s}}_{\mathbf{r},\mu} \cdot 
  \left(
  J_c \hat{\mathbf{s}}_{\mathbf{r}+\mathbf{e}_{\mu}^{c,\mathrm{odd}}\!\!\!\!\!\!\!\!,\,\,\,\,\,\,\mu} 
  + J_c' \hat{\mathbf{s}}_{\mathbf{r}+\mathbf{e}_{\mu}^{c,\mathrm{odd}}\!\!\!\!\!\!\!\!,\,\,\,\,\,\,\mu+2}
  + J_c'' \hat{\mathbf{s}}_{\mathbf{r}+\mathbf{e}_{\mu+1}^{c,\mathrm{odd}}\!\!\!,\,\,\mu+1}
  \right),
\end{align}
as shown in Fig.~\ref{FIG_2026_SI8}{b}. 

In the 1/3-magnetization plateau regime, we derive the effective interaction by projecting the above expressions onto the twofold-degenerate $S = S_z = 1/2$ subspace for each trimer. In this regime, each spin operator is projected as
\begin{align}
  \hat{\mathcal{P}}_{1/3}\, \hat{\mathbf{s}}_{\mathbf{r},\mu} \hat{\mathcal{P}}_{1/3}
  &= \frac{\hat{z}}{6} \left( 1 - 4 \hat{\bm{\tau}}_{\mathbf{r}} \cdot {\mathbf{n}}_\mu \right),
\end{align}
where $\hat{\mathcal{P}}_{1/3}$ denotes the projection operator and ${\mathbf{n}}_\mu$ is defined earlier. Using this formula, we obtain the effective Hamiltonians
\begin{align}
    \hat{\mathcal{H}}^\mathrm{c,\,even-odd}_{\mathbf{r},1/3}
    &= \frac{4}{9} \sum_{0 \le \mu \le 2} \bm{\tau}_{\mathbf{r}} \cdot {\mathbf{n}}_\mu
    \left(
        J_c \bm{\tau}_{\mathbf{r} + \mathbf{e}_{\mu}^{c,\mathrm{even}}} \cdot {\mathbf{n}}_\mu
        + J'_c \bm{\tau}_{\mathbf{r} + \mathbf{e}_{\mu}^{c,\mathrm{even}}} \cdot {\mathbf{n}}_{\mu+1}
        + J''_c \bm{\tau}_{\mathbf{r} + \mathbf{e}_{\mu+2}^{c,\mathrm{even}}} \cdot {\mathbf{n}}_{\mu+2}
    \right),
\end{align}
and
\begin{align}
    \hat{\mathcal{H}}^\mathrm{c,\,odd-even}_{\mathbf{r},1/3}
    &= \frac{4}{9}\sum_{0 \le \mu \le 2}     \bm{\tau}_{\mathbf{r}} \cdot {\mathbf{n}}_\mu
    \left(
    J_c \bm{\tau}_{\mathbf{r} + \mathbf{e}_{\mu}^{c,\mathrm{odd}}} \cdot {\mathbf{n}}_\mu
    + J'_c \bm{\tau}_{\mathbf{r} + \mathbf{e}_{\mu}^{c,\mathrm{odd}}} \cdot {\mathbf{n}}_{\mu+2}
    + J''_c \bm{\tau}_{\mathbf{r} + \mathbf{e}_{\mu+1}^{c,\mathrm{odd}}} \cdot {\mathbf{n}}_{\mu+1}
    \right),
\end{align}
for the even-odd and odd-even interlayer interactions, respectively.

In the following, we assume that each layer exhibits in-plane ferro-orbital order stabilized by intralayer effective orbital interactions. We denote the orbital order parameter in the $j$-th layer as
$\langle{\bm{\tau}}\rangle_j = (\cos\phi_j, \sin\phi_j, 0)$. For a site $\mathbf{r}$ in the ($2j$)-th layer, the mean-field expectation value of the interlayer interaction is given by
\begin{align}
    &\left\langle \hat{\mathcal{H}}^\mathrm{c,even-odd} _{\mathbf{r}} \right\rangle_\mathrm{MF}
    = \frac{4}{9}
    \sum_{0 \le \mu \le 2}
    \langle{\bm{\tau}}\rangle_{2j} \cdot \mathbf{n}_\mu
    \left[
        \langle{\bm{\tau}}\rangle_{2j+1} \cdot 
        \left(
            J_c \mathbf{n}_\mu
            + J'_c \mathbf{n}_{\mu+1}
            + J''_c \mathbf{n}_{\mu+2}
        \right)
    \right]
    \notag \\
    &\hspace{50pt}= \frac{2}{3}
    \left[
        \left( J_c - \frac{J'_c}{2} - \frac{J''_c}{2} \right) \cos(\phi_{2j+1} - \phi_{2j}) 
        - \frac{\sqrt{3}}{2} (J'_c - J''_c) \sin(\phi_{2j+1} - \phi_{2j})
    \right].
\end{align}
For a site $\mathbf{r}'$ in the $(2j+1)$-th layer, a similar expression is obtained:
\begin{align}
    &\left\langle \hat{\mathcal{H}}^\mathrm{c,odd-even} (\mathbf{r}') \right\rangle_\mathrm{MF}
     = \frac{4}{9}
    \sum_{0 \le \mu \le 2}
    \langle{\bm{\tau}}\rangle_{2j+1} \cdot \mathbf{n}_\mu
    \left[
        \langle{\bm{\tau}}\rangle_{2j+2} \cdot 
        \left(
            J_c \mathbf{n}_\mu
            + J'_c \mathbf{n}_{\mu+2}
            + J''_c \mathbf{n}_{\mu+1}
        \right)
    \right]
    \notag \\
    &\hspace{50pt} = \frac{2}{3}
    \left[
        \left( J_c - \frac{J'_c}{2} - \frac{J''_c}{2} \right) \cos(\phi_{2j+2} - \phi_{2j+1}) 
        + \frac{\sqrt{3}}{2} (J'_c - J''_c) \sin(\phi_{2j+2} - \phi_{2j+1})
    \right].
\end{align}
By introducing the complex phase,
\begin{align}
    \varphi = \arg \left[
        \left( J_c - \frac{J'_c}{2} - \frac{J''_c}{2} \right)
        - i \frac{\sqrt{3}}{2}(J'_c - J''_c)
    \right],
\end{align}
we find that the interlayer interaction is minimized when
\begin{align}
    \phi_{2j+1} - \phi_{2j} &= \varphi + \pi,
    \notag\\
    \phi_{2j+2} - \phi_{2j+1} &= -\varphi + \pi,
\end{align}
which corresponds to a two-sublattice canted antiferroelectric structure with a net dipole moment.
Thus, the relative angle between the in-plane ferroelectric polarizations of even and odd layers generally becomes nonzero when $J'_c,\, J''_c \ne 0$. In most cases, the system may retain a net ferroelectric moment, but the partial cancellation can reduce the total magnitude.


\end{document}